\documentclass[11pt,a4paper]{article}
\pdfoutput=1

\usepackage{dsfont}
\usepackage{jheppub}
\usepackage{amsmath, latexsym, amssymb, graphicx, color, multirow, makecell, diagbox}
\usepackage{changepage}

\usepackage{tabularx,arydshln}
\usepackage[T1]{fontenc}

\usepackage[acronym]{glossaries-extra}
\makeglossaries

\setabbreviationstyle[acronym]{long-short}

\glssetcategoryattribute{acronym}{nohyperfirst}{true}

\newacronym{uv}{UV}{ultraviolet}
\newacronym{ckm}{CKM}{Cabibbo–Kobayashi–Maskawa}
\newacronym{rg}{RG}{renormalisation group}
\newacronym{mc}{MC}{multi-channeling}
\newacronym{nrqcd}{NRQCD}{non-relativistic QCD}
\newacronym{he}{HE}{High-Energy}
\newacronym{bfkl}{BFKL}{Balitsky-Fadin-Kuraev-Lipatov}
\newacronym{pnrqcd}{pNRQCD}{potential NRQCD}
\newacronym{pdf}{PDF}{parton distribution function}
\newacronym{ibp}{IBP}{Integration-By-Parts}
\newacronym{bphz}{BPHZ}{Bogoliubov–Parasiuk–Hepp–Zimmermann}
\newacronym{msbar}{$\overline{\text{MS}}$}{modified minimal subtraction}
\newacronym{fs}{FS}{Forward-Scattering}
\newacronym{kln}{KLN}{Kinoshita–Lee–Nauenberg}

\newacronym{ltd}{LTD}{Loop-Tree Duality}
\newacronym{cff}{CFF}{Cross-Free Family}
\newacronym{ir}{IR}{infrared}
\newacronym{qcd}{QCD}{quantum chromodynamics}
\newacronym{nlo}{NLO}{next-to-leading order}
\newacronym{nnlo}{NNLO}{next-to-next-to-leading order}
\newacronym{lo}{LO}{ leading order}
\newacronym{lu}{LU}{Local Unitarity}
\newacronym{ew}{EW}{electroweak}
\newacronym{nlp}{NLP}{next-to-leading power}
\newacronym{lp}{LP}{leading power}
\newacronym{lep}{LEP}{Large Electron-Positron Collider}
\newacronym{upc}{UPC}{ultraperipheral collisions}
\newacronym{lhc}{LHC}{Large Hadron Collider}
\newacronym{epa}{EPA}{equivalent photon approximation}
\newacronym{sm}{SM}{Standard Model}
\newacronym{bsm}{BSM}{Beyond the Standard Model}
\newacronym{cern}{CERN}{Conseil Européen pour la Recherche Nucléaire,first-style}
\glsunset{cern}
\glsunset{qcd}
\glsunset{lep}
\usepackage{dcolumn}
\usepackage{xspace}
\usepackage{comment}
\usepackage{booktabs} 
\usepackage{listings}
\lstdefinestyle{mystyle}{mathescape,basicstyle=\small\ttfamily,frame=leftline,aboveskip=4mm,belowskip=4mm,xleftmargin=20pt,framexleftmargin=10pt,numbers=none,framerule=2pt,abovecaptionskip=0.0mm,belowcaptionskip=3.5mm}
\usepackage{enumitem}
\usepackage{float}
\usepackage{epsfig}
\usepackage{longtable}
\usepackage{rotating}
\usepackage{ulem}
\usepackage{upgreek}
\usepackage{xcolor}
\usepackage{mathtools}
\usepackage{tabularx}
\usepackage{booktabs}
\usepackage{appendix}
\usepackage{pifont}% http://ctan.org/pkg/pifont
\usepackage{comment}

\usepackage{soul} % Striking text
\usepackage[capitalize]{cleveref}
\definecolor{nicered}{rgb}{.7,.1,.1}
\definecolor{nicegreen}{rgb}{.1,.5,.1}
\definecolor{darkblue}{rgb}{0,0,.5}
\hypersetup{colorlinks, citecolor=nicegreen,linkcolor=nicered, urlcolor=darkblue}
\usepackage{multirow}
\usepackage{slashed}
\usepackage{verbatim}

\usepackage{tikz}
\usetikzlibrary{positioning,arrows}
\usetikzlibrary{decorations.pathmorphing}
\usetikzlibrary{decorations.markings}
\usetikzlibrary{shapes.geometric}
\tikzset{
    vector/.style={decorate, decoration={snake}, draw},
    provector/.style={decorate, decoration={snake,amplitude=2.5pt}, draw},
    antivector/.style={decorate, decoration={snake,amplitude=-2.5pt}, draw},
    fermion/.style={draw=black, line width=0.4mm,
      postaction={decorate},decoration={markings,mark=at position .55
        with {\arrow[draw=black]{>}}}},
    fermionbar/.style={draw=black, postaction={decorate},
                       decoration={markings,mark=at position .55 with {\arrow[draw=black]{<}}}},
    fermionnoarrow/.style={draw=black},
    top/.style={draw=black, line width=1mm,
      postaction={decorate},decoration={markings,mark=at position .55
        with {\arrow[draw=black]{>}}}},
    gluon/.style={decorate, draw=black,decoration={coil,amplitude=4pt, segment length=5pt}},
    photon/.style={decorate, draw=red,decoration={snake,amplitude=5pt, segment length=9pt}},
    scalar/.style={dashed,draw=black,
      postaction={decorate},decoration={markings,mark=at position .55
        with {\arrow[draw=black]{>}}}},
    scalarbar/.style={dashed,draw=black,
      postaction={decorate},decoration={markings,mark=at position .55
        with {\arrow[draw=black]{<}}}},
    scalarnoarrow/.style={dashed,draw=black},
    electron/.style={draw=black,
      postaction={decorate},decoration={markings,mark=at position .55
        with {\arrow[draw=black]{>}}}},
    bigvector/.style={decorate, decoration={snake,amplitude=4pt}, draw},
}

\def\ttt#1{\texttt{\small #1}}

\newcommand{\red}[1]{\textcolor[rgb]{1,0,0}{#1}}

\newcommand{\halfcmark}{\textcolor{orange}{\ding{51}}{\small\textcolor{orange}{\kern-0.7em\ding{55}}}}

\newcommand{\pp}{p-p}

\providecommand{\pA}{p-A}
\providecommand{\pPb}{p-Pb}
\providecommand{\AaAa}{A-A}
\providecommand{\PbPb}{Pb-Pb}

\newcommand{\epem}{e^+e^-}

\newcommand{\mumu}{\mu^+\mu^-}

\newcommand{\ttbar}{t\bar{t}}
\newcommand{\bbbar}{b\bar{b}}
\newcommand{\ccbar}{c\bar{c}}

\newcommand{\gaga}{\gamma\gamma}
\newcommand{\QQ}{Q\bar{Q}}

\def\beq{\begin{equation}}
\def\eeq{\end{equation}}
\def\beqn{\begin{eqnarray}}
\def\eeqn{\end{eqnarray}}

\newcommand{\K}[1]{K_{\mathrm{A}_1\mathrm{A}_2\rightarrow {#1}}}

\usepackage{float}

\newcommand{\sqrtsnn}{\sqrt{s_{_\text{NN}}}}

\newcommand{\helaconia}{\textsc{HELAC-Onia}}
\newcommand{\madgraph}{\textsc{MadGraph5\_aMC@\gls{nlo}}}
\newcommand{\mgshort}{\textsc{MG5\_aMC}}
\newcommand{\gammaUPC}{\textsc{gamma-\gls{upc}}}
\newcommand{\phique}{\textsc{PHIQUE}}
\newcommand{\alphaLoop}{\textsc{$\alpha$Loop}}
\newcommand{\gammaLoop}{\textsc{$\gamma$Loop}}
\newcommand{\feyngen}{\textsc{FeynGen}}
\newcommand{\symbolica}{\textsc{Symbolica}}

\usepackage{xspace}
\newcommand*{\eg}{e.g.,\@\xspace}
\newcommand*{\ie}{i.e.,\@\xspace}

\newcommand*{\cm}{c.m.\@\xspace}

\renewcommand\arraystretch{1.5}% Tabular row height (1.0 is for standard spacing)

\allowdisplaybreaks
\begin{document}

%\title{Heavy-quark pair production in photon fusion at NNLO \gls{qcd} and \gls{nlo} \gls{ew} accuracy}
\title{
%Heavy quark pair production from photon fusion at
NNLO QCD corrections to $\gaga\to \QQ$ from Local Unitarity combined with Coulomb resummation and NLO EW effects
}
%\title{Local Unitarity for NNLO \gls{qcd} corrections to the Coulomb-resummed process $\gaga\to \QQ$ with \gls{nlo} \gls{ew} contributions}

\author{Zeno Capatti $^{a}$, Mathijs Fraaije $^{a}$, Valentin Hirschi $^{a}$, Lucien Huber $^{a}$, Ben Ruijl $^{b}$, Hua-Sheng Shao $^{c}$}
\emailAdd{zeno.capatti@unibe.ch, mathijsfraaije@gmail.com, valentin.hirschi@gmail.com, mail@lucien.ch, ben@symbolica.io, huasheng.shao@lpthe.jussieu.fr}
\affiliation{$^a$Institute for Theoretical Physics, University of Bern, Sidlerstrasse 5, 3012 Bern, Switzerland
\\$^b$ Ruijl Research, Chamerstrasse 117, 6300 Zug, Switzerland
\\$^c$Laboratoire de Physique Th\'eorique et Hautes Energies (LPTHE), UMR 7589, Sorbonne Universit\'e et CNRS, 4 place Jussieu, 75252 Paris Cedex 05, France
}

\preprint{}

% ********** Abstract **********

\abstract{
The \gls{lu} formalism provides a constructive, integrand‐level realisation of the \gls{kln} theorem, by combining loop and phase-space integrals appearing in scattering cross-sections in such a way that their final-state infrared singularities cancel before integration. Supplemented with localised ultraviolet renormalisation, it enables the direct Monte Carlo integration of cross sections at arbitrary perturbative order in four-dimensional spacetime.
In this paper, we present its application to the \gls{nnlo} \gls{qcd} total cross sections for heavy-quark pair production in direct photon fusion, involving the contribution from 138 distinct forward-scattering diagrams where external photons couple only to heavy quarks.
By combining \gls{nnlo} \gls{qcd} with \gls{nlo} \gls{ew} corrections and \gls{nlp} Coulomb resummation, we obtain state-of-the-art predictions for top-, bottom-, and charm-quark production in ultraperipheral hadron collisions and at $\epem$ colliders.
}

\maketitle

\newpage
\section{Introduction}

Heavy quark pair production is one of the most fundamental and extensively studied processes in high-energy collider physics. Since the heavy quark mass $m_Q$ is much larger than the intrinsic \gls{qcd} scale $\Lambda_{\mathrm{\gls{qcd}}}$, the corresponding production cross section is believed to be reliably computed within perturbative \gls{qcd}, owing to asymptotic freedom~\cite{Gross:1973id,Politzer:1973fx}.

In this paper, we study heavy-quark pair production processes in photon-photon collisions, $\gaga\to \QQ$, where the heavy quark $Q$ can be a charm ($c$), bottom ($b$), or top ($t$) quark. The colliding photons can behave either as a point-like particles or exhibit hadronic structure. In the latter case, the photon fluctuates into a hadronic state, and one of its partonic constituents, a (anti-)quark or a gluon, participates in the hard interaction. Accordingly, three production channels are identified: direct (both photons are point-like), single-resolved (one point-like and one hadron-like), and double-resolved (both photons are hadron-like). At higher orders in perturbation theory, however, these channels cannot be distinguished sharply, as quantum corrections induce mixing among them. As a consequence, the distinction between direct and resolved contributions becomes scheme-dependent, and thus unphysical, when considering incoming photons with finite virtuality.

The production of open charm and bottom quarks through photon fusion, $\gaga\to \ccbar$ and $\gaga\to \bbbar$, has been extensively studied at high-energy $\epem$ colliders, most notably at \gls{cern}'s \gls{lep}. Measurements were carried out by the JADE~\cite{JADE:1986pku}, TASSO~\cite{TASSO:1990api}, TPC/$2\gamma$~\cite{TPCTwo-Gamma:1990lmu}, TOPAZ~\cite{TOPAZ:1993pxd,TOPAZ:1994olc,TOPAZ:1994fti}, VENUS~\cite{VENUS:1993erd}, AMY~\cite{AMY:1995rmf,AMY:1996tcc}, ALEPH~\cite{ALEPH:1995xym,ALEPH:2003ycd,ALEPH:2007pfw}, DELPHI~\cite{DaSilva:2004km}, L3~\cite{L3:1998hya,L3:1999vqn,L3:2000oud,L3:2000rpp,L3:2002itq,L3:2005jyf}, and OPAL~\cite{OPAL:1999peo} collaborations at center-of-mass (\cm) energies $\sqrt{s_{\epem}}$ ranging from $29$ to $209$ GeV. At LEP energies, direct and single-resolved channels are predicted to contribute comparably to the heavy-quark production cross section~\cite{Drees:1992eh}, while at lower energies the direct process dominates. Contributions from double-resolved processes are expected to be small. Since the single-resolved process is dominated by photon-gluon ($\gamma g$) fusion, cross-section measurements provide sensitivity to the gluon content of the photon. 
Bottom-quark production is expected to be suppressed by about two orders of magnitude relative to charm production, due to its larger mass and smaller absolute charge ($|e_Q|$, with $e_c=2/3$ versus $e_b=-1/3$).

At LEP, however, the measured cross section for $\gaga\to \bbbar$, including both direct and single-resolved contributions, was found by the L3~\cite{L3:2000oud,L3:2005jyf} and DELPHI~\cite{DaSilva:2004km} collaborations to be roughly three times larger than the \gls{nlo} \gls{qcd} prediction~\cite{Drees:1992eh}, corresponding to a discrepancy of about three standard deviations. Moreover, tensions exceeding two standard deviations exist between the most recent L3~\cite{L3:2005jyf} and ALEPH~\cite{ALEPH:2007pfw} measurements, with the latter being consistent with the \gls{nlo} \gls{qcd} calculation. This discrepancy has motivated further studies, including proposals~\cite{Chyla:2003tg} to evaluate the direct photon contribution at \gls{nnlo} in \gls{qcd}\footnote{Similar computations for hadroproduction exist in the literature~\cite{Czakon:2013goa,Czakon:2015owf,Czakon:2016ckf,Czakon:2016dgf,Catani:2019iny,Catani:2019hip,Catani:2020tko,Catani:2020kkl}.}.
It would therefore be highly interesting to revisit these measurements at future $\epem$ colliders with significantly higher luminosities.

Another avenue for studying heavy-quark production in two-photon collisions is provided by hadron colliders. The electromagnetic field of a relativistic charged particle, such as a proton or a heavy ion at the \gls{lhc}, can be interpreted as a flux of quasi-real photons~\cite{Budnev:1975poe}. The photon energy $E_\gamma$ scales with the Lorentz factor, while the photon number density $N_\gamma$ scales with the squared charge of the beam particle, $Z^2$. Studies of $\gaga\to\QQ$ processes can therefore be carried out in proton-proton (\pp), proton-nucleus (\pA), and nucleus-nucleus (\AaAa) \gls{upc}, namely when, qualitatively, the transverse separation between the two colliding hadron beams exceeds the sum of their radii. In this regime, the beam particles remain intact--aside from possible ion excitations leading to forward neutron emission~\cite{Crepet:2024dvv}. As a result, \gls{upc}s yield
exceptionally clean event topologies, with final-state particles produced centrally and nearly empty forward regions. These forward rapidity gaps can be tagged using Roman Pots and Zero Degree Calorimeters for protons and heavy ions, respectively. Since the quasi-real photons emitted in \gls{upc}s have negligible virtualities compared to the energy scale of the hard process, the \gls{epa}~\cite{Fermi:1924tc,vonWeizsacker:1934nji,Williams:1934ad} provides a reliable description.
The results we present in this paper are based on this approximation.

The motivations for studying $\gaga\to\QQ$ in \gls{upc}s are multifold. For charm and bottom quarks, this process provides a clean testing ground for heavy-flavour jet tagging algorithms and for testing the universality of heavy-quark fragmentation functions. The latter has attracted attention recently, given the observed baryon-to-meson enhancements~\cite{ALICE:2020wla,ALICE:2020wfu,ALICE:2021psx,ALICE:2021dhb,ALICE:2021rzj,ALICE:2021npz,ALICE:2022cop,ALICE:2022exq,ALICE:2023sgl,ALICE:2024lon,ALICE:2025wrq} for charmed hadrons in \pp\ collisions at the \gls{lhc} relative to those in $\epem$ and $ep$ collisions. For top quarks, it offers a unique opportunity to probe colour-singlet exchange in an unexplored kinematic regime. For example, spin correlations in $\gaga\to \ttbar$ differ from those in the inclusive reaction $pp\to \ttbar+X$, due to the very different initial-state partons. Near threshold, Coulomb-gluon enhancement is dominated by the colour-singlet channel, in contrast to the inclusive case~\cite{ATLAS:2023fsd,CMS:2024pts,CMS:2025kzt,ATLAS:2025mvr,CMS:2025dzq}. The process also delivers complementary information to inclusive reactions such as $pp\to \ttbar\gamma+X$ and $pp\to \ttbar\gaga+X$, useful for testing the interaction between top quarks and photons and for determining top-quark properties, including the electric charge and electromagnetic dipole moments~\cite{Fayazbakhsh:2015xba}. In addition, $\gaga\to \QQ$ is of particular interest in searches for  \gls{bsm} physics. It provides a sensitive probe of flavour-changing neutral current interactions~\cite{Howarth:2020uaa}, anomalous top-photon couplings (\eg\ quartic dimension-$8$ anomalous interactions)~\cite{Baldenegro:2022kaa}, new resonances~\cite{Baldenegro:2022kaa}, and extra dimension models~\cite{Inan:2011zz}. Furthermore, $\gaga\to\QQ$ processes constitute irreducible backgrounds to \gls{qcd} pomeron-induced production, whose theoretical modelling carries large uncertainties~\cite{Howarth:2020uaa}.

A first measurement of $\gaga\to \ttbar$ was carried out by the CMS-TOTEM collaboration~\cite{CMS:2023naq} at a \cm\ energy of $13$ TeV in \pp\ collisions at the \gls{lhc}. Tagging two forward intact protons with the precision proton spectrometer yields only an upper bound on the production cross section, based on data with an integrated luminosity of 29.4 fb$^{-1}$. The experimental observation of $\gaga\to \ttbar$ is to be expected at the high-luminosity \gls{lhc}~\cite{CMS:2021ncv,Goncalves:2020saa,Martins:2022dfg}. 

In this paper, we restrict ourselves to the direct production channel $\gaga\to\QQ$. As mentioned earlier, this contribution ceases to be a physical observable starting from \gls{nnlo} in \gls{qcd}. Nevertheless, if one ignores the interaction between photons and massless quarks, it remains unambiguously defined and does not mix with the resolved contributions at \gls{nnlo}. The actual size of the resolved contributions will depend on the energy of the scattering process considered. At the LHC, there are however experimental ways to mitigate contamination from
resolved-photon contributions by vetoing events with higher particle multiplicities in the forward detectors. Overall, a full assessment of the magnitude of the resolved contributions depends on
the details of the differential observables considered and is therefore beyond the scope of this work.

Finally, we discuss the novelty of the  \gls{lu} computational approach, first introduced in refs.~\cite{Capatti:2020xjc,Capatti:2021bsm,Capatti:2022tit} and applied here to the \gls{nnlo} \gls{qcd} corrections to $\gaga\to\QQ$.
The traditional semi-analytical workflow treats loop and phase-space integrals separately\footnote{Even numerical approaches such as using \texttt{pySecDec}~\cite{Borowka:2017idc} consider loop amplitudes separately, \eg\ see ref.~\cite{Heinrich:2023til}}. This standard route dimensionally regularises loop integrals, reduces them via \gls{ibp} identities, and solves the ensuing differential equations (often semi-numerically). Phase-space integrals are then integrated by Monte Carlo methods after their infrared singularities are subtracted by suitable counter-terms.
Instead, \gls{lu} targets the full cross-section directly through the momentum-space Monte Carlo integration of a single integrand:

\begin{itemize}
\item The phase-space integral of the squared amplitude
is expressed in terms of interference diagrams, by applying cutting rules to  \gls{fs} diagrams. The \gls{fs} diagram is obtained by gluing together the final states of the Feynman
diagrams arising from the amplitude and its conjugate. Cross-sections are then organised in terms of infrared-finite classes of these interference diagrams, each class being defined by the \gls{fs} diagram common to all the interference diagrams contained within it.
\item Loop and phase-space integrations of interference diagrams are expressed in terms of a common measure over all spatial loop momenta of the \gls{fs} diagram, leveraging the \gls{ltd}~\cite{LTDRodrigoOrigin2008, LTDRodrigoMultiLoop2010, Capatti:2019ypt, Runkel:2019yrs, Capatti:2020ytd, JesusAguilera-Verdugo:2020fsn, Aguilera-Verdugo:2020set, Berghoff:2020bug,Capatti:2022mly}.
\item Final-state \gls{ir} singularities cancel locally in each class~\cite{Capatti:2020xjc}, without dimensional regularisation or subtraction counter-terms, by the \gls{kln} theorem~\cite{Kinoshita:1962ur, Lee:1964is}.
\item \Gls{uv} singularities are subtracted locally using the \\
\gls{bphz}~\cite{Bogoliubov:1957gp, Zimmermann:1969jj, Hepp:1966eg, Herzog:2017jgk} formalism; the resulting cross-section is directly renormalised in the \gls{msbar} scheme for couplings and the On-Shell (OS) scheme for massive fermions~\cite{Capatti:2022tit}.
\item Threshold singularities in \gls{lu} cancel locally for completely\footnote{By ``completely'' inclusive we mean that all Cutkosky cuts of each \gls{fs} graph are included without any fiducial cut.} inclusive quantities, a property we exploit for the non-singlet \gls{nnlo} \gls{qcd} contributions to $\gaga\to\QQ$. For the singlet contribution, and more in general, for (semi-)differential cross-sections, the \gls{lu} integrand requires a form of threshold regularisation~\cite{ContourDeformation2020,Kermanschah:2021wbk,Vicini:2024ecf}.
\end{itemize}
The above highlights \gls{lu}'s departure from the traditional paradigm: it shares with state-of-the-art methods only the endpoint, namely the fixed-order prediction.
One core differentiating factor of \gls{lu} is that it yields finite, albeit unphysical, cross-sections for individual \gls{fs} graphs (see appendix~\ref{sec:graph_distribution}), so that discrete importance sampling can apply to their sum. 
Due to its innovative aspects, \gls{lu} has required substantial advances in both theory~\cite{ContourDeformation2020,Capatti:2020xjc} and implementation~\cite{Capatti:2022tit}. Proof-of-principle results—\eg\ partial next-to-\gls{nnlo} (N$^3$LO) corrections to the R-ratio~\cite{Capatti:2022tit} and two-loop \gls{nlo} corrections\footnote{In this case, the computation of the inclusive cross-section for the process $\gamma \gamma \rightarrow \gamma \gamma$ only involves a single Cutkosky cut per \gls{fs} graph, so that it is more akin to simple amplitude calculation, even though loop and phase-space degrees of freedom were still integrated over simultaneously using the causal flow.
In particular, we note that unlike for the process computed in this work, $\gamma \gamma \rightarrow \gamma \gamma$ does not feature the mechanism of local cancellation of infrared singularities across multiple cuts of the same \gls{fs} graph.} to $\gamma \gamma \rightarrow \gamma \gamma$~\cite{AH:2023kor}—demonstrate soundness, but do not alone establish viability as a predictive tool at today’s computational frontier.
This work concludes the proof of feasibility consisting in applying \gls{lu}, for the first time, to the previously unknown \gls{nnlo} \gls{qcd} corrections to $\gaga\to\QQ$, a semi-inclusive process that stress-tests the core \gls{lu} mechanisms.
In parallel, we are developing a new high-performance automated implementation~\cite{Borinsky:2025asc, spenso, Symbolica} in a new public code dubbed $\gamma\text{Loop}$~\cite{gammaloop}, and a first investigation of \gls{lu} extensions to initial-state singularities~\cite{Capatti:2025gqj}. 

The rest of the paper is organised as follows. Section~\ref{sec:notation} introduces the general formalism for computing cross-sections for the scattering of two photons emitted from two charged sources. Section~\ref{sec:LU} discusses the application of the \gls{lu} method to compute the \gls{nnlo} \gls{qcd} corrections to $\gaga \to \QQ$. In particular, section~\ref{sec:LU_general} gives a brief and general review of the Local Unitarity formalism and its application to $2\rightarrow N$ scattering processes. We then specialise the discussion to the construction of the Local Unitarity integrand for the computation of \gls{nnlo} \gls{qcd} corrections to $\gaga\to \QQ$ in sections~\ref{sec:LU_graph_enum} and~\ref{sec:LU_2_body}. We discuss the tests we performed to ensure the correcteness of the \gls{lu} construction in section~\ref{sec:tests}. In section~\ref{sec:results}, we present and discuss the results for the Coulomb-resummed cross-section for $\gaga\to \QQ$ including the \gls{nnlo} \gls{qcd} corrections computed within \gls{lu} as well as the \gls{nlo} \gls{ew} results computed within \madgraph~\cite{Alwall:2014hca,Frederix:2018nkq,Shao:2025bma}. In the appendices, we
show the visualisation of the relative contribution of individual \gls{fs} diagrams for each value of the \cm\ energy of the colliding photon pair and discuss runtime performance (appendix~\ref{sec:graph_distribution}), then provide a discussion of the supplementary material attached with this paper (appendix~\ref{sec:supplementary_material}), and finally give a brief overview of the \phique~\cite{phique} code that collects all the components of the partonic cross-section and convolves them with appropriate photon fluxes (appendix~\ref{sec:phique}).

\section{Photon fluxes and notations\label{sec:notation}}

We consider the production of a heavy quark-antiquark pair (or a $\QQ$ pair) from the collision of two photons emitted from two distinct sources generically denoted as $\mathrm{A}_1$ and $\mathrm{A}_2$. In the photon collinear factorisation formalism, the physical cross section can be written as
\begin{equation}
\sigma(\mathrm{A}_1\mathrm{A}_2\overset{\gaga}{\to} \QQ+X)=\int^1_{4m_Q^2/S}{\mathrm{d}\tau \mathcal{L}^{(\mathrm{A}_1\mathrm{A}_2)}_{\gaga}(\tau)\hat{\sigma}_{\gaga\to \QQ+X_p}(s=\tau S,m_Q^2)}\,,\label{eq:xsinAB}
\end{equation}
where $\hat{\sigma}_{\gaga\to \QQ+X_p}$ is the partonic cross section, $\mathcal{L}^{(\mathrm{A}_1\mathrm{A}_2)}_{\gaga}(\tau)$ is the two-photon luminosity function, and $\sqrt{S}$ and $\sqrt{s}$ denote the \cm\ energies of the initial charged particles $\mathrm{A}_1$ and $\mathrm{A}_2$ and of the two initial-state photons, respectively. The symbol $X_p$ represents any parton radiation in the final state.

The precise definition of $\mathcal{L}^{(\mathrm{A}_1\mathrm{A}_2)}_{\gaga}(\tau)$ depends on the properties of $\mathrm{A}_1$ and $\mathrm{A}_2$. In this paper, we consider the following three possible cases:
\begin{itemize}
\item \textbf{Photon beams as initial particles}: If the beam particles are just the initial-state photons, then
\begin{equation}
\mathcal{L}^{(\gaga)}_{\gaga}(\tau)=\delta(1-\tau)\,.
\end{equation}
In this case, eq.~\eqref{eq:xsinAB} reduces to $X=X_p$.
\item \textbf{Charged hadron beams in \gls{upc}s}: Here the beam particles are charged hadrons (\eg\ protons and ions) with a transverse separation greater than twice their radii. The corresponding luminosity function~\cite{Shao:2022cly} reads
\begin{equation}
\mathcal{L}^{(\mathrm{A}_1\mathrm{A}_2)}_{\gaga}(\tau)=\int_0^1{\frac{\mathrm{d}x_1}{x_1}\frac{\mathrm{d}x_2}{x_2}\delta(x_1x_2-\tau)\frac{\mathrm{d}^2N^{(\mathrm{A}_1\mathrm{A}_2)}_{\gamma_1/\mathrm{Z}_1,\gamma_2/\mathrm{Z}_2}}{\mathrm{d}E_{\gamma_1}\mathrm{d}E_{\gamma_2}}}\,.\label{eq:lum4UPC}
\end{equation}
where $E_{\gamma_i}$ is the energy of the $i$-th photon, $Z_{1,2}$ are the charge numbers of $\mathrm{A}_{1,2}$, and the longitudinal fractions are $x_i=A_i E_{\gamma_i}/E_{\mathrm{A}_i}$ with $A_i$ ($E_{\mathrm{A}_i}$) the atomic mass number (total energy) of $\mathrm{A}_i$. Using this notation, the nucleon-nucleon \cm\ energy is $\sqrt{S}=\sqrtsnn\approx \sqrt{4E_{\mathrm{A}_1}E_{\mathrm{A}_2}/A_1A_2}$. The effective photon-photon luminosity can be expressed as a convolution of the two photon number densities, $N_{\gamma_1/\mathrm{Z}_1}(E_{\gamma_1},\pmb{b}_1)$ and $N_{\gamma_2/\mathrm{Z}_2}(E_{\gamma_2},\pmb{b}_2)$, at impact parameters $\pmb{b}_{1,2}$ with respect to the centers of hadrons $\mathrm{A}_1$ and $\mathrm{A}_2$:
\begin{equation}
\frac{\mathrm{d}^2N^{(\mathrm{A}_1\mathrm{A}_2)}_{\gamma_1/\mathrm{Z}_1,\gamma_2/\mathrm{Z}_2}}{\mathrm{d}E_{\gamma_1}\mathrm{d}E_{\gamma_2}} =  \int{\mathrm{d}^2\pmb{b}_1\mathrm{d}^2\pmb{b}_2\,P_\text{no\,inel}(\pmb{b}_1,\pmb{b}_2)\,N_{\gamma_1/\mathrm{Z}_1}(E_{\gamma_1},\pmb{b}_1)N_{\gamma_2/\mathrm{Z}_2}(E_{\gamma_2},\pmb{b}_2)}\,,\label{eq:2photonintegral}
\end{equation}
where $P_\text{no\,inel}(\pmb{b}_1,\pmb{b}_2)$ accounts for the probability of no inelastic hadronic interactions between A$_1$ and A$_2$. In this paper, we use the \gammaUPC\ code~\cite{Shao:2022cly} to evaluate the two-photon luminosity in \gls{upc}s. The code provides two types of coherent photon fluxes as functions of the impact parameter: the electric-dipole (EDFF) and charge (ChFF) form factors for protons and nuclei. The ChFF flux is preferred for at least two reasons. First, the EDFF photon number density diverges at vanishing impact parameters, necessitating an arbitrary cutoff (typically $b=|\pmb{b}|\gtrsim R_{\mathrm{A}}$, with $R_{\mathrm{A}}$ the radius of $\mathrm{A}$) to regulate the integral in eq.~\eqref{eq:2photonintegral}. In contrast, the ChFF flux remains well behaved for all $b$, making it more realistic. Second, cross sections computed with the ChFF flux agree better with experimental results for dilepton production processes such as $\gaga\to \epem$~\cite{CMS:2024bnt} and $\gaga\to \mumu$~\cite{Shao:2024dmk}. Therefore, we adopt the ChFF flux throughout this paper for \gls{upc}s\footnote{Since the main difference between ChFF and EDFF photon fluxes is that the former allows the hard reaction to occur inside the proton or nucleus, it would be interesting to see from experiments whether some additional hadronic interaction effects, such as interactions between the beam particle and final-state quarks, potentially break the proton/ion apart. These effects go beyond the beam hadron survival factor $P_\text{no\,inel}(\pmb{b}_1,\pmb{b}_2)$ implemented in eq.~\eqref{eq:2photonintegral}. We thank David d'Enterria for pointing this out.}. 
Finally, since the two beam particles are intact (aside from possible ion excitations leading to forward neutral hadron emission~\cite{Crepet:2024dvv}), we have $X=\mathrm{A}_1~X_p~\mathrm{A}_2$ in eq.~\eqref{eq:xsinAB}.
We note that the parametric uncertainties in the photon flux modelling have been assessed to be at the percent level in ref.~\cite{Shao:2024dmk}, and are shown in figure 2 therein.

\item \textbf{Lepton beams}: Motivated by past, running, and future electron-positron ($\epem$) colliders and a possible future muon collider, the colliding particles can also be two same-flavour, opposite-sign leptons $\ell^+\ell^-$, where $\ell=e$ or $\mu$. In this case, we have $X=\ell^+~X_p~\ell^-$ in eq.~\eqref{eq:xsinAB}, and the luminosity function is
\begin{equation}
\mathcal{L}_{\gaga}^{(\ell^+\ell^-)}(\tau)=\int_0^1{\mathrm{d}x_1\mathrm{d}x_2 \delta(x_1x_2-\tau)f_\gamma^{(\ell^+)}(x_1,Q_{\mathrm{max}}^2)f_\gamma^{(\ell^-)}(x_2,Q_{\mathrm{max}}^2)}\,.
\end{equation}
The photon \gls{pdf} of an elementary lepton can be approximated in the improved Weizs\"acker-Williams (iWW) form~\cite{Frixione:1993yw}:
\begin{equation}
f_\gamma^{(\ell^{\pm})}(x,Q_{\mathrm{max}}^2)=\frac{\alpha}{2\pi}\Bigg[\frac{1+(1-x)^2}{x}\log{\left(\frac{(1-x)Q_{\mathrm{max}}^2}{x^2m_\ell^2}\right)}+\bigg(\frac{2x m_\ell^2}{Q_{\mathrm{max}}^{2}}-\frac{2(1-x)}{x}\bigg)\Bigg]\,,\label{eq:WWPDF4lepton}
\end{equation}
where $\alpha$ is the fine-structure constant and $m_\ell$ is the lepton mass. Here, $\sqrt{S}=\sqrt{s_{\ell^+\ell^-}}$. Since the photon flux decreases rapidly with photon virtuality, we set the maximal photon virtuality to $Q_{\mathrm{max}}=1$ GeV in this work\footnote{The value of $Q_{\mathrm{max}}$ can be converted into angular cuts on the outgoing lepton beams (cf. eq.~(26) in ref.~\cite{Frixione:1993yw}).}. More accurate photon \gls{pdf}s for electrons and muons can be found in refs.~\cite{Frixione:2019lga,Bertone:2019hks,Bertone:2022ktl,Frixione:2023gmf,Stahlhofen:2025hqd,Schnubel:2025ejl}. The resulting uncertainties could be quantitatively relevant and should be the object of future work. Our calculations, however, use eq.~\eqref{eq:WWPDF4lepton}, as \gls{qcd} $K$ factors are expected to be largely insensitive to the precise choice of photon \gls{pdf}.
\end{itemize}

The partonic cross section can be expanded in a double series in the small couplings $\alpha_s$ and $\alpha$ of the \gls{sm}:
\begin{equation}
\hat{\sigma}_{\gaga\to \QQ+X_p}(s,m_Q^2)=\hat{\sigma}_Q^{(0,2)}+\hat{\sigma}_Q^{(1,2)}+\hat{\sigma}_Q^{(0,3)}+\hat{\sigma}_Q^{(2,2)}+\mathcal{O}(\alpha^4,\alpha_s\alpha^3,\alpha_s^3\alpha^2)\,,\label{eq:pxsexpansion}
\end{equation}
where $\hat{\sigma}_Q^{(i,j)}$ denotes the contribution proportional to $\alpha_s^i \alpha^j$. The \gls{lo} cross section $\hat{\sigma}_Q^{(0,2)}$ is purely \gls{ew} and independent of $\alpha_s$ (cf. eq.~\eqref{eq:LOpxs}). The \gls{nlo} \gls{qcd} and \gls{ew} corrections, $\hat{\sigma}_Q^{(1,2)}$ and $\hat{\sigma}_Q^{(0,3)}$, are known from refs.~\cite{Drees:1992eh,Cacciari:1995ej,Kamal:1998wz,Frixione:1999if,Kramer:2001gd,Kramer:2003cw,Kniehl:2009kh,Shao:2022cly} and ref.~\cite{Shao:2025bma}, respectively. Their implementation for \gls{upc}s have been automated in ref.~\cite{Shao:2025bma} within the \madgraph\ (\mgshort\ hereafter) framework~\cite{Alwall:2014hca,Frederix:2018nkq}. Taking top-quark pair production in \pp\ \gls{upc}s at the \gls{lhc} as an example, the \gls{nlo} \gls{qcd} correction ($\hat{\sigma}_Q^{(1,2)}$) enhances the \gls{lo} total cross section by about $20\%$~\cite{Shao:2022cly}, while the \gls{nlo} \gls{ew} correction ($\hat{\sigma}_Q^{(0,3)}$) reduces it by approximately $5\%$~\cite{Shao:2025bma}.
The main novelty of this paper is the computation of the \gls{nnlo} \gls{qcd} corrections, $\hat{\sigma}_Q^{(2,2)}$, which were previously unknown and now established to be increasing the cross-section by $6\%$ (see table~\ref{tab:ttbarxsQCD} in section~\ref{sec:topres}).

%\newpage

In the following, we also introduce shorthand notations to present our results:
\begin{eqnarray}
\sigma_{\mathrm{A}_1\mathrm{A}_2\to\QQ}^{(\mathrm{\gls{lo}})}&=&\int^1_{4m_Q^2/S}{\mathrm{d}\tau \mathcal{L}_{\gaga}^{(\mathrm{A}_1\mathrm{A}_2)}(\tau)\hat{\sigma}_Q^{(0,2)}}\,,\nonumber\\
\Delta\sigma_{\mathrm{A}_1\mathrm{A}_2\to\QQ}^{(\mathrm{\gls{nlo}~\gls{qcd}})}&=&\int^1_{4m_Q^2/S}{\mathrm{d}\tau \mathcal{L}_{\gaga}^{(\mathrm{A}_1\mathrm{A}_2)}(\tau)\hat{\sigma}_Q^{(1,2)}}\,,\nonumber\\
\Delta\sigma_{\mathrm{A}_1\mathrm{A}_2\to\QQ}^{(\mathrm{\gls{nlo}~\gls{ew}})}&=&\int^1_{4m_Q^2/S}{\mathrm{d}\tau \mathcal{L}_{\gaga}^{(\mathrm{A}_1\mathrm{A}_2)}(\tau)\hat{\sigma}_Q^{(0,3)}}\,,\nonumber\\
\Delta\sigma_{\mathrm{A}_1\mathrm{A}_2\to\QQ}^{(\mathrm{\gls{nnlo}~\gls{qcd}})}&=&\int^1_{4m_Q^2/S}{\mathrm{d}\tau \mathcal{L}_{\gaga}^{(\mathrm{A}_1\mathrm{A}_2)}(\tau)\hat{\sigma}_Q^{(2,2)}}\,,\nonumber\\
\sigma_{\mathrm{A}_1\mathrm{A}_2\to\QQ}^{(\mathrm{\gls{nlo}~\gls{qcd}})}&=&\sigma_{\mathrm{A}_1\mathrm{A}_2\to\QQ}^{(\mathrm{\gls{lo}})}+\Delta\sigma_{\mathrm{A}_1\mathrm{A}_2\to\QQ}^{(\mathrm{\gls{nlo}~\gls{qcd}})}\,,\nonumber\\
\sigma_{\mathrm{A}_1\mathrm{A}_2\to\QQ}^{(\mathrm{\gls{nnlo}~\gls{qcd}})}&=&\sigma_{\mathrm{A}_1\mathrm{A}_2\to\QQ}^{(\mathrm{\gls{lo}})}+\Delta\sigma_{\mathrm{A}_1\mathrm{A}_2\to\QQ}^{(\mathrm{\gls{nlo}~\gls{qcd}})}+\Delta\sigma_{\mathrm{A}_1\mathrm{A}_2\to\QQ}^{(\mathrm{\gls{nnlo}~\gls{qcd}})}\,,\nonumber\\
\sigma_{\mathrm{A}_1\mathrm{A}_2\to\QQ}^{(\mathrm{\gls{nlo}~\gls{ew}})}&=&\sigma_{\mathrm{A}_1\mathrm{A}_2\to\QQ}^{(\mathrm{\gls{lo}})}+\Delta\sigma_{\mathrm{A}_1\mathrm{A}_2\to\QQ}^{(\mathrm{\gls{nlo}~\gls{ew}})}\,,\nonumber\\
\sigma_{\mathrm{A}_1\mathrm{A}_2\to\QQ}^{(\mathrm{\gls{nlo}~\gls{qcd}+\gls{ew}})}&=&\sigma_{\mathrm{A}_1\mathrm{A}_2\to\QQ}^{(\mathrm{\gls{lo}})}+\Delta\sigma_{\mathrm{A}_1\mathrm{A}_2\to\QQ}^{(\mathrm{\gls{nlo}~\gls{qcd}})}+\Delta\sigma_{\mathrm{A}_1\mathrm{A}_2\to\QQ}^{(\mathrm{\gls{nlo}~\gls{ew}})}\,,\nonumber\\
\sigma_{\mathrm{A}_1\mathrm{A}_2\to\QQ}^{(\mathrm{\gls{nnlo}~\gls{qcd}+\gls{ew}})}&=&\sigma_{\mathrm{A}_1\mathrm{A}_2\to\QQ}^{(\mathrm{\gls{lo}})}+\Delta\sigma_{\mathrm{A}_1\mathrm{A}_2\to\QQ}^{(\mathrm{\gls{nlo}~\gls{qcd}})}+\Delta\sigma_{\mathrm{A}_1\mathrm{A}_2\to\QQ}^{(\mathrm{\gls{nlo}~\gls{ew}})}+\Delta\sigma_{\mathrm{A}_1\mathrm{A}_2\to\QQ}^{(\mathrm{\gls{nnlo}~\gls{qcd}})}\, \nonumber \\
\K{\QQ}^{(\ldots)}&=&\sigma_{\mathrm{A}_1\mathrm{A}_2\to \QQ}^{(\ldots)}\;/\;\sigma_{\mathrm{A}_1\mathrm{A}_2\to \QQ}^{(\text{\gls{lo}})}\, \label{eq:xsdefs1}.
%\Knlo{\QQ}&=&\frac{\sigma_{\mathrm{A}_1\mathrm{A}_2\to \QQ}^{(\mathrm{\gls{nlo}}\ \mathrm{\gls{qcd}})}}{\sigma_{\mathrm{A}_1\mathrm{A}_2\to \QQ}^{(\mathrm{\gls{lo}})}}\, \nonumber \\
%\Knnlo{\QQ}&=&\displaystyle\frac{\sigma_{\mathrm{A}_1\mathrm{A}_2\to \QQ}^{(\mathrm{\gls{nnlo}}\ \mathrm{\gls{qcd}})}}{\sigma_{\mathrm{A}_1\mathrm{A}_2\to \QQ}^{(\mathrm{\gls{lo}})}}\nonumber \\
%\KnloEW{\QQ}&=&\frac{\sigma_{\mathrm{A}_1\mathrm{A}_2\to \QQ}^{(\mathrm{\gls{nlo}}\ \mathrm{\gls{qcd}+\gls{ew}})}}{\sigma_{\mathrm{A}_1\mathrm{A}_2\to \QQ}^{(\mathrm{\gls{lo}})}}\, \nonumber \\
%\KnnloEW{\QQ}&=&\displaystyle\frac{\sigma_{\mathrm{A}_1\mathrm{A}_2\to \QQ}^{(\mathrm{\gls{nnlo}}\ \mathrm{\gls{qcd}+\gls{ew}})}}{\sigma_{\mathrm{A}_1\mathrm{A}_2\to \QQ}^{(\mathrm{\gls{lo}})}}.\label{eq:xsdefs1}
\end{eqnarray}

%\section{Computational approach: Local Unitarity}
\section{Setup of the computation within the Local Unitarity formalism}
\label{sec:LU}
% The old tex file
%\input{LU.tex}

% The one amended by HSS
We compute the \gls{nnlo} \gls{qcd} correction term $\hat{\sigma}_Q^{(2,2)}$ to the $\gaga \to \QQ$ cross section fully numerically using \gls{lu}~\cite{Capatti:2020xjc,Capatti:2022tit} as implemented in a custom version of the code \alphaLoop~\cite{alphaloop}, which has already been used in refs.~\cite{Capatti:2022tit,AH:2023kor}. In this section, we present the \gls{lu} formalism in generality (section~\ref{sec:LU_general}), then we discuss the details of the construction of the \gls{lu} integrand for the \gls{nnlo} \gls{qcd} corrections to $\gaga \to \QQ$ (section~\ref{sec:LU_2_body}) and finally describe how we tested it (section~\ref{sec:tests}).

%The core {\color{red} BR: unfinished sentence}

\subsection{Local Unitarity formalism for $2\to N$ processes}
\label{sec:LU_general}

We start by presenting a short review of the Local Unitarity formalism. It allows to write the integrand for a generic $2\to N$ process in a way that manifests the cancellation of final-state singularities predicted by the \gls{kln} theorem at the integrand level. Initial-state singularities remain unregulated but, as we will discuss in the next section, they depend on the photon collinear factorisation scheme. Our discussion starts by reviewing the well-known fact that the computation of a cross-section can be re-framed in terms of the cuts of \gls{fs} diagrams (section~\ref{sec:from_amp_to_FS}). We then proceed in section~\ref{sec:LU_FS} to construct the \gls{lu} representation for the sum of cuts of a \gls{fs} diagram, which is locally infrared-finite provided the observable is infrared-safe.

\subsubsection{From amplitudes to forward-scattering diagrams}
\label{sec:from_amp_to_FS}

Let us consider a generic scattering process:
\begin{equation}
\mathrm{P}_1(p_1) \mathrm{P}_2(p_2) \to \mathrm{M}(q_1,\ldots,q_{|\mathrm{M}|}) + X(\tilde{q}_1,...,\tilde{q}_{|X|})\,,
\end{equation}
where $\mathrm{M}$ is any fixed collection of resolved particles with momenta $q_1,...,q_{|\mathrm{M}|}$, $X$ denotes a collection of massless unresolved particles, whose momenta labeled as $\tilde{q}_1,...,\tilde{q}_{|X|}$~\footnote{For an N$^l$\gls{lo} computation, one has to include all possible contributions with $|X|\leq l$.}, and $\mathrm{P}_{1}$ and $\mathrm{P}_{2}$ denote the initial-state massless particles with momenta $p_1$ and $p_2$. We have used the notation $|\mathrm{M}|$ and $|X|$ to represent the number of partons in the particle sets $\mathrm{M}$ and $X$, respectively. For an physical observable $\mathcal{O}$, the (partonic) differential cross section in $\mathcal{O}$ is given by
\begin{equation}
\label{eq:x_sec}
\frac{\mathrm{d}\sigma}{\mathrm{d}\mathcal{O}}=\frac{1}{2(p_1+p_2)^2}\sum_X\int \mathrm{d}\Pi_{\mathrm{M}\cup X} |\mathcal{A}(\mathrm{P}_1 \mathrm{P}_2\rightarrow \mathrm{M} + X)|^2 \mathds{O}_{\mathcal{O}}(\{q_j\}_{j=1}^{|\mathrm{M}|},\{\tilde{q}_i\}_{i=1}^{|X|})\,.
\end{equation}
The phase-space integration involves the momenta of all final-state particles:
\begin{align}
\mathrm{d}\Pi_{\mathrm{M}\cup X}=\prod_{j=1}^{|\mathrm{M}|}\frac{\mathrm{d}^4q_j}{(2\pi)^3}&\delta^{(+)}\left(q_j^2-m_j^2\right)\prod_{i=1}^{|X|}\frac{\mathrm{d}^4\tilde{q}_i}{(2\pi)^3}\delta^{(+)}\left(\tilde{q}_i^2\right) \nonumber\\ &\times(2\pi)^4\delta^{(4)}\left(p_1+p_2-\sum_{j=1}^{|\mathrm{M}|}{q_j}-\sum_{i=1}^{|X|}{\tilde{q}_i}\right)\,, 
\label{eq:LUmaster}
\end{align}
where $\delta^{(+)}(q^2-m^2)=\delta(q^2-m^2)\Theta(q^0)$, with $\Theta$ denoting the Heaviside step function.
Colour, helicity, and other quantum-number summmations, as well as final-state symmetry and initial-state averaged factors, are suppressed in eq.~\eqref{eq:x_sec} for ease of notations. $\mathds{O}_{\mathcal{O}}$ is a distribution that implements the observable $\mathcal{O}$ in terms of external momenta of final-state particles. 

In perturbation theory, the amplitude $\mathcal{A}$ is represented as a sum over Feynman diagrams. Letting $\Gamma_{\text{proc}}$ be the set of all Feynman graphs at the amplitude-level contributing to the process, we have
\begin{equation}
\mathcal{A}(\mathrm{P}_1 \mathrm{P}_2\rightarrow \mathrm{M} + X)=\sum_{G\in\Gamma_{\text{proc}}}\mathds{D}_G\left(\{p_i\}_{i=1}^2;\{q_j\}_{j=1}^{|\mathrm{M}|},\{\tilde{q}_i\}_{i=1}^{|X|}\right)\,.
\end{equation}
$\mathds{D}_G$ is the Feynman diagram $G$, obtained by ordinary Feynman rules. Plugging this into eq.~\eqref{eq:x_sec}, we see that the cross section can be written as a double sum over Feynman diagrams, one for the amplitude and one for the complex-conjugated amplitude, on top of the sum over $X$:  
\begin{equation}
\frac{\mathrm{d}\sigma}{\mathrm{d}\mathcal{O}}=\frac{1}{2(p_1+p_2)^2}\sum_X\sum_{G_1,G_2\in \Gamma_{\text{proc}}}\int \mathrm{d}\Pi_{\mathrm{M}\cup X} \mathds{D}_{G_1}\mathds{D}_{G_2}^\star \mathds{O}_{\mathcal{O}}(\{q_j\}_{j=1}^{|\mathrm{M}|},\{\tilde{q}_i\}_{i=1}^{|X|})\,.
\end{equation}
It is well known~\cite{Cutkosky:1960sp} that this triple sum over $X$, $G_1$, and $G_2$ can be traded by a double sum instead: one over \gls{fs} diagrams, and one over the Cutkosky cuts that can be operated on the \gls{fs} diagrams compatible with the process under investigation. 

It is possible that some of the Cutkosky cuts of a given \gls{fs} diagram are not compatible with the process definition, which is implemented through a vanishing observable measure $\mathds{O}_{\mathcal{O}}$. Letting $\Gamma_{\text{proc}}^{\text{\gls{fs}}}$ be the set of all graphs compatible with the \gls{fs} process, namely $\mathrm{P}_1(p_1)\mathrm{P}_2(p_2)\rightarrow \mathrm{P}_1(p_1) \mathrm{P}_2(p_2)$, we have
\begin{eqnarray}
\frac{\mathrm{d}\sigma}{\mathrm{d}\mathcal{O}}&=&\sum_{G\in \Gamma_{\text{proc}}^{\text{\gls{fs}}}}{\frac{\mathrm{d}\sigma_G}{\mathrm{d}\mathcal{O}}}\,, \nonumber\\
\frac{\mathrm{d}\sigma_G}{\mathrm{d}\mathcal{O}}&=&\frac{1}{2(p_1+p_2)^2}\sum_{c\in \mathcal{C}_{\text{proc}}^G}\int \mathrm{d}\Pi_{\mathrm{M}\cup X} \mathds{D}_{G_1^c}\mathds{D}_{G_2^c}^\star\mathds{O}_{\mathcal{O}}(\{q_j\}_{j=1}^{|\mathrm{M}|},\{\tilde{q}_i\}_{i=1}^{|X|})\,.
\end{eqnarray}
In the above equation, $\mathcal{C}_{\text{proc}}^G$ contains all Cutkosky cuts $c\in \mathcal{C}_{\text{proc}}^G$ of a given \gls{fs} diagram $G$ compatible with the process definition. The Cutkosky cut $c$ is a collection of edges such that, upon deletion, the \gls{fs} graph $G$ is divided into two connected components $G_1^c$ and $G_2^c$ and such that $c$ contains all the resolved particles in $\mathrm{M}$ and massless unresolved particles in $X$. 

We conventionally denote the labels $e_1^{\pm}$ and $e_2^{\pm}$ as the external edges of the \gls{fs} diagram $G$ carrying four-momenta $\pm p_1$ and $\pm p_2$, respectively, where we have used the all-incoming convention. For example, in the inclusive reaction $\gaga\to \QQ+X$, $\mathds{O}_{\mathcal{O}}$ evaluates to the ones for all cuts that are compatible with the process definition; otherwise, it is zero for the incompatible cuts. 

% We will set $\mathds{O}_{\mathcal{O}}$ to take these values from now on.

\subsubsection{Local Unitarity representation}
\label{sec:LU_FS}
In virtue of the \gls{kln} theorem~\cite{Kinoshita:1962ur,Lee:1964is}, $\mathrm{d}\sigma_G/\mathrm{d}\mathcal{O}$ is \gls{ir} finite provided the observable $\mathcal{O}$ is \gls{ir}-safe. The \gls{lu} formalism~\cite{Capatti:2020xjc,Capatti:2022tit} provides an integral representation for $\mathrm{d}\sigma_G/\mathrm{d}\mathcal{O}$ that is free of \gls{ir} singularities (also called pinched thresholds). In order to write it down explicitly, let us first consider the $L$-loop \gls{fs} diagram $G$ and its three-dimensional representation~\cite{Catani:2008xa,Bierenbaum:2010cy,PhysRevLett.122.111603,Capatti:2019ypt,Capatti:2020ytd,JesusAguilera-Verdugo:2020fsn}
\begin{equation}
\mathds{D}_G(\{p_i\}_{i=1}^2;\{p_i\}_{i=1}^2)=\int \left[\prod_{i=1}^L \frac{\mathrm{d}^3\vec{k}_i}{(2\pi)^3}\right] f_{\text{3d},G}(\vec{\mathbf{K}};p_1,p_2)\,,
\label{eq:LU_representation}
\end{equation}
and the integrand is given by
\begin{equation}
\label{eq:3d_rep}
f_{\text{3d},G}(\vec{\mathbf{K}};p_1,p_2)=\int \left[\prod_{i=1}^L \frac{\mathrm{d}k_i^0}{2\pi}\right] \frac{\mathcal{N}_G(\mathbf{K};p_1,p_2)}{\prod_{e\in\mathcal{E}_G}(q_e^2-m_e^2+i0^+)}\,,
\end{equation}
where $\mathbf{K}\in\mathbb{R}^{4L}$ and $\vec{\mathbf{K}}\in\mathbb{R}^{3L}$  are the concatenation of the $L$-loop four momenta $k_1,...,k_L$ and their spatial counterparts $\vec{k}_1,...,\vec{k}_L$, respectively. In general, we will denote by $\vec{\mathbf{V}}$ the concatenation of $L$ vectors $\vec{v}_1,...,\vec{v}_L$. $\mathcal{N}_G$ is the standard numerator in the covariant representation as derived using Feynman rules, $\mathcal{E}_G$ is the set of internal propagators in $G$, and $q_e$ can be written as a linear combination of the loop momenta and external momenta:
\begin{equation}
q_e=\sum_{j=1}^{2}{r_{ej}p_j}+\sum_{i=1}^L s_{ei}k_i\,,
\end{equation}
where $s_{ei},r_{ej}\in\{-1,0,1\}$ are the unique coefficients of the decomposition (see, for example, ref.~\cite{Capatti:2023omc} for a detailed account on how to determine them). %In order to obtain this decomposition, one should first choose a spanning tree $T$ of the graph $G$. The edges in $\{e_1,...,e_L\}=\mathcal{E}_G\setminus T$ are then conventionally labelled by loop momenta $k_1,...,k_L$, with $L=|\mathcal{E}_G\setminus T|$. Each edge in $e_i=\mathcal{E}_G\setminus T$ is associated to a cycle $c_i$ obtained by adding to $e_i$ the unique path that is entirely contained in the spanning tree and that connects its ends.   

There are many ways to perform the energy-component integrations in eq.~\eqref{eq:3d_rep} analytically. The final result can be expressed into a closed form, which can be generally written as (see refs.~\cite{Capatti:2022mly,Capatti:2023shz}):
\begin{equation}
\label{eq:3d_rep2}
f_{\text{3d},G}(\vec{\mathbf{K}};p_1,p_2)=\frac{F_G(\vec{\mathbf{K}};p_1,p_2)}{ \prod_{c\in\mathcal{C}^G}\eta_{c}(\vec{\mathbf{K}};p_1,p_2)^{n_c}}\,,
\end{equation}
where $n_c$ is a positive integer, $F_G$ is an integrable function except for theories plagued by pathological soft singularities (\eg\ the massless $\phi^3$ theory), and 
\begin{equation}
\eta_{c}(\vec{\mathbf{K}};p_1,p_2)\equiv\sum_{e\in c}E_e-p_1^0-p_2^0\, .
\end{equation}
Importantly, the surface $\eta_{c}(\vec{\mathbf{K}};p_1,p_2)=0$ is \textit{convex and bounded}. 
The on-shell energies are defined as
\begin{equation}
E_e=\left\{\begin{array}{rl}
\sqrt{|\vec{q}_e|^2+m_e^2}, & \text{if } e\notin\{e_1^+,e_1^-,e_2^+,e_2^-\} \\
p_1^0, & \text{if } e=e^{\pm}_1\\
p_2^0, & \text{if } e=e^{\pm}_2
\end{array}\right.\, .\label{eq:Eedef}
\end{equation}
We assume that the external momenta satisfy $p_1^0,p_2^0>0$. $\mathcal{C}^G$ is the set of all possible Cutkosky cuts that divide the \gls{fs} graph $G$ into two connected components upon deletion of edges in the cut\footnote{These cuts are allowed to cross external particles and not be compatible with the process definition, contrary to those in $\mathcal{C}_{\text{proc}}^G$.}. 

In line with the work of ref.~\cite{Capatti:2019edf}, we define the overlap structure:
\begin{equation}
\label{eq:overlap}
\text{O}^G\equiv\{C\subseteq\mathcal{C}_{\text{proc}}^G \, | \, \exists \vec{\mathbf{S}}_C\in\mathbb{R}^{3L} \ \text{s.t.} \  \eta_{c}(\vec{\mathbf{S}}_C;p_1,p_2)<0, \ \forall c \in C\}\,.
\end{equation}
The maximal overlap structure is instead given by
\begin{equation}
\label{eq:maximal_overlap}
\text{O}^G_{\text{max}}=\{C\in\text{O}^G \, | \, \nexists C' \in \text{O}^G \text{ with } C\subset C' \}\,.
\end{equation}
Note that the definition of the overlap structure depends on the kinematics of the initial-state particles, $p_1$ and $p_2$. From a physical point of view, one can think of the interior of the surface $\eta_c=0$ as identifying the kinematic region in which the on-shell momenta of the particles flowing through the cut $c$ sum up to a momentum with an invariant mass larger than $s=(p_1 + p_2)^2$. In other words, if $p_e=(E_e,\vec{p}_e)$ are the momenta of the edges $e\in c$ (with consistent orientation with respect to the cut), then, using energy-momentum conservation, one finds:
\begin{equation}
p_c^2=\left(\sum_{e\in c}p_e\right)^2=(\eta_c+p_1^0+p_2^0-|\vec{p}_1+\vec{p}_2|)(\eta_c+p_1^0+p_2^0+|\vec{p}_1+\vec{p}_2|)>s
\end{equation}
if $\eta_c>0$. An element $C\in \text{O}^G$ identifies collections of cuts such that there exists a kinematic point that corresponds to all the quantities $p_c^2-s$, $c\in C$ being positive simultaneously. 

For each element of the maximal overlap structure, we associate the flow
\begin{equation}
\label{eq:causal_flow}
\vec{\boldsymbol{\Phi}}_C(\vec{\mathbf{K}}; t)\equiv e^t \vec{\mathbf{K}}+\vec{\mathbf{S}}_C,
\end{equation}
which is simply a radial field centered at a choice of a ``source'' $\vec{\mathbf{S}}_C$ with radius $e^t$, $t\in(-\infty,+\infty)$. $\vec{\boldsymbol{\Phi}}_C(\vec{\mathbf{K}}; t)$ satisfies a flow ordinary differential equation system:
\begin{equation}
\begin{cases}
\partial_t\vec{\boldsymbol{\Phi}}_C(\vec{\mathbf{K}}; t)=(\vec{\boldsymbol{\Phi}}_C(\vec{\mathbf{K}}; t)-\vec{\mathbf{S}}_C) \\
\vec{\boldsymbol{\Phi}}_C(\vec{\mathbf{K}}; 0)=\vec{\mathbf{K}}+\vec{\mathbf{S}}_C
\end{cases}
\end{equation}
and is dubbed the \textit{causal flow}. The chosen source $\vec{\mathbf{S}}_C$ has the same definition as the interior point introduced in eq.~\eqref{eq:overlap}:
\begin{equation}
\eta_c(\vec{\mathbf{S}}_C;p_1,p_2)<0, \quad \forall c\in C.
\label{eq:sourceDefinition}\end{equation}
This property, together with the convexity of the surface $\eta_{c}(\vec{\mathbf{K}};p_1,p_2)=0$, allows one to conclude that:
\begin{itemize}
\item The equation
\begin{equation}
\eta_c(\vec{\boldsymbol{\Phi}}_C(\vec{\mathbf{K}}; t);p_1,p_2)=0\,,
\end{equation}
for any fixed $\vec{\mathbf{K}}\in\mathbb{R}^{3L}$, has a single solution, $t=t_c(\vec{\mathbf{K}})$, which can be found in a polynomial time, for example, by using the Newton's bisection method.
\item The expansion of $\eta_c(\vec{\boldsymbol{\Phi}}_C(\vec{\mathbf{K}},t);p_1,p_2)$ about $t=t_c(\vec{\mathbf{K}})$ reads
\begin{align}
\eta_c(\vec{\boldsymbol{\Phi}}_C(\vec{\mathbf{K}}; t);p_1,p_2)=(t-t_c(\vec{\mathbf{K}}))&\partial_t \eta_c(\vec{\boldsymbol{\Phi}}_C(\vec{\mathbf{K}}; t);p_1,p_2)|_{t=t_c(\vec{\mathbf{K}})} \nonumber\\
&+\mathcal{O}\left(\left(t-t_c(\vec{\mathbf{K}})\right)^{2}\right)\label{eq:etac_exp2}
\end{align}
with
\begin{equation}
\label{eq:t_star}
\partial_t \eta_c(\vec{\boldsymbol{\Phi}}_C(\vec{\mathbf{K}}; t);p_1,p_2)|_{t=t_c(\vec{\mathbf{K}})}= e^{t_c(\vec{\mathbf{K}})}\vec{\mathbf{K}} \cdot \nabla_{\vec{\mathbf{K}}'}\eta_c(\vec{\mathbf{K}}';p_1,p_2)\Bigg|_{\vec{\mathbf{K}}'=\vec{\boldsymbol{\Phi}}_C(\vec{\mathbf{K}}; t_c(\vec{\mathbf{K}}))}>0.
\end{equation}
In particular, the left-hand side of eq.~\eqref{eq:t_star} is \textit{positive definite and never vanishes} for any values of $\vec{\mathbf{K}}$. 
\end{itemize}
Eqs.~\eqref{eq:maximal_overlap}, ~\eqref{eq:causal_flow}, ~\eqref{eq:etac_exp2}, and ~\eqref{eq:t_star} allow us to construct a parametrisation for the phase space associated with the final-state particles that correlates their \gls{ir} singularities. Given our interpretation of the elements of the overlap structure as identifying regions of kinematic space in which a number of cuts have an overall momentum $p_c$ flowing through them with $p_c^2>s$, the causal flow can be interpreted as a way to transport a kinematic point from such regions to those in which some or all of these overall cut momenta now satisfy $p_c^2<s$ by crossing thresholds. In other words, it provides us with a one-dimensional projection of the integration space in which we can compute residues of the thresholds.

With the above constructions, we are ready to define the \gls{lu} representation of the differential cross section. It reads, for any graph $ G\in\Gamma_{\text{proc}}^{\mathrm{\gls{fs}}}$, as
\begin{align}
\frac{\mathrm{d}\sigma_G}{\mathrm{d}\mathcal{O}}=\frac{1}{2(p_1+p_2)^2}\int &\left[\prod_{i=1}^L \frac{\mathrm{d}^3\vec{k}_i}{(2\pi)^3}\right]\sum_{C\in\mathrm{O}^G_{\text{max}}}\sum_{c\in C}\frac{2\pi }{(n_c-1)!}\lim_{t\rightarrow t_c(\vec{\mathbf{K}}) } \frac{\mathrm{d}^{n_c-1}}{\mathrm{d}t^{n_c-1}}\Big[\left(t-t_c(\vec{\mathbf{K}})\right)^{n_c} \nonumber\\
e^{3L t} h(t) 
&\text{\acrshort{mc}}_C(\vec{\boldsymbol{\Phi}}_C(\vec{\mathbf{K}}; t);p_1,p_2) f_{\text{3d},G}(\vec{\boldsymbol{\Phi}}_C(\vec{\mathbf{K}}; t);p_1,p_2)\Big]\,,\label{eq:LU_2to2}
\end{align}
where the \gls{mc} factor allowing for the separation of threshold singularities from each element of the maximal overlap structure can be written as
\begin{equation}
\label{eq:mc_factor}
\text{\gls{mc}}_C(\vec{\mathbf{K}};p_1,p_2)\equiv\frac{\prod_{c\notin C} |\eta_c(\vec{\mathbf{K}};p_1,p_2)|^{m}}{\sum_{C^\prime\in\mathrm{O}^G_{\mathrm{max}}} \prod_{c\notin C^\prime} |\eta_c(\vec{\mathbf{K}};p_1,p_2)|^{m}}\,
\end{equation}
where $m\geq 1$ and $h(t)$ is an auxiliary function, normalised to unity, $\int_{{-\infty}}^{+\infty}{dt h(t)}=1$, used to solve the energy part of the momentum conservation Dirac delta function of eq.~\eqref{eq:LUmaster} using a common variable for all cuts which is crucial to align the measure between real and virtual contributions\footnote{Note that there is a change of variable $t_\star=e^{t}$ and a redefinition of the function $h(t)=t_\star h_\star(t_\star)=e^{t} h_\star(e^{t})$, where $t_\star$ and $h_\star$ denote the auxiliary variable $t$ and function $h$ used in ref.~\cite{Capatti:2020xjc}.}. For the results presented in this paper, we have set $m=1$. The value of $m$ seems to impact the Monte Carlo convergence only marginally, as long as it is not taken too large.
%For $G\notin\Gamma^{\mathrm{\gls{fs}}}_{\text{proc}}$, we set $\mathrm{d}\sigma_G/\mathrm{d}\mathcal{O}=0$. 

Note that the causal flow problem is completely analogous to the computation of the integrated threshold counter-terms of refs.~\cite{Kermanschah:2021wbk,Kermanschah:2024utt}, although the formulation of eq.~\eqref{eq:LU_2to2} also supports generic observables. If there are no initial-state singularities, $\mathrm{d}\sigma_G/\mathrm{d}\mathcal{O}$ is \gls{ir} finite. Furthermore, \gls{uv} singularities can be regulated by the automated application of the \gls{bphz} renormalisation formalism~\cite{Bogoliubov:1957gp,Hepp:1966eg,Zimmermann:1969jj} as outlined in ref.~\cite{Capatti:2022tit}. On the other hand, when initial-state singularities are present, the formula above is not enough to obtain an integrable function and additional strategies need to be put in place, such as counter-term subtraction (\eg\ refs.~\cite{Anastasiou:2018rib,Anastasiou:2020sdt,Anastasiou:2025cvy} for the subtraction of virtual contributions and typical phase-space subtraction methods for the subtraction of singularities associated with real emission from initial states) or by extending the summation over the initial states to include all degenerate massless configurations~\cite{Capatti:2025gqj}.

\subsection{Enumeration of forward-scattering graphs for $\gaga\to\QQ$\label{sec:FSGenumerate}}

\label{sec:LU_graph_enum}

A direct approach to generating interference diagrams consists in building all amplitude graphs, expanding their squared complex norm, and combining the resulting terms into topologically equivalent ones. Beyond \gls{lo}, this method becomes cumbersome as it involves considering various amplitude loop counts and definitions of final-state contents.

Instead, we choose a method more directly aligned with the \gls{lu} construction that considers the direct enumeration of topologically inequivalent \gls{fs} graphs with a subsequent filtering based on their Cutkosky cuts. This construction eventually yields the set $\Gamma_{\text{proc}}^{\text{\gls{fs}}}$.
We built a general-purpose tool to this end, called \feyngen, based on the graph generation capabilities of the computer algebra library \symbolica~\cite{Symbolica}, and distributed as part of the \gammaLoop\ code~\cite{gammaloop} under development. Our implementation also supports accounting for the symmetrisation of initial states of \gls{fs} graphs, as well as the ``complex conjugation'' symmetry, corresponding to the interchange of the external states on the left- and right-hand sides of the Cutkosky cut\footnote{The hermitian nature of the Lagrangian, or equivalently the condition that the $S$-matrix be unitary, guarantees the validity of this symmetry.}.
Moreover, \feyngen\ automatically performs a numerator analysis of the \gls{fs} graphs to combine equivalent contributions (\eg\ from different massless quark flavours), apply Furry-like cancellations~\cite{Furry:1937zz}, and discard identical zeros originating from group-theoretical structures (\eg\ $\text{Tr}\left[t^a\right]=0$ and $\text{Tr}\left[\gamma^\mu \gamma^\nu \gamma^\rho\right]=0$).
In addition, it can apply custom filters to remove special topologies, such as tadpoles, which are modded out in the perturbative expansion of the $S$-matrix. Further details on steering graph generation are available on the \feyngen\ wiki page~\cite{FeynGenWiki}. 

\begin{figure}[!bt]
    \centering
    \includegraphics[width=12cm]{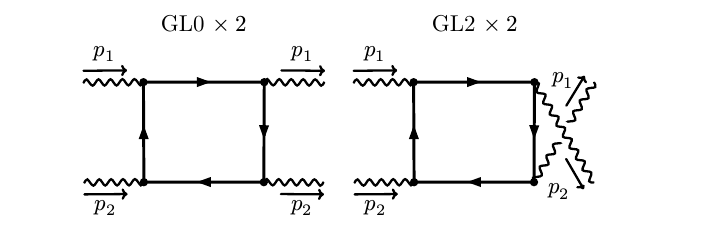}
    \caption{The two \gls{lo} \gls{fs} graphs contributing to the process $\gaga \to \QQ$. Note that these graphs only involve a single Cutkosky cut each, and are assigned a multiplicity factor of $2$ stemming from the symmetrisation of the initial states.}
    \label{fig:LO_Graphs}
\end{figure}

At any perturbative order, the \gls{fs} graphs\footnote{This also holds for the singlet \gls{fs} graphs of class C where the \gls{fs} graph is deemed \textit{direct} if there exists a spanning $2$-forest separately containing $p_1$ and $p_2$ only respectively (\texttt{GL172} in figure~\ref{fig:NNLO_singlet_graphs} being the only crossed one).} for $\gaga \to \QQ$ can be separated into two classes depending on the external ordering when following the outer heavy-quark loop. We call ``\textit{direct graphs}'' those with ordering $\{p_1, p_1, p_2, p_2\}$, like \texttt{GL0} in figure~\ref{fig:LO_Graphs}, and ``\textit{crossed graphs}'' the ones like \texttt{GL2} with ordering $\{p_1, p_2, p_1, p_2\}$. Each \gls{fs} graph topology with a single heavy-quark loop appears twice: once for each of these two orderings. 
 
At \gls{lo}, the $\gaga \to \QQ$ process receives two Feynman diagrams at the amplitude level, which can be called as $t$- and $u$-channels, respectively, where $t$ and $u$ are Mandelstam variables defined for the $2\to 2$ scattering kinematics.
The resulting four interference diagrams can be combined into two \gls{fs} graphs as shown in figure~\ref{fig:LO_Graphs}, after symmetrising over the initial states ($p_1 \leftrightarrow p_2$). We choose to work in the rest frame of the two initial-state photons, and set the \gls{fs} kinematics to be $p_1=(E_{\gamma},0,0,E_{\gamma})$ and $p_2=(E_{\gamma},0,0,-E_{\gamma})$.
The single Cutkosky cut of the symmetrised \gls{lo} direct \gls{fs} graph reproduces the complex norm of individual \gls{lo} amplitude diagrams, giving contributions to the \gls{lo} cross section proportional to $t^{-2}$ and $u^{-2}$, whereas the single cut of the symmetrised crossed \gls{fs} graph captures the interference terms between the two \gls{lo} amplitude graphs, combining into a cross-section contribution proportional to $(t u)^{-1}$.

\begin{figure}[!bt]
    \includegraphics[width=\textwidth]{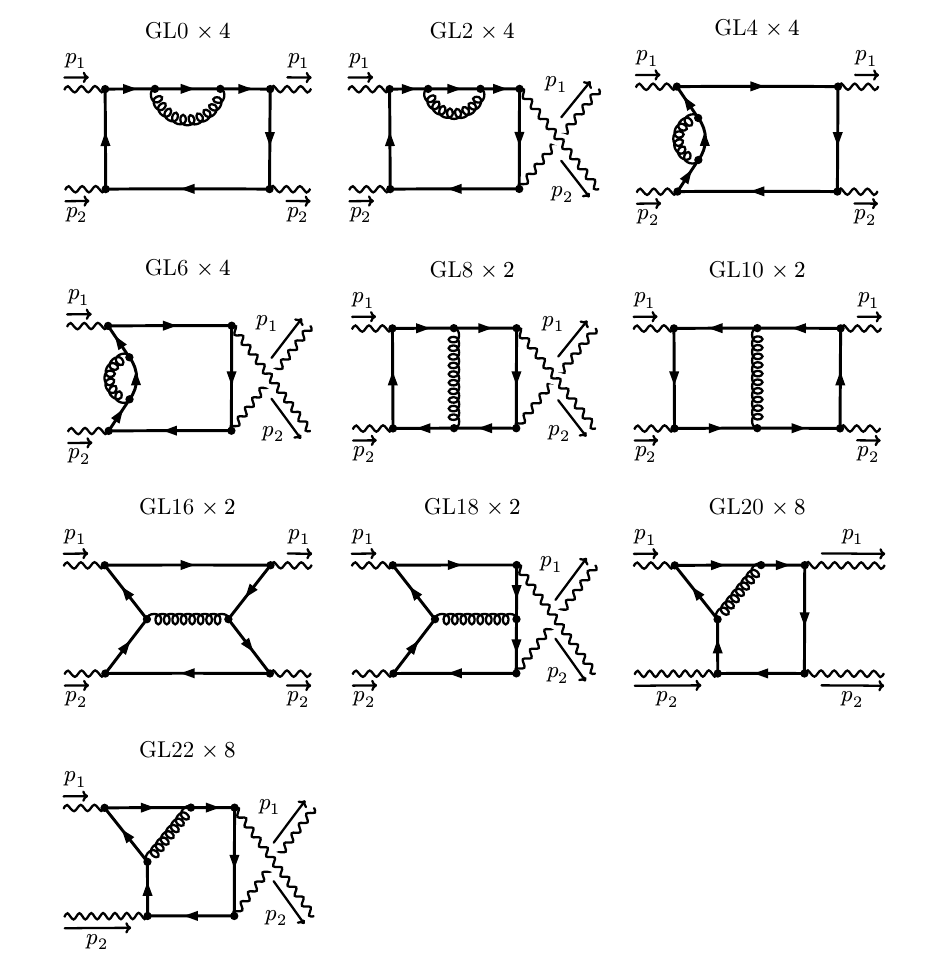}
    \caption{The ten \gls{fs} graphs contributing to the \gls{nlo} \gls{qcd} correction to the process $\gaga \to \QQ$. Multiplicity factors arise from the combination of isomorphic graphs when accounting for the symmetry stemming from swapping initial states as well as complex conjugation symmetry.}
    \label{fig:NLO_Graphs}
\end{figure}

At \gls{nlo}, the \gls{qcd} corrections can be obtained by attaching a gluon to each possible pair of heavy-quark propagators of the \gls{lo} direct and crossed boxes, or by correcting any heavy-quark propagator. This procedure yields $20$ graphs which are combined into only $10$ graphs shown in figure~\ref{fig:NLO_Graphs}, when considering both initial-state symmetrisation and complex-conjugation symmetrisation.

At \gls{nnlo} \gls{qcd}, there are contributions featuring a massless quark loop connected to the external photons. In the context of this paper, we ignore these and focus solely on the dominant gauge invariant set of graphs factorising four powers of the photon coupling to the heavy quarks, \ie\ $\mathcal{O}(e_Q^4)$.

\begin{figure}[!bt]
    \centering
    \includegraphics[width=11cm,height=5cm, page=1]{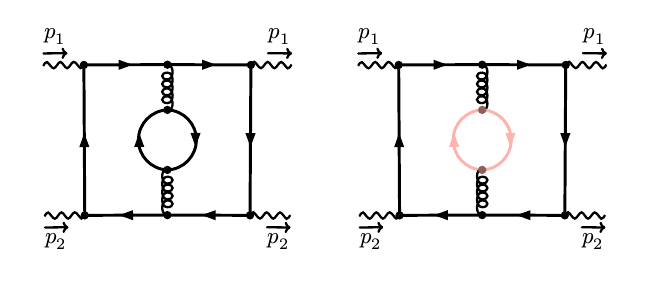}
    \caption{Two example diagrams contributing to the classes A and B of section~\ref{sec:FSGenumerate}. Pink-coloured fermion lines denote massless quarks. }
    \label{fig:nf_light_and_heavy}
\end{figure}

\begin{figure}[!bt]
    \centering
    \includegraphics[width=12cm,height=10cm, page=1]{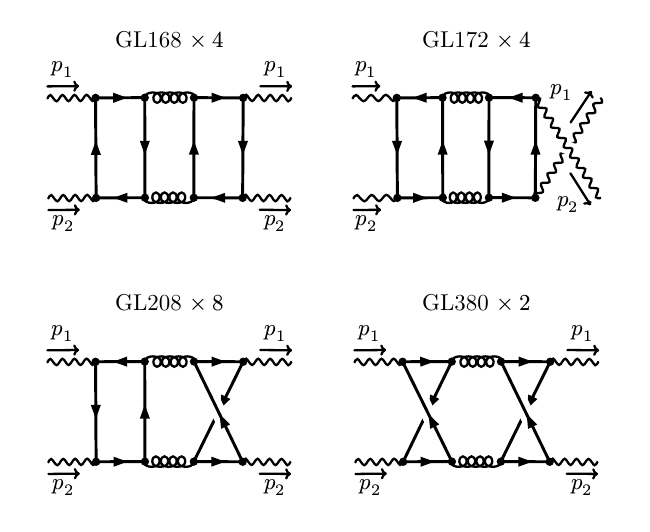}
    \caption{The four singlet \gls{fs} graph contributions.  Multiplicity factors arise from the combination of isomorphic graphs when accounting for the symmetry stemming from swapping initial states as well as complex conjugation symmetry.}
    \label{fig:NNLO_singlet_graphs}
\end{figure}

\begin{figure}[!bt]
    \centering
    \includegraphics[trim={0 0 0 0},clip,width=.8\textwidth, page=1]{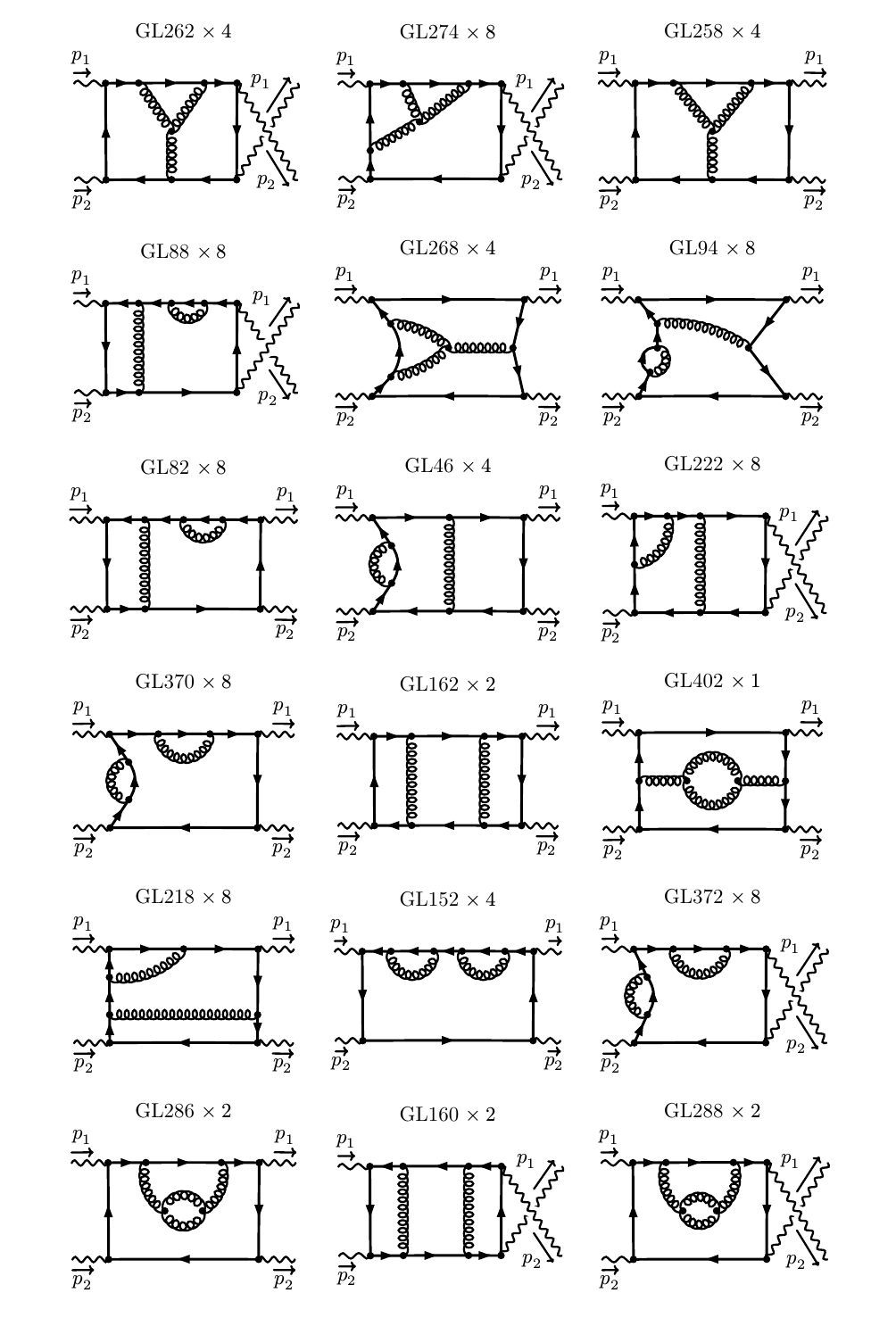}
    \caption{Selection of the 18 (amongst 138) \gls{nnlo} \gls{fs} diagrams highlighted in figure~\ref{fig:SupergraphHierarchy}, sorted according to the absolute value of their contribution for $\gaga \to t\bar{t}$ at $\sqrt{s}=500\;\text{GeV}$.
    All of them belong to class D in the classification of section~\ref{sec:FSGenumerate}. Multiplicity factors arise from the combination of isomorphic graphs when accounting for the symmetry stemming from swapping initial states as well as complex conjugation symmetry.
    }
    \label{fig:NNLO_other_graphs_selection}
\end{figure}

\newpage
We can then separate these contributions into four gauge-invariant classes totaling 138 distinct topologies:

\begin{enumerate}[label=Class \Alph*:]
\item Graphs depending on the number of massless quark flavours, $n_q$ (10 graphs): contributions obtained by dressing the gluon propagator of each of the 10 \gls{nlo} graphs with $n_q$ grouped copies of a massless quark loop (3 copies for $\gaga \to \ccbar$, 4 copies for $\gaga \to \bbbar$, and 5 copies for $\gaga \to \ttbar$). An example diagram belonging to this class is the right one in figure~\ref{fig:nf_light_and_heavy}.
\item Graphs involving $n_h$ heavy quarks in the gluon wavefunction correction (10 graphs): contributions analogous to the $n_q$ one, but obtained by dressing the gluon propagator with a massive heavy-quark loop. We will only use $n_h=1$, with the heavy quark $Q$ set to the one appearing in the final state of $\gaga \to \QQ$. An example diagram belonging to this class is the left one in figure~\ref{fig:nf_light_and_heavy}.
\item The ``singlet'' contributions (4 graphs): contributions featuring two closed heavy-quark loops, each connected to two photons and two gluons, see figure~\ref{fig:NNLO_singlet_graphs}.
\item All other contributions (114 graphs), where graphs with a closed ghost\footnote{In line with section 4.2 of ref.~\cite{Capatti:2020xjc} and unitarity arguments, we account for the presence of ghosts and unphysical gluon polarisations in the final state. This is required in order to achieve infrared finiteness at a local level.} and a closed gluon loop are counted separately. See figure~\ref{fig:NNLO_other_graphs_selection} for some example diagrams in this class.
\end{enumerate}

Our motivation for introducing these classes
is solely to structure the discussion of our results and grouping gauge-invariant classes of \gls{fs} diagrams does not impact performance. For example, our treatment of threshold singularities is purely based on the topology of an individual graph.

We show a selection of \gls{nnlo} graphs in figure~\ref{fig:NNLO_other_graphs_selection}. In total, we compute the contribution of 138 distinct \gls{fs} diagrams. We construct the \gls{lu} representation of eq.~\eqref{eq:LU_2to2} for each of them. Since the integrand of each \gls{fs} diagram is separately infrared-finite, each contribution can be integrated independently. This enables us to significantly optimise the Monte Carlo integration by summing over them using discrete importance sampling (see appendix~\ref{sec:graph_distribution}).

\clearpage
\newpage
\subsection{Details of the Local Unitarity  construction for $\gaga\to \QQ$\label{sec:detailsinLU}}

\begin{figure}[!bt]
    \centering
    \includegraphics[width=10cm,height=8cm, page=1]{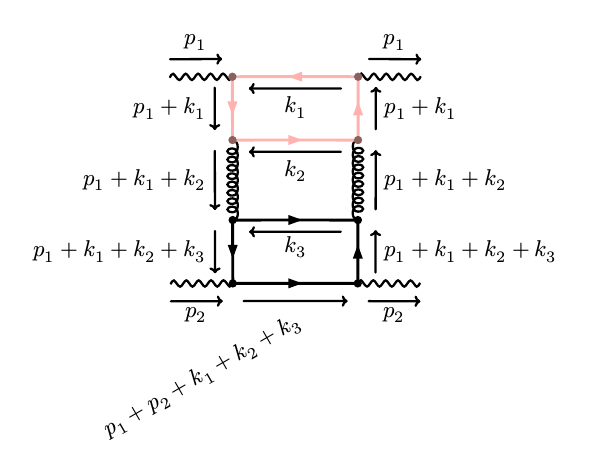}
    \caption{A forward-scattering diagram that depends on the photon collinear factorisation scheme 
and that involves coupling of photons to light quarks. Light quarks are coloured in pink. It is excluded from our \gls{nnlo} \gls{qcd} computation. This forward-scattering diagram has initial-state collinear singularities, \eg\ at $k_1=-x p_1$ and thus cannot be computed within the \gls{lu} formalism as presented in this paper, which only regulates final-state singularities.}
    \label{fig:IR_singular_FS}
\end{figure}

\begin{figure}[!bt]
    \centering
    \includegraphics[width=12cm,height=10cm, page=1]{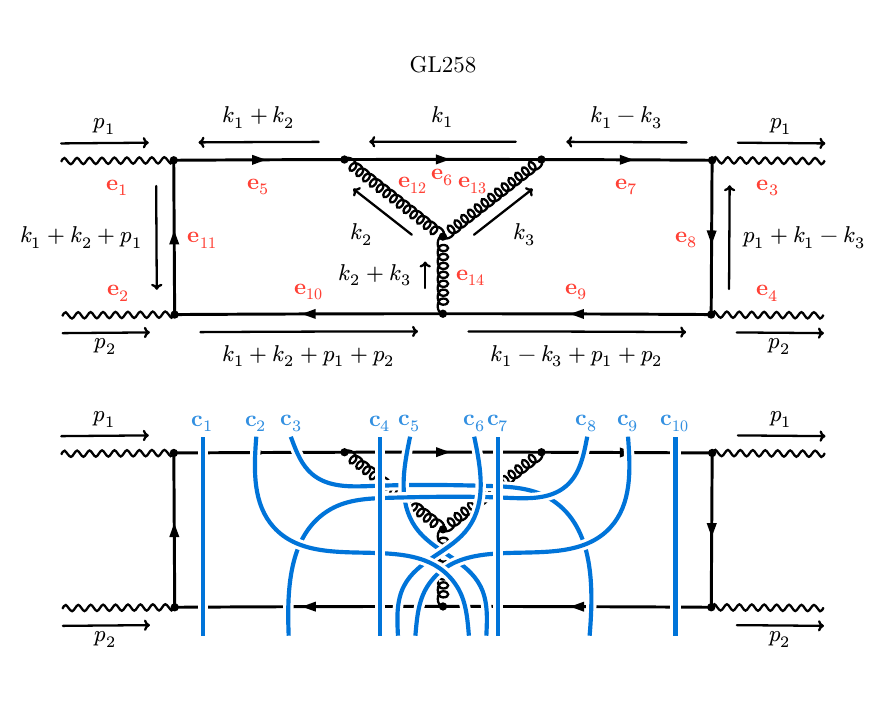}
    \caption{The existing cuts and routing of \texttt{GL258} for $s>4m_Q^2$. All the existing cuts of this diagram are compatible with the process definition.}
    \label{fig:GL258_cut_and_routed}
\end{figure}

\begin{figure}[!bt]
    \centering
    \includegraphics[width=12cm,height=10cm, page=1]{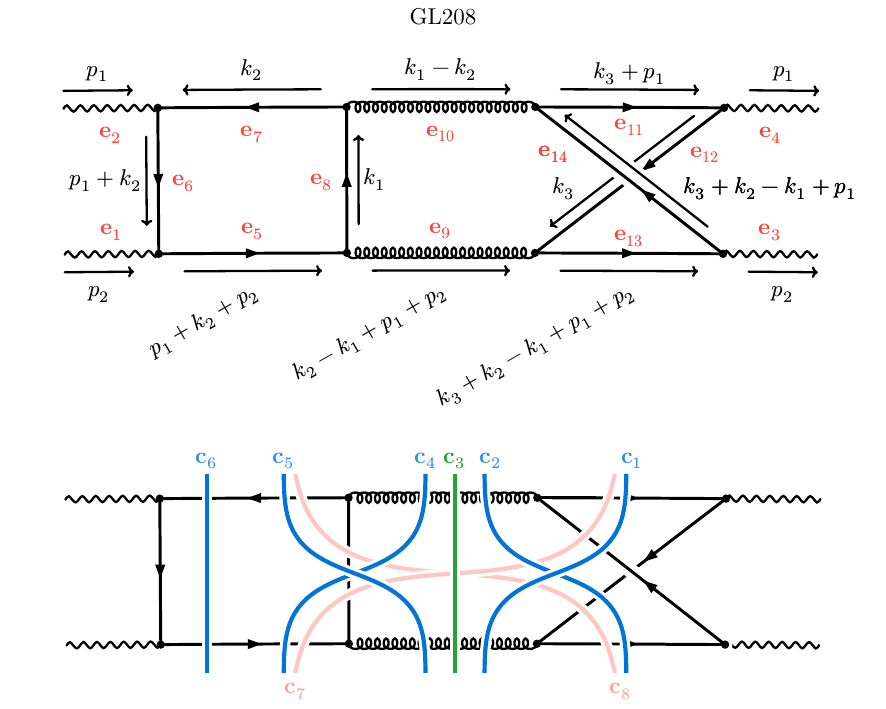}
    \caption{The existing cuts and routing of \texttt{GL208} (one of the singlet diagrams) for $s>4m_Q^2$. We colour Cutkosky cuts compatible with the process definition (namely, cutting $\QQ$ or $\QQ j$) in blue. Cutkosky cuts cutting four massive quarks are drawn in pink, while the double gluon cut is green.
    }
    \label{fig:GL208_cut_and_routed}
\end{figure}

\label{sec:LU_2_body}

In this section we present the details of the construction of the \gls{lu} representation of eq.~\eqref{eq:LU_2to2} for the process at hand. We discuss, in particular, two aspects: a) the construction of the overlap structure (eq.~\eqref{eq:maximal_overlap}) and the causal flow (eq~\eqref{eq:causal_flow}); b) the lack of need for additional threshold regularisation procedures (specifically, contour deformation~\cite{Capatti:2019edf} or threshold subtraction~\cite{Kermanschah:2021wbk}). %In order to simplify the derivation of the \gls{lu} representation of eq.~\eqref{eq:LU_2to2}, and in particular the construction of the causal flow, we always boost to the rest frame of the colliding photon pair. 

\subsubsection{Causal flow}
Equation~\eqref{eq:LU_2to2} is general enough to accommodate the Monte Carlo integration for the $\mathcal{O}(e_Q^4)$ \gls{nnlo} \gls{qcd} corrections to the $\gaga\to \QQ$ process. All the \gls{fs} diagrams contributing in $\Gamma_{\text{proc}}^{\text{\gls{fs}}}$ are indeed free of initial-state \gls{ir} singularities, enabling the computation of the corresponding interference diagrams within the \gls{lu} method. This is not the case for the \gls{fs} diagrams in which a photon couples to a massless quark, since these have initial-state collinear singularities that need to be renormalised in the photon PDF and mix with the resolved photon contributions as mentioned in the introduction section. We give an instance of such divergent \gls{fs} diagrams in figure~\ref{fig:IR_singular_FS}.

For the \gls{fs} diagrams in $\Gamma_{\text{proc}}^{\text{\gls{fs}}}$ that are proportional to $e_Q^4$, $\mathrm{d}\sigma_G/\mathrm{d}\mathcal{O}$, as defined in eq.~\eqref{eq:LU_2to2}, is \gls{ir} finite. For completeness, let us provide some details of its construction for the classes identified in section \ref{sec:FSGenumerate} in the rest frame of the colliding photon pair. In this frame, $\vec{p}_1+\vec{p}_2=\vec{0}$; for the example of figure~\ref{fig:GL258_cut_and_routed}, this means that $E_{10}$ and $E_9$, the on-shell energies associated with the edges $e_{10}$ and $e_9$, read $E_{10}=\sqrt{|\vec{k}_1+\vec{k}_2|^2+m_Q^2}$ and $E_{9}=\sqrt{|\vec{k}_1-\vec{k}_3|^2+m_Q^2}$. In particular, they do not depend on external kinematics. This in turn considerably simplifies the overlap structure defined in eq.~\eqref{eq:maximal_overlap}. Given this choice, the following holds:

\begin{itemize}
\item For the \gls{fs} diagrams in classes A, B, and D, the maximal overlap structure contains a single element, $\mathcal{O}^G_{\text{max}}=\{\mathcal{C}^G_{\text{proc}}\}$, independently of the values of $p_1$ and $p_2$, provided that $s>4m_Q^2$ (below threshold, it is empty). Furthermore, for all graphs we find a loop momentum basis, in which the origin is a valid source, $\vec{\mathbf{S}}_C=\vec{\mathbf{0}}$. The maximal value of $n_c$ is $n_c=3$, corresponding to dressing one of the $s$-channel edges of the one-loop heavy-quark contribution to $\gaga\to\gaga$ (figure~\ref{fig:LO_Graphs}) with two consecutive one-loop self-energies.

An example is the graph \texttt{GL258}, shown in figure~\ref{fig:GL258_cut_and_routed}. The existing cuts for this \gls{fs} diagram, when $s>4m_Q^2$, are all those shown in the figure, $c_1,...,c_{10}$, which are all compatible with the final-state definition for the process. The source $\vec{\mathbf{S}}_C=(\vec{k}_1=\vec{0},\vec{k}_1=\vec{0},\vec{k}_1=\vec{0})$ is inside all the surfaces associated with the cuts $c_1,...,c_{10}$. The causal flow simply reads $\partial_t\vec{\boldsymbol{\Phi}}_C(\vec{\mathbf{K}}; t)=e^t \vec{\mathbf{K}}$ and the multi-channelling factor is one, $\text{\gls{mc}}_C(\vec{\mathbf{K}};p_1,p_2)=1$.

\item For the \gls{fs} diagrams in class C, the maximal overlap structure can contain up to four elements, depending on the specific \gls{fs} graph and on the values of $p_1$ and $p_2$. When $s/m_Q^2>16/3$, the overlap structure contains a single element for all graphs in the class. However, in this case, a loop momentum basis does not always exist such that the origin is a valid source $\vec{\mathbf{S}}_C$. Nevertheless, it is possible to obtain a parametric form of the source in terms of $\vec{p}_1$. For $4<s/m_Q^2<16/3$, the \gls{fs} graph \texttt{GL208} (bottom-left in figure~\ref{fig:NNLO_singlet_graphs}) has a maximal overlap structure with two elements, whereas the graph \texttt{GL380} (bottom-right in the same figure) has one with four elements. In both cases, the sources $\vec{\mathbf{S}}_C$ can be expressed parametrically in terms of $\vec{p}_1$.

For \texttt{GL208} (figure~\ref{fig:GL208_cut_and_routed}), for example, the cuts compatible with the process definition are $\mathcal{C}_{\text{proc}}^G=\{c_1,c_2,c_4,c_5,c_6\}$,
which, for $4<s/m_Q^2<16/3$, define two overlap groups in the maximal overlap structure,
\begin{equation}
\mathcal{O}^G_{\text{max}}=\{C_1=\{c_1,c_4,c_5,c_6\},C_2=\{c_2,c_4,c_5,c_6\}\}\,.
\end{equation}
With the routing shown in figure~\ref{fig:GL208_cut_and_routed}, a valid source associated with $C_1$ is $\vec{\mathbf{S}}_{C_1}=(\vec{k}_1=\vec{0},\vec{k}_2=\vec{0},\vec{k}_3=-\vec{p}_1)$, while for $C_2$ a valid source is $\vec{\mathbf{S}}_{C_2}=(\vec{k}_1=\vec{0},\vec{k}_2=\vec{0}, \vec{k}_3=\vec{0})$. The existence of two overlaps also implies that the multi-channelling factors, after simplifications, read:
\begin{align}
&\text{\gls{mc}}_{C_1}(\vec{\mathbf{K}};p_1,p_2)=\frac{|\eta_{c_2}(\vec{\mathbf{K}};p_1,p_2)|^m}{|\eta_{c_1}(\vec{\mathbf{K}};p_1,p_2)|^m+|\eta_{c_2}(\vec{\mathbf{K}};p_1,p_2)|^m}\,, \\
&\text{\gls{mc}}_{C_2}(\vec{\mathbf{K}};p_1,p_2)=\frac{|\eta_{c_1}(\vec{\mathbf{K}};p_1,p_2)|^m}{|\eta_{c_1}(\vec{\mathbf{K}};p_1,p_2)|^m+|\eta_{c_2}(\vec{\mathbf{K}};p_1,p_2)|^m}\,.
\end{align}
For $s/m_Q^2>16/3$, by contrast, there is only one overlap group, $\mathcal{O}^G_{\text{max}}=\{\mathcal{C}^G_{\text{proc}}\}$, and the associated source is $\vec{\mathbf{S}}_{C}=(\vec{k}_1=\vec{0},\vec{k}_2=\vec{0}, \vec{k}_3=-\vec{p}_1/2)$.

\end{itemize}

\subsubsection{Threshold singularities\label{sec:thresinginLU}}

The \gls{lu} representation allows one to construct an integrand for the cross section that is free of \gls{ir} singularities (\ie\ collinear or soft, also known as pinched thresholds). However, non-pinched threshold singularities generally remain uncancelled and must be regularised through a contour deformation procedure. The amount and structure of threshold singularities requiring regularisation, as well as the specific regions of phase space in which this is necessary, depend on the process definition and the observable.

One way to study the threshold regularisation problem for the \gls{lu} integrand associated with a specific \gls{fs} diagram $G$ is as follows. First, construct the set of all Cutkosky cuts of the \gls{fs} diagram, $\mathcal{C}^G$, and identify the subset $\mathcal{C}^G_{\text{proc}}\subset\mathcal{C}^G$ that is compatible with the process definition. If there exists a cut $c\in \mathcal{C}^G\setminus \mathcal{C}^G_{\text{proc}}$ such that the surface $\eta_c(\vec{\mathbf{K}};p_1,p_2)=0$ is non-empty, contains a point $\vec{\mathbf{S}}$ (\ie\ $\eta_c(\vec{\mathbf{S}};p_1,p_2)<0$), and intersects at least one of the surfaces $\eta_{c^\prime}(\vec{\mathbf{K}};p_1,p_2)=0$ for $c^\prime\in \mathcal{C}^G_{\text{proc}}$, then the cut $c$ corresponds to a non-pinched threshold that requires regularisation. In the presence of a complicated observable, all cuts $c\in \mathcal{C}^G_{\text{proc}}$ may require threshold regularisation, depending on the region of phase space. For example, in N$^l$\gls{lo} \gls{qcd} corrections to the inclusive process $e^+e^-\rightarrow jj+X$, one has $\mathcal{C}^G=\mathcal{C}^G_{\text{proc}}$, and all threshold singularities cancel across the entire phase space. In this case, no additional threshold regularisation is needed: the \gls{lu} representation is finite from the start and can be directly fed into the Monte Carlo integrator. For the process at the heart of this paper, $\gaga\to \QQ+X$, however, this is not the case. Much like before, the discussion breaks-down per classes of \gls{fs} diagrams:

\begin{itemize}
 \item One can easily show that all \gls{fs} diagrams belonging to classes A, B, and D (as defined in section \ref{sec:FSGenumerate}) yield \gls{lu} integrands free of threshold singularities. Indeed, all the existing (namely, such that the associated surface has support) Cutkosky cuts of the \gls{fs} diagrams in these three classes are of the form $\QQ$, $\QQ j$ or $\QQ \QQ$, where the jet $j$ denotes any multiplicity of gluons and massless quarks. Since the process definition involves only two heavy quarks, all cuts containing four heavy quarks are excluded. One might worry that these cuts correspond to threshold singularities that require regularisation. However, this is not the case: the $\QQ$ and $\QQ j$ thresholds do not intersect with the $\QQ \QQ$ thresholds. A simple argument to see that this must hold is the following: if the two types of cuts intersected, then the \gls{lo} $\QQ \QQ$ process would also require threshold regularisation (such as a finite decay width) for all surfaces associated with the cuts of types $\QQ$ and $\QQ j$.

\item However, the \gls{fs} diagrams in class C, in addition to the type of cuts listed above, feature a new type of cut: the two-gluon $gg$ cut (see $c_3$ in figure~\ref{fig:GL208_cut_and_routed}). In particular, any diagram in class C has cuts of type: $\QQ$, $\QQ g$, $\QQ\QQ$ or $gg$, and only $\QQ$ and $\QQ g$ cuts are compatible with the process definition. While the Cutkosky cuts involving four heavy quarks ($c_7$ and $c_8$ in the same figure) again do not pose a problem, the double-gluon cuts represent a potential issue. In fact, these cuts correspond precisely to non-pinched thresholds of the \gls{lu} integrand that require regularisation. One way to address this issue is to note that the $gg$ cuts of the \gls{fs} diagrams in class C are exactly those contributing to the \gls{lo} loop-induced $\gaga\rightarrow gg$ process mediated by a closed heavy-quark loop. The problem of regularising the thresholds of the \gls{lu} integrand for the real-virtual $\gaga\to \QQ,\QQ g$ contribution can be circumvented as follows. Let $\sigma_G(\gaga\to \QQ,\QQ g)$ denote the contribution to the cross section from a \gls{fs} diagram $G$, where $G$ is one of the four singlet diagrams in class C (figure \ref{fig:NNLO_singlet_graphs}). Let $\sigma_G(\gaga\to \QQ, \QQ g, gg)$ denote the contribution from the same diagram including all Cutkosky cuts with $\QQ$, $\QQ g$ and $gg$ final states; this quantity can be computed in the \gls{lu} formalism without threshold regularisation. Finally, let $\sigma_G(\gaga\to gg)$ be the contribution from $G$ to the \gls{lo} $\gaga\to gg$ process mediated by a heavy-quark loop, computed using \mgshort. Then, the quantity of interest can be obtained as
\begin{align}
\sum_{G\in \text{class C}}\sigma_G(\gaga\to \QQ,\QQ g)=&\sum_{G\in \text{class C}}\underbrace{\sigma_G(\gaga\to \QQ,\QQ g, gg)}_{\text{\gls{lu}}} \nonumber\\
&-\underbrace{\sum_{G\in \text{class C}}\sigma_G(\gaga \to gg)}_{\mgshort}\,.
\label{eq:singlet_unitarity_relation}
\end{align}
With this rewriting, $\sigma_G(\gaga\to \QQ ,\QQ g)$ can be computed entirely through the \gls{lu} formalism, without the need for threshold regularisation. 

For \texttt{GL208} (figure~\ref{fig:GL208_cut_and_routed}), the \gls{lu} computation, corresponding to the first term on the right-hand side of eq.~\eqref{eq:singlet_unitarity_relation}, includes the contributions of the cuts $c_1$, $c_2$, $c_3$, $c_4$, $c_5$ and $c_6$. The contribution of $c_3$ is then subtracted via the second term on the right-hand side of eq.~\eqref{eq:singlet_unitarity_relation}, ensuring that the overall cross section on the left-hand side receives only the contributions consistent with the process definition.

\end{itemize}
Before presenting the results, we now discuss the tests we performed to verify the correctness of our implementation.

\subsection{Tests of our Local Unitarity implementation\label{sec:tests}}

To validate our implementation, we performed a series of cross-checks and tests, satisfied within our Monte-Carlo accuracy (percent level or better). On top of the agreement of our numerical results in the Coulomb limit shown in section~\ref{sec:threshold_region}, we verified the construction of the \gls{lu} integrand by:
\begin{enumerate}
\item[1)] Checking the \gls{lo} inclusive cross section for $\gaga\to t\bar{t}t\bar{t}$, which involves \gls{fs} diagrams in classes B, C, and D, against \mgshort. 
\item[2)] Computing the \gls{lo} inclusive process $\gaga\to gg$ mediated by a top-quark loop below threshold ($\sqrt{s}<2m_t$) which involves the computation of double-gluon cuts in all and only \gls{fs} diagrams in class C, and comparing against \mgshort.
\end{enumerate}
We also tested the correctness of the \gls{uv} subtraction and renormalisation procedures by:
\begin{enumerate}
\item[3)] Verifying that each \gls{fs} diagram is independent of the auxiliary scale $M_{\text{\gls{uv}}}$ used to construct the \gls{uv} renormalisation counter-terms.
\item[4)] Confirming that the sum of diagrams in class C does not exhibit any renormalisation scale dependence.
\item[5)] Checking that the sum of diagrams in classes A, B, and D reproduces the expected renormalisation group flow. Since the electric charge does not receive \gls{qcd} corrections, the \gls{nlo} \gls{qcd} correction term $\hat{\sigma}^{(1,2)}_Q$ has the renormalisation scale $\mu_R$ dependence only through $\alpha_s$. In contrast, the \gls{nnlo} \gls{qcd} correction $\hat{\sigma}^{(2,2)}_Q$ receives explicit scale dependence through one-loop quantum corrections to \gls{qcd} vertices and propagators. Remaining logarithms from one-loop quark self-energies combine with those from one-loop gluon self-energies and one-loop \gls{qcd} vertex corrections to give the one-loop evolution of the strong coupling. Overall, this yields
%\begin{equation}
%\hat{\sigma}^{(2,2)}_Q(\mu_R)=\hat{\sigma}^{(2,2)}_Q(\mu_R^\prime)+\beta_0\frac{\alpha_s(\mu_R^\prime)}{4\pi}\log{\left(\frac{\mu^2_R}{\mu_R^{\prime2}}\right)}\hat{\sigma}_Q^{(1,2)}(\mu_R^\prime)+\mathcal{O}(\alpha_s^3)\,,\label{eq:RGeq4sigma22}
%\end{equation}
\begin{equation}
\frac{\hat{\sigma}^{(2,2)}_Q(\mu_R)}{\alpha_s^{2}(\mu_R)}=\frac{\hat{\sigma}^{(2,2)}_Q(\mu_R^\prime)}{\alpha_s^{2}(\mu_R^\prime)}+\frac{\beta_0}{4\pi}\log{\left(\frac{\mu^2_R}{\mu_R^{\prime2}}\right)}\frac{\hat{\sigma}_Q^{(1,2)}(\mu_R^\prime)}{\alpha_s(\mu_R^\prime)}\,,\label{eq:RGeq4sigma22}
\end{equation}
where 
\begin{equation}
\beta_0\equiv \frac{11}{3}C_A-\frac{4}{3}T_Fn_q\,,\label{eq:beta0def}
\end{equation}
$C_A=3$, and $T_F=1/2$ in \gls{qcd}.
\end{enumerate}
Finally, we performed additional tests concerning the computation of diagrams in class C, which require the most advanced technology due to multiple overlaps:
\begin{itemize}
\item[6)] We compared two choices of photon polarisation spin-sum rules: $\sum_{\lambda} \varepsilon^\mu_\lambda \varepsilon_\lambda^{\nu\star} \rightarrow -g^{\mu\nu}$ and $\sum_\lambda \varepsilon_\lambda^\mu(k) \varepsilon_\lambda^{\nu\star}(k) \rightarrow -g^{\mu\nu} + \frac{k^\mu n^\nu+ n^\mu k^\nu}{k\cdot n} - \frac{k^\mu k^\nu}{ (k\cdot n)^2}$, with $n = (1,\vec{0})$, and verified that we obtained the same cross-section for the sum of diagrams in class C, although the values of individual diagrams differ.
\item[7)] We confirmed that the result for each \gls{fs} diagram in class C is independent of the choice of source $\vec{\mathbf{S}}_C$ entering eq.~\eqref{eq:causal_flow}, as long as $\eta_c(\vec{\mathbf{S}}_C;p_1,p_2)<0$ for all $c\in C$. Furthermore, results are independent of the choice of power $m$ in the multi-channelling factor of eq.~\eqref{eq:mc_factor}.
\end{itemize}
We now proceed to present our results.

\section{Results\label{sec:results}}

In this section, we discuss the results for the Coulomb-resummed \gls{nnlo} \gls{qcd} and \gls{nlo} \gls{ew} cross-section for $\gaga \to \QQ$. We start by deriving a Regge-motivated fit of the high-energy limit of the \gls{nnlo} \gls{qcd} corrections and then proceed to describe how its Coulomb enhancements can be resummed using tools from \gls{nrqcd}~\cite{Bodwin:1994jh} and \gls{pnrqcd}~\cite{Pineda:1997bj, Brambilla:1999xf, Beneke:1999zr, Beneke:1999qg}. Then, we provide a detailed analysis of the full results, including their general features and phenomenological implications, the values of the \gls{sm} parameters used for the computation, scale choice and variation and the estimation of theoretical uncertainties. We specialise our presentation to the production $Q=t$, $Q=b$, and $Q=c$ in $\epem$ scattering as well as ultraperipheral collisions of protons and heavy ions at a variety of different collider-motivated values of the \cm\ energy of the scattering charged particles.

\subsection{High-energy limit\label{sec:HElimit}}

The \gls{he} behaviour of the inclusive total cross section for $\gamma(p_1)\gamma(p_2)\to Q(p_Q)\bar{Q}(p_{\bar{Q}})+X$ in perturbative \gls{qcd} is dominated by the Regge limit, namely the kinematic region in which $s\sim -u\gg -t, m_Q^2$ or $s\sim -t\gg -u, m_Q^2$. We can use a Regge-theory-motivated ansatz to fit and extrapolate the high-energy behaviour. Unlike the reggeised gluon in the \gls{bfkl} approach~\cite{Lipatov:1976zz,Kuraev:1976ge,Kuraev:1977fs,Balitsky:1978ic}, the reggeised quark gives rise to double logarithms~\cite{Ermolaev:2017uhy}. Our \gls{he} fit of the complete \gls{nlo} \gls{qcd} calculation yields:%Although there is a possibility to predict the high-energy behaviour of the inclusive cross section for $\gamma(p_1)\gamma(p_2)\to Q(p_Q)\bar{Q}(p_{\bar{Q}})+X$ in perturbative \gls{qcd} by using reggeised quark, where the phase-space integrated cross section is dominated by the Regge limit where $s\sim -u\gg -t, m_Q^2$ or $s\sim -t\gg -u, m_Q^2$. The Mandelstam variables are defined as $t=(p_1-p_Q)^2$ and $u=(p_1-p_{\bar{Q}})^2$. We, however, do not attempt to use reggeised quark to predict the high-energy asymptotic behaviour here. Instead, we take the Regge theory motivated Ansatz to fit the high-energy behaviour to allow us to do extrapolation. Unlike the reggeised gluon in \gls{bfkl}, the reggeised quark gives rise double logarithms~\cite{Ermolaev:2017uhy}. Our fitted high-energy (\gls{he}) limit from the complete \gls{nlo} \gls{qcd} calculation yields
\begin{equation}
\hat{\sigma}_{Q,\mathrm{\gls{he}}}^{(1,2)}=\hat{\sigma}_{Q}^{(0,2)}\frac{\alpha_s}{\pi^2}\log{(2)}\left[\log^2{\left(\frac{s}{m_Q^2}\right)}-\frac{1}{2}\log{\left(\frac{s}{m_Q^2}\right)}\right]\,.\label{eq:NLOHE}
\end{equation}

\begin{figure}
\centering
    \includegraphics[width=0.45\textwidth]{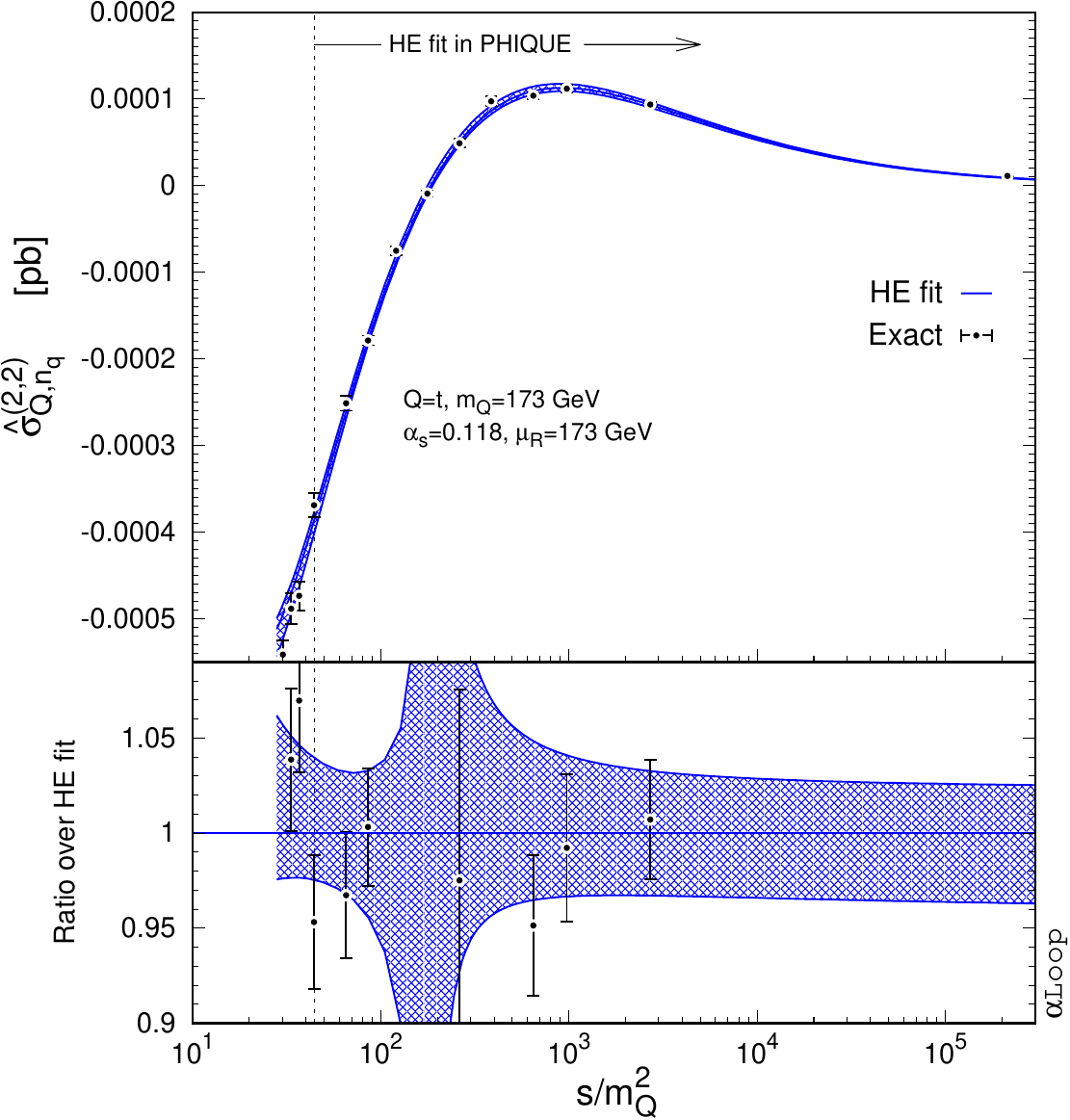}
    \includegraphics[width=0.45\textwidth]{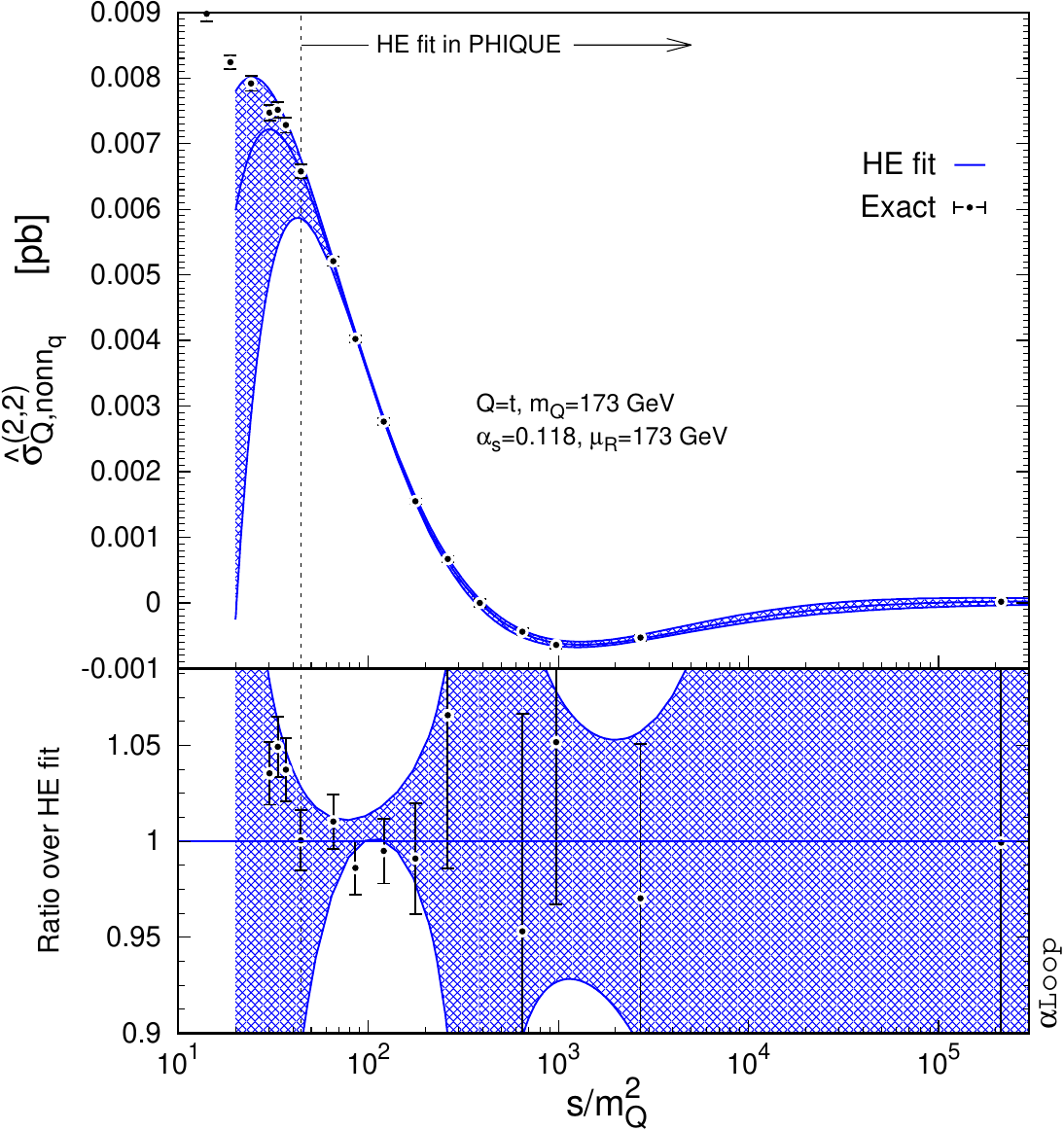}
    \caption{Comparison of the exact \gls{nnlo} \gls{qcd} corrections $\hat{\sigma}_{Q,n_q}^{(2,2)}$ and  $\hat{\sigma}_{Q,\mathrm{non}n_q}^{(2,2)}$ (black error bars) with the high-energy fits $\hat{\sigma}_{Q,n_q,\mathrm{\gls{he}}}^{(2,2)}$ and $\hat{\sigma}_{Q,\mathrm{non}n_q,\mathrm{\gls{he}}}^{(2,2)}$ (blue bands) as functions of $s/m_Q^2$. The top quark $Q = t$ is used as a representative case. The lower panel shows the ratio $\hat{\sigma}_{Q,n_q}^{(2,2)}/\hat{\sigma}_{Q,n_q,\mathrm{\gls{he}}}^{(2,2)}$ (left) and $\hat{\sigma}_{Q,\mathrm{non}n_q}^{(2,2)}/\hat{\sigma}_{Q,\mathrm{non}n_q,\mathrm{\gls{he}}}^{(2,2)}$ (right). The vertical dashed lines indicate that the central \gls{he} fit is used in \phique\ for $s/m_Q^2 > 44.18$.}
    \label{fig:DNNLOvssomQ2HE}
\end{figure}

Concerning the \gls{he} limit of the \gls{nnlo} \gls{qcd} correction, $\hat{\sigma}_Q^{(2,2)}$, we decompose it into $n_q$-dependent and $n_q$-independent parts:
%On the other hand, the \gls{nnlo} \gls{qcd} correction, $\hat{\sigma}_Q^{(2,2)}$, can be decomposed into $n_q$ dependent and $n_q$ independent parts:
\begin{equation}
\hat{\sigma}_Q^{(2,2)}=\hat{\sigma}_{Q,\mathrm{non}n_q}^{(2,2)}+n_q\hat{\sigma}_{Q,n_q}^{(2,2)}\,.\label{eq:sigmaQ22intonq}
\end{equation}
%\draftnoteZC{with a \gls{he} limit given by}
Their \gls{he} approximating fits are
%\begin{eqnarray}
%\hat{\sigma}_{Q,n_q,\mathrm{\gls{he}}}^{(2,2)}&=&\hat{\sigma}_{Q}^{(0,2)}\left(\frac{\alpha_s}{2\pi}\right)^2\Bigg\{\log^2{(2)}\bigg[\frac{8}{3}\log^2{\left(\frac{s}{m_Q^2}\right)}-\frac{39}{2}\log{\left(\frac{s}{m_Q^2}\right)}\bigg]+14\Bigg\}\nonumber\\
%&&-\frac{1}{3}\hat{\sigma}_{Q}^{(1,2)}\frac{\alpha_s}{2\pi}\log{\left(\frac{\mu_R^2}{m_Q^2}\right)}\,,\label{eq:NNLOnqHE}\\
%%\hat{\sigma}_{Q,\mathrm{non}n_q,\mathrm{\gls{he}}}^{(2,2),\mathrm{nosinglet}}&=&\hat{\sigma}_{Q}^{(0,2)}\left(\frac{\alpha_s}{2\pi}\right)^2\Bigg\{\log^2{(2)}\bigg[\frac{1}{3}\log^4{\left(\frac{s}{m_Q^2}\right)}-\frac{5}{2}\log^3{\left(\frac{s}{m_Q^2}\right)}-\frac{277}{6}\log^2{\left(\frac{s}{m_Q^2}\right)}\nonumber\\
%%&&+\frac{1265}{3}\log{\left(\frac{s}{m_Q^2}\right)}\bigg]-358\Bigg\}+\frac{11}{6}C_A\hat{\sigma}_{Q}^{(1,2)}\frac{\alpha_s}{2\pi}\log{\left(\frac{\mu_R^2}{m_Q^2}\right)}\,,\label{eq:NNLOnonnqHEnosinglet}\\
%\hat{\sigma}_{Q,\mathrm{non}n_q,\mathrm{\gls{he}}}^{(2,2)}&=&\hat{\sigma}_{Q}^{(0,2)}\left(\frac{\alpha_s}{2\pi}\right)^2\Bigg\{\log^2{(2)}\bigg[\frac{1}{3}\log^4{\left(\frac{s}{m_Q^2}\right)}-\frac{8}{3}\log^3{\left(\frac{s}{m_Q^2}\right)}-\frac{167}{4}\log^2{\left(\frac{s}{m_Q^2}\right)}\nonumber\\
%&&+\frac{775}{2}\log{\left(\frac{s}{m_Q^2}\right)}\bigg]-328\Bigg\}+\frac{11}{6}C_A\hat{\sigma}_{Q}^{(1,2)}\frac{\alpha_s}{2\pi}\log{\left(\frac{\mu_R^2}{m_Q^2}\right)}\,.\label{eq:NNLOnonnqHE}
%\end{eqnarray}
\begin{eqnarray}
\hat{\sigma}_{Q,n_q,\mathrm{\gls{he}}}^{(2,2)}&=&\hat{\sigma}_{Q}^{(0,2)}\left(\frac{\alpha_s}{2\pi}\right)^2\Bigg\{\log^2{(2)}\bigg[c_{n_q,2}\log^2{\left(\frac{s}{m_Q^2}\right)}+c_{n_q,1}\log{\left(\frac{s}{m_Q^2}\right)}\bigg]+c_{n_q,0}\Bigg\}\nonumber\\
&&-\frac{1}{3}\hat{\sigma}_{Q}^{(1,2)}\frac{\alpha_s}{2\pi}\log{\left(\frac{\mu_R^2}{m_Q^2}\right)}\,,\label{eq:NNLOnqHE}\\
\hat{\sigma}_{Q,\mathrm{non}n_q,\mathrm{\gls{he}}}^{(2,2)}&=&\hat{\sigma}_{Q}^{(0,2)}\left(\frac{\alpha_s}{2\pi}\right)^2\Bigg\{\log^2{(2)}\bigg[c_{\mathrm{non}n_q,4}\log^4{\left(\frac{s}{m_Q^2}\right)}+c_{\mathrm{non}n_q,3}\log^3{\left(\frac{s}{m_Q^2}\right)}\nonumber\\
&&+c_{\mathrm{non}n_q,2}\log^2{\left(\frac{s}{m_Q^2}\right)}
+c_{\mathrm{non}n_q,1}\log{\left(\frac{s}{m_Q^2}\right)}\bigg]+c_{\mathrm{non}n_q,0}\Bigg\}\nonumber\\
&&+\frac{11}{6}C_A\hat{\sigma}_{Q}^{(1,2)}\frac{\alpha_s}{2\pi}\log{\left(\frac{\mu_R^2}{m_Q^2}\right)}\,\,\label{eq:NNLOnonnqHE}
\end{eqnarray}
where the fitted coefficients are given by
\begin{eqnarray}
c_{n_q,2}&=&\left\{2.67,2.73,2.55\right\}\,,\quad c_{n_q,1}=\left\{-19.5,-19.9,-18.4\right\}\,,\nonumber\\
c_{\mathrm{non}n_q,4}&=&\left\{0.33,0.92,-0.60\right\}\,,\quad c_{\mathrm{non}n_q,3}=\left\{-2.67,-15.5,20.4\right\}\,,\nonumber\\
c_{\mathrm{non}n_q,2}&=&\left\{-41.8,64.1,-252\right\}\,,\quad c_{\mathrm{non}n_q,1}=\left\{388,4.52,1226\right\}\,,\nonumber\\
c_{n_q,0}&=&\left\{14.0,14.4,12.6\right\}\,,\quad c_{\mathrm{non}n_q,0}=\left\{-328,-81.2,-918\right\}\,,
\label{eq:coeffvalues}
\end{eqnarray}
corresponding to the central, maximal, and minimal values, respectively (obtained by varying the interpolated points by their corresponding Monte-Carlo errors).
 Whereas the renormalisation scale $\mu_R$ dependence is fixed by the renormalisation group, the $\log^k{\left(s/m_Q^2\right)}$ terms and constant contributions in eqs.~\eqref{eq:NLOHE}, \eqref{eq:NNLOnqHE}, and \eqref{eq:NNLOnonnqHE} are determined by fitting to our numerical data, which results in percent-level agreement for $s/m_Q^2\in [44,2700]$ (see figure \ref{fig:DNNLOvssomQ2HE}). In the implementation within the \phique\ code (cf. appendix~\ref{sec:phique}), we use the central \gls{he} fit for $s/m_Q^2 > 44.18$. Given the limited precision of our numerical data in the \gls{he} regime, we emphasise that an exact calculation based on Regge theory could yield different coefficients.
 
 We stress that the contribution to the total cross section from the \gls{he} region is suppressed by $1/s$, so that uncertainties from the fitting procedure above $s/m_Q^2 = 44.18$ induce errors well below our Monte Carlo resolution.
 Moreover, the coefficients given in eq.~\eqref{eq:coeffvalues} are highly correlated, so their large individual uncertainties translate into relatively moderate fitting errors, as shown in figure~\ref{fig:DNNLOvssomQ2HE}.

%initial width=0.45\textwidth
%\begin{figure}
%\centering
%    \includegraphics[width=0.45\textwidth]{figures/DNNLOnf_vs_somQ2_HE_top-crop.pdf}
%   % \includegraphics[width=0.32\textwidth]{figures/DNNLOnonnf_vs_somQ2_HE_top-crop.pdf}
%    \includegraphics[width=0.45\textwidth]{figures/DNNLOnonnf_withsinglet_vs_somQ2_HE_top-crop.pdf}
%    \caption{Comparison of the exact \gls{nnlo} \gls{qcd} corrections $\hat{\sigma}_{Q,n_q}^{(2,2)}$ and  $\hat{\sigma}_{Q,\mathrm{non}n_q}^{(2,2)}$ (black error bars) with the high-energy fits $\hat{\sigma}_{Q,n_q,\mathrm{\gls{he}}}^{(2,2)}$ and $\hat{\sigma}_{Q,\mathrm{non}n_q,\mathrm{\gls{he}}}^{(2,2)}$ (blue lines) as functions of $s/m_Q^2$. The top quark $Q = t$ is used as a representative case. The lower panel shows the ratio $\hat{\sigma}_{Q,n_q}^{(2,2)}/\hat{\sigma}_{Q,n_q,\mathrm{\gls{he}}}^{(2,2)}$ (left) and $\hat{\sigma}_{Q,\mathrm{non}n_q}^{(2,2)}/\hat{\sigma}_{Q,\mathrm{non}n_q,\mathrm{\gls{he}}}^{(2,2)}$ (right).}
%    \label{fig:DNNLOvssomQ2HE}
%\end{figure}

\newpage
\subsection{Coulomb resummation in the threshold region\label{sec:threshold_region}}

It is well known that the cross section for heavy-quark pair production (\ie\ in the threshold limit $s\to 4m_Q^2$) is plagued by Coulomb singularities. Near threshold, the N$^k$\gls{lo} contributions scale with respect to the \gls{lo} cross section as the $k$-th power of the expansion parameter $\alpha_s/\beta_Q$,~\footnote{Strictly speaking, according to the \gls{lp} Coulomb potential function~\cite{Fadin:1987wz,Fadin:1990wx,Beneke:2010da}, the leading Coulomb singularity of the form $\alpha_s^k/\beta_Q^k$ at N$^k$\gls{lo} vanishes for odd $k$ when $k\geq 3$.} where $\beta_Q$ denotes the relative velocity of the heavy quark in the partonic rest frame and is defined as
\begin{equation}
\beta_Q=\sqrt{1-\frac{4m_Q^2}{s}}\,.
\end{equation}
Analogous to heavy-quark pair hadroproduction and electroproduction~\cite{Fadin:1987wz,Fadin:1990wx,Jezabek:1992np,Jezabek:1992iy,Sumino:1992ai,Beneke:1999qg,Pineda:2006ri,Kiyo:2008bv,Hagiwara:2008df,Beneke:2009ye,Beneke:2010da,Beneke:2011mq,Ju:2019mqc,Ju:2020otc,Shao:2025dzw}, the production of a heavy-quark pair $\QQ$ in two-photon fusion can, in the threshold region ($\beta_Q\to 0$), be factorised into a short-distance cross section, describing the creation of the pair at a hard scale $\mu_R\sim m_Q$, and a potential function that accounts for the exchanges of Coulomb-type virtual gluons between the quark and antiquark~\footnote{The \gls{lp} Coulomb resummation for S-wave and P-wave top-quark production in $\gaga\to\ttbar$ is presented in ref.~\cite{Bigi:1992pr}.}. These
Coulombic contributions can be systematically described within \gls{nrqcd}~\cite{Bodwin:1994jh} and \gls{pnrqcd}~\cite{Pineda:1997bj, Brambilla:1999xf, Beneke:1999zr, Beneke:1999qg}, up to \gls{nlp} in the expansion around $\beta_Q\sim 0$. The phase-space-integrated cross section for $\gaga\to \QQ+X_p$ in the threshold region then factorises in \gls{pnrqcd} as
\begin{equation}
\hat{\sigma}_{\gaga\to \QQ+X_p}(s,m_Q^2)=\frac{d\hat{\sigma}_{Q,[1]}}{ds}J^{[1]}(E)+\mathcal{O}\left(\alpha^2\frac{\alpha_s^k}{\beta_Q^{k-3}},\alpha^3\right)\,,\label{eq:xsCoulresum}
\end{equation}
where $E\equiv \sqrt{s}-2m_Q=2m_Q\left(\frac{1}{\sqrt{1-\beta_Q^2}}-1\right)=m_Q\left(\beta_Q^2+\mathcal{O}(\beta_Q^4)\right)$ is the binding energy. Here, $J^{[1]}(E)$ denotes the colour-singlet long-distance potential function (computed up to \gls{nlp} in $\beta_Q$ in ref.~\cite{Beneke:2010da}) that resums multiple Coulomb-gluon exchanges, while $d\hat{\sigma}_{Q,[1]}/ds$ is the short-distance cross section for
the production of a colour-singlet $\QQ$ pair. At \gls{nnlo} in \gls{qcd}, the Coulomb-gluon enhancement may appear to yield \gls{nlp} enhancements $1/\beta_Q$ for colour-octet $\QQ$ production in the real-virtual contribution. However, since we consider only the phase-space-integrated cross section (rather than fully differential distributions), the short-distance cross section for a colour-octet $Q\bar{Q}$ pair is strongly suppressed near threshold and does not contribute at \gls{nlp}.
However, we stress that all ingredients necessary for including colour-octet contributions in the Coulomb resummation are already available (see, \eg\ refs. \cite{Hagiwara:2008df,Kiyo:2008bv,Beneke:2009ye}).

The potential function $J^{[1]}(E)$ is given by the imaginary part of the \gls{nrqcd} Green function $G^{[1]}(\vec{0},\vec{0};E)$ for the heavy-quark pair, evaluated at the spatial origin~\cite{Beneke:2010da}:
\begin{equation}
J^{[1]}(E)=2\text{Im}[G^{[1]}(\vec{0},\vec{0};E)]=\underbrace{2\text{Im}[G^{[1],\mathrm{\gls{lp}}}(\vec{0},\vec{0};E)]}_{\equiv J^{[1],\mathrm{\gls{lp}}}(E)}+\underbrace{2\text{Im}[G^{[1],\mathrm{\gls{nlp}}}(\vec{0},\vec{0};E)]}_{\equiv J^{[1],\mathrm{\gls{nlp}}}(E)}\,.\label{eq:potentialJ}
\end{equation}
Expressions for the \gls{lp} and \gls{nlp} Green functions at the origin can be found in the literature~\cite{Fadin:1987wz,Fadin:1990wx,Beneke:2010da,Beneke:2011mq}. 

\noindent Their expressions are:
\begin{eqnarray}
G^{[1],\mathrm{\gls{lp}}}\left(\vec{0},\vec{0};E\right)&=&-\frac{m_Q^2}{4\pi}\Bigg\{\sqrt{-\frac{E}{m_Q}}-D_1\alpha_s(\mu_C)\bigg[\frac{1}{2}\log{\left(-\frac{4m_QE}{\mu_C^2}\right)}-\frac{1}{2}+\gamma_E\nonumber\\
&+&\psi\left(1+\frac{D_1\alpha_s(\mu_C)}{2\sqrt{-E/m_Q}}\right)\bigg]\Bigg\}\,,\label{eq:LPG00CS}\\
G^{[1],\mathrm{\gls{nlp}}}(\vec{0},\vec{0};E)&=&-\frac{m_Q^2}{4\pi}D_{1}\frac{\alpha_s^2(\mu_C)}{4\pi}\Bigg\{a_1\bigg[-\frac{1}{2}\log{\left(-\frac{4m_QE}{\mu_C^2}\right)}+\lambda\psi^{(1)}(1-\lambda)-\psi(1-\lambda)-\gamma_E\bigg]\nonumber\\
&&+\beta_0\bigg[\frac{1}{4}\log^2{\left(-\frac{4m_QE}{\mu_C^2}\right)}-\log{\left(-\frac{4m_QE}{\mu_C^2}\right)}\left(\lambda \psi^{(1)}(1-\lambda)-\psi(1-\lambda)-\gamma_E\right)\nonumber\\
&&+4{}_4F_3(1,1,1,1;2,2,1-\lambda;1)+\lambda\psi^{(2)}(1-\lambda)-2\lambda\psi^{(1)}(1-\lambda)\left(\psi(1-\lambda)+\gamma_E\right)\nonumber\\
&&-3\psi^{(1)}(1-\lambda)+\left(\psi(1-\lambda)+\gamma_E\right)^2-\zeta_2\bigg]\Bigg\}\,,\label{eq:NLPG00CS}
\end{eqnarray}
where $D_1=-C_F=-4/3$, $\gamma_E$ is the Euler-Mascheroni constant, $a_1\equiv 31/9C_A-20/9T_Fn_q$, $\lambda\equiv -D_1\alpha_s(\mu_C)/(2\sqrt{-E/m_Q})$, $\zeta_n$ is the Riemann zeta value $\zeta_n\equiv \zeta(n)$, and $\mu_C$ is the Coulomb scale. The special functions appearing in the Green function expressions include the digamma function, $\psi(x)=d\log{\Gamma(x)}/dx$, its $n$th derivative, $\psi^{(n)}(x)=d^n\psi(x)/dx^n$, and the generalised hypergeometric function ${}_4F_3(\ldots)$. The generalised hypergeometric function can be expressed in terms of (nested) harmonic sums $S_n(-\lambda)$ and $S_{2,1}(-\lambda)$~\cite{Vermaseren:1998uu,Blumlein:1998if} (cf. appendix A.1 in ref.~\cite{Beneke:2011mq}):
\begin{equation}
\begin{aligned}
{}_4F_3(1,1,1,1;2,2,1-\lambda;1)=&\zeta_2-S_2(-\lambda)-\lambda\big[\zeta_3+S_3(-\lambda) \\
&-S_1(-\lambda)\left(\zeta_2-S_2(-\lambda)\right)-S_{2,1}(-\lambda)\big]\,\label{eq:HF4F3inS}
\end{aligned}
\end{equation}
for $\lambda\in\mathbb{C}$. For numerical evaluations of these harmonic sums, we follow the approach of ref.~\cite{Albino:2009ci}. Our implementation of eq.~\eqref{eq:HF4F3inS} in the \phique\ code (cf. appendix~\ref{sec:phique}) has been numerically validated against the built-in {\sc\small Mathematica} function \texttt{HypergeometricPFQ} for over 12,000 complex values of $\lambda$.
Given eqs.~\eqref{eq:LPG00CS} and \eqref{eq:NLPG00CS}, a natural choice for the Coulomb scale is $\mathcal{O}(\sqrt{4m_Q E})$. Equations \eqref{eq:potentialJ}, \eqref{eq:LPG00CS}, and \eqref{eq:NLPG00CS} can be analytically continued to $E<0$ and to non-zero heavy-quark width; however, this regime is not considered in this paper. It would be relevant only if we wanted to include contributions from bound states of the two heavy quarks. For the top quark, as we will show later, the Coulomb effect is quite small in the total cross section. For bottom and charm quarks, we include only the cross sections for open-bottom and open-charm production, rather than for bottomonium or charmonium production.

The partonic short-distance cross section $d\hat{\sigma}_{Q,[1]}/ds$ admits a perturbative expansion in $\alpha_s$:
\begin{equation}
\frac{d\hat{\sigma}_{Q,[1]}}{ds}=\frac{d\hat{\sigma}^{(0,2)}_{Q,[1]}}{ds}+\frac{d\hat{\sigma}^{(1,2)}_{Q,[1]}}{ds}+\mathcal{O}(\alpha_s^2\alpha^2,\alpha^3)\,,
\end{equation}
where $d\hat{\sigma}_{Q,[1]}^{(i,j)}/ds \propto \alpha_s^i \alpha^j$. 

\noindent The terms of the series are expressed as phase-space-integrated partonic cross sections. Specifically,
\begin{equation}
\frac{d\hat{\sigma}^{(0,2)}_{Q,[1]}}{ds}=\frac{2\pi(1-\beta_Q^2)}{\beta_Q m_Q^2}\hat{\sigma}_Q^{(0,2)}\,,\label{eq:SDLOexp}
\end{equation}
where the \gls{lo} cross section reads
\begin{equation}
\hat{\sigma}_Q^{(0,2)}=\frac{3\pi e_Q^4\alpha^2}{m_Q^2}\left(1-\beta_Q^2\right)\left[\frac{3-\beta_Q^4}{2}\log{\left(\frac{1+\beta_Q}{1-\beta_Q}\right)}-\beta_Q\left(2-\beta_Q^2\right)\right]\,.\label{eq:LOpxs}
\end{equation}
Here, $e_Q$ is the heavy-quark electric charge in units of the positron charge. In the threshold region $s\to 4m_Q^2$, where $\beta_Q\to 0$, the \gls{lo} cross section simplifies to
\begin{equation}
\hat{\sigma}_Q^{(0,2)}=\frac{3\pi e_Q^4\alpha^2}{m_Q^2}\beta_Q+\mathcal{O}(\beta_Q^3)\,.\label{eq:LOxsinbetaQexp0}
\end{equation}
The prefactor in eq.~\eqref{eq:SDLOexp} originates from the $1\to2$ phase space. The \gls{nlo} short-distance cross section $d\hat{\sigma}_{Q,[1]}^{(1,2)}/ds$ can be expressed in terms of the non-Coulomb part of the \gls{nlo} \gls{qcd} partonic cross section. The \gls{nlo} \gls{qcd} cross section is naturally decomposed into Coulomb and non-Coulomb parts:
\begin{equation}
\begin{aligned}
\hat{\sigma}_Q^{(1,2)}=\hat{\sigma}_{Q,\mathrm{Coul}}^{(1,2)}+\hat{\sigma}_{Q,\mathrm{nonCoul}}^{(1,2)}\,,
\end{aligned}
\end{equation}
where the Coulomb contribution is predicted by the \gls{lp} Coulomb resummation:
\begin{equation}
\begin{aligned}
\hat{\sigma}_{Q,\mathrm{Coul}}^{(1,2)}=&\left.\frac{d\hat{\sigma}_{Q,[1]}^{(0,2)}}{ds}J^{[1],\mathrm{\gls{lp}}}(E)\right|_{\mathcal{O}(\alpha_s\alpha^2),\mu_C\to \mu_R}\\
=&\frac{C_F\pi\alpha_s(\mu_R)(1-\beta_Q^2)}{2\beta_Q}\hat{\sigma}_Q^{(0,2)}=\frac{3C_F\pi^2e_Q^4\alpha^2\alpha_s(\mu_R)}{2m_Q^2}+\mathcal{O}(\beta_Q)\,.\label{eq:NLOQCDxsCoulapprox}
\end{aligned}
\end{equation}
Hence, the \gls{nlo} non-Coulomb contribution reads
\begin{equation}
\frac{d\hat{\sigma}_{Q,[1]}^{(1,2)}}{ds}=\frac{2\pi(1-\beta_Q^2)}{\beta_Q m_Q^2}\hat{\sigma}_{Q,\mathrm{nonCoul}}^{(1,2)}\label{eq:SDNLOexp}\,.
\end{equation}

\begin{figure}
    \includegraphics[trim=0.0cm -1.2cm 0cm 0cm,clip,width=1.0\textwidth]{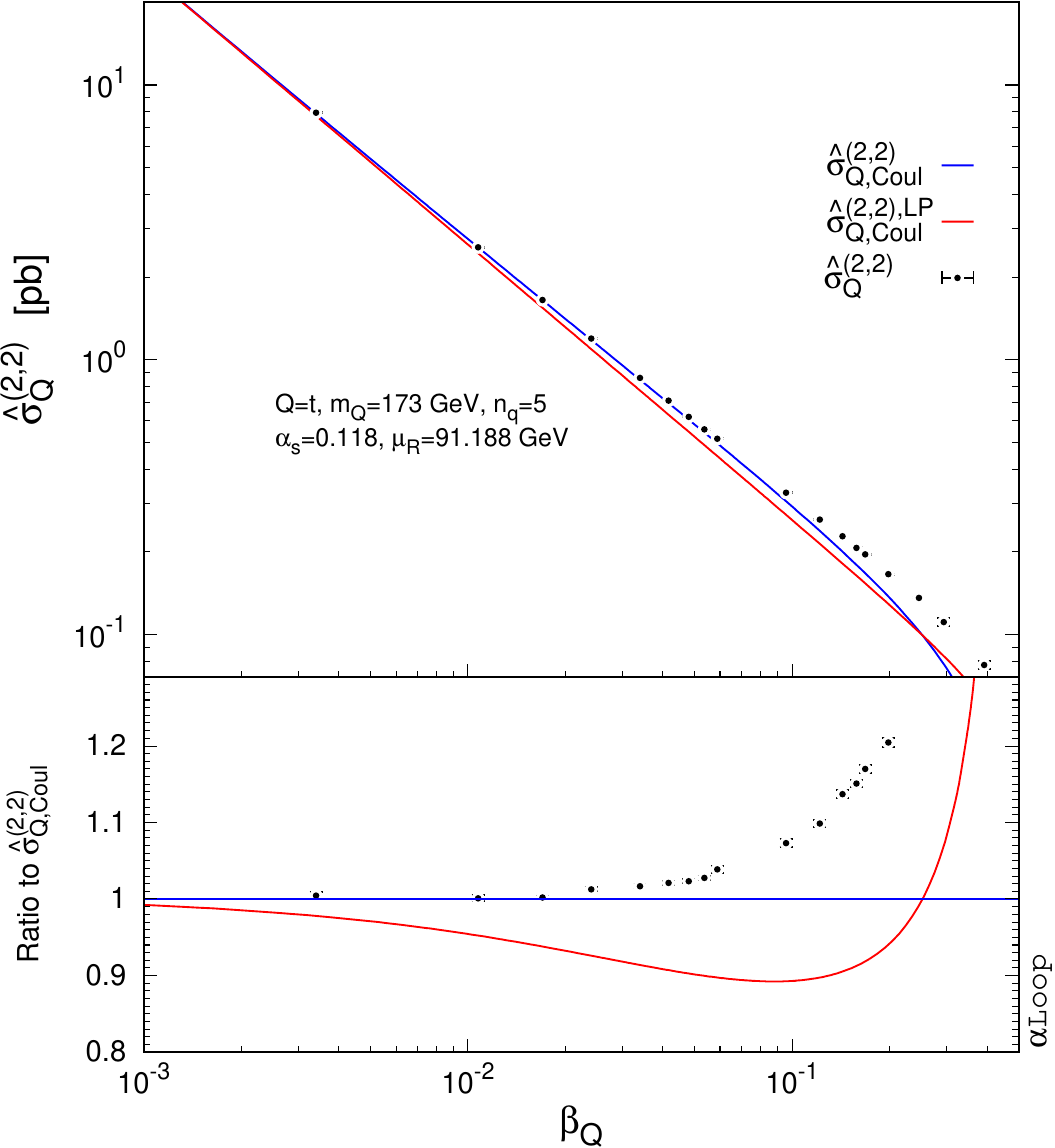}
    \caption{Comparison of the exact \gls{nnlo} \gls{qcd} correction $\hat{\sigma}_{Q}^{(2,2)}$ (black error bars) with the Next-to-Leading-Power (\gls{nlp}) Coulomb approximation $\hat{\sigma}_{Q,\mathrm{Coul}}^{(2,2)}$ (blue line) and the Leading-Power (\gls{lp}) Coulomb approximation $\hat{\sigma}_{Q,\mathrm{Coul}}^{(2,2),\mathrm{\gls{lp}}}$ (red line) as a function of $\beta_Q$. We consider the top quark ($Q=t$) as a representative case. The lower panel shows the ratios $\hat{\sigma}_{Q}^{(2,2)}/\hat{\sigma}_{Q,\mathrm{Coul}}^{(2,2)}$ and $\hat{\sigma}_{Q,\mathrm{Coul}}^{(2,2),\mathrm{\gls{lp}}}/\hat{\sigma}_{Q,\mathrm{Coul}}^{(2,2)}$.}
    \label{fig:DNNLOvsbetaCoul}
\end{figure}

Similarly, the \gls{nnlo} \gls{qcd} partonic cross section can be split into Coulomb and non-Coulomb parts:
\begin{equation}
\hat{\sigma}_Q^{(2,2)}=\hat{\sigma}_{Q,\mathrm{Coul}}^{(2,2)}+\hat{\sigma}_{Q,\mathrm{nonCoul}}^{(2,2)}\,,
\end{equation}
where the Coulomb part, predicted from eq.~\eqref{eq:xsCoulresum} up to \gls{nlp} in $\beta_Q$, is
\begin{align}
\hat{\sigma}_{Q,\mathrm{Coul}}^{(2,2)}=&\left.\frac{d\hat{\sigma}_{Q,[1]}}{ds}J^{[1]}(E)\right|_{\mathcal{O}(\alpha_s^2\alpha^2),\mu_C\to \mu_R}\nonumber \\
=&\left\{\frac{d\hat{\sigma}_{Q,[1]}^{(0,2)}}{ds}\left[J^{[1],\mathrm{\gls{lp}}}(E)+J^{[1],\mathrm{\gls{nlp}}}(E)\right]+\frac{d\hat{\sigma}_{Q,[1]}^{(1,2)}}{ds}J^{[1],\mathrm{\gls{lp}}}(E)\right\}_{\mathcal{O}(\alpha_s^2\alpha^2),\mu_C\to\mu_R}\nonumber\\
=&\underbrace{\alpha_s^2(\mu_R)\frac{C_F^2\pi^2(1-\beta_Q^2)^{\frac{3}{2}}}{12\beta_Q^2}\hat{\sigma}_Q^{(0,2)}}_{\equiv \hat{\sigma}_{Q,\mathrm{Coul}}^{(2,2),\mathrm{\gls{lp}}}}\nonumber\\
&+\underbrace{\alpha_s^2(\mu_R)\frac{C_F\left(a_1-\beta_0\log{\left(\frac{8m_Q^2}{\mu_R^2}\frac{1-\sqrt{1-\beta_Q^2}}{1-\beta_Q^2}\right)}\right)}{8\beta_Q}\hat{\sigma}_Q^{(0,2)}+\frac{C_F\pi\alpha_s(\mu_R)(1-\beta_Q^2)}{2\beta_Q}\hat{\sigma}_{Q,\mathrm{nonCoul}}^{(1,2)}}_{\equiv \hat{\sigma}_{Q,\mathrm{Coul}}^{(2,2),\mathrm{\gls{nlp}}}}\,.\label{eq:DNNLOCoul}
\end{align}
Figure \ref{fig:DNNLOvsbetaCoul} shows the comparison of the exact \gls{nnlo} \gls{qcd} correction $\hat{\sigma}_{Q}^{(2,2)}$ (black error bars) with the Coulomb approximation $\hat{\sigma}_{Q,\mathrm{Coul}}^{(2,2)}$ (blue line) and its \gls{lp} approximation $\hat{\sigma}_{Q,\mathrm{Coul}}^{(2,2),\mathrm{\gls{lp}}}$ (red line). The Coulomb approximation agrees with the exact computation at the percent level for $\beta_Q<0.03$. Together with renormalisation scale variation, this provides a powerful cross-check of the fully numerical computation of the \gls{nnlo} cross section presented in section~\ref{sec:results}.

Analogous to eq.~\eqref{eq:xsdefs1}, we define the Coulomb-resummed cross sections as
\begin{eqnarray}
\sigma_{\mathrm{A}_1\mathrm{A}_2\to\QQ}^{(\mathrm{\gls{nnlo}~\gls{qcd}+\gls{lp}})}&=&\sigma_{\mathrm{A}_1\mathrm{A}_2\to\QQ}^{(\mathrm{\gls{nnlo}~\gls{qcd}})}+\int^1_{4m_Q^2/S}{\mathrm{d}\tau \mathcal{L}_{\gaga}^{(\mathrm{A}_1\mathrm{A}_2)}(\tau)\left[\hat{\sigma}_{Q,\mathrm{\gls{lp}}}-\hat{\sigma}^{(0,2)}_Q-\hat{\sigma}_{Q,\mathrm{Coul}}^{(1,2)}-\hat{\sigma}^{(2,2),\mathrm{\gls{lp}}}_{Q,\mathrm{Coul}}\right]}\,,\nonumber\\
\sigma_{\mathrm{A}_1\mathrm{A}_2\to\QQ}^{(\mathrm{\gls{nnlo}~\gls{qcd}+\gls{nlp}})}&=&\sigma_{\mathrm{A}_1\mathrm{A}_2\to\QQ}^{(\mathrm{\gls{nnlo}~\gls{qcd}})}+\int^1_{4m_Q^2/S}{\mathrm{d}\tau \mathcal{L}_{\gaga}^{(\mathrm{A}_1\mathrm{A}_2)}(\tau)\left[\hat{\sigma}_{Q,\mathrm{\gls{nlp}}}-\hat{\sigma}^{(0,2)}_Q-\hat{\sigma}_{Q,\mathrm{Coul}}^{(1,2)}-\hat{\sigma}^{(2,2)}_{Q,\mathrm{Coul}}\right]}\,.\nonumber
\end{eqnarray}
The \gls{lp} and \gls{nlp} resummed partonic cross sections are
\begin{eqnarray}
\hat{\sigma}_{Q,\mathrm{\gls{lp}}}&=&\frac{d\hat{\sigma}_{Q,[1]}^{(0,2)}}{ds}J^{[1],\mathrm{\gls{lp}}}(E)\,,\nonumber\\
\hat{\sigma}_{Q,\mathrm{\gls{nlp}}}&=&\frac{d\hat{\sigma}_{Q,[1]}^{(0,2)}}{ds}\left[J^{[1],\mathrm{\gls{lp}}}(E)+J^{[1],\mathrm{\gls{nlp}}}(E)\right]+\frac{d\hat{\sigma}_{Q,[1]}^{(1,2)}}{ds}J^{[1],\mathrm{\gls{lp}}}(E)\,.
\end{eqnarray}
After combining with the \gls{nlo} \gls{ew} corrections, we define
\begin{align}
\sigma_{\mathrm{A}_1\mathrm{A}_2\to\QQ}^{(\mathrm{\gls{nnlo}~\gls{qcd}+\gls{nlp}+\gls{ew}})}=\sigma_{\mathrm{A}_1\mathrm{A}_2\to\QQ}^{(\mathrm{\gls{nnlo}~\gls{qcd}+\gls{nlp}})}+\Delta\sigma_{\mathrm{A}_1\mathrm{A}_2\to\QQ}^{(\mathrm{\gls{nlo}~\gls{ew}})}\, .
\end{align}
The Coulomb scale $\mu_C$ appears only in the potential functions $J^{[1],\mathrm{\gls{lp}}}(E)$ and $J^{[1],\mathrm{\gls{nlp}}}(E)$, while other ingredients, including fixed-order, short-distance, and Coulomb approximated cross sections, depend on the renormalisation scale $\mu_R$. Given the behaviour of the high-energy cross section as discussed in section \ref{sec:HElimit}, we choose the renormalisation scale as\footnote{At high energies, $\sqrt{s} \gg m_Q$, the typical off-shellness of the internal propagators is of order $\mathcal{O}(m_Q)$ rather than $\mathcal{O}(\sqrt{s})$.}
\begin{equation}
\mu_R=\mathrm{max}\left(\xi_Rm_Q,1~\mathrm{GeV}\right)\,,\label{eq:muRchoice}
\end{equation}
where $\xi_R$ is an $\mathcal{O}(1)$ variable varied up and down by a factor of two around unity. For the Coulomb scale, a natural choice from the potential function is $\mu_C\sim 2\beta_Qm_Q$. To prevent $\mu_C$ from hitting the Landau pole, we adopt the strategy suggested in ref.~\cite{Beneke:2010da}, where $\mu_C$ is frozen at the scale related to the inverse Bohr radius of the first $\QQ$ bound state, $C_F\alpha_s m_Q$. Specifically, we first solve
\begin{equation}
\mu_B=C_F\alpha_s(\mu_B)m_Q\,,
\end{equation}
and then define the Coulomb scale as
\begin{equation}
\mu_C=\mathrm{max}\left(\mathrm{min}\left(\xi_R\times\mathrm{max}\left(2\beta_Qm_Q,\mu_B\right),\mu_R\right),1~\mathrm{GeV}\right)\,.\label{eq:muCchoice}
\end{equation}
In other words, $\mu_C$ interpolates from $\mu_R$ in the relativistic regime ($\beta_Q>1/2$) to $\xi_R \mu_B$ in the ultra-nonrelativistic regime ($\beta_Q<\mu_B/(2m_Q)$). In the intermediate nonrelativistic regime ($\mu_B/(2m_Q)<\beta_Q<1/2$), the Coulomb scale behaves as $\mu_C=\xi_R (2\beta_Q m_Q)$, varying linearly with $\beta_Q$.
All results presented in this work that include Coulomb resummation employ the scale choice $\mu_C$ defined in eq.~\eqref{eq:muCchoice}. 
Additionally, we fully correlate the renormalisation and Coulomb scales through the common variation factor $\xi_R$. Although, in principle, one could vary $\mu_C$ and $\mu_R$ independently, we believe that fully correlating them is a more suitable choice to ensure the correct asymptotic limit $\mu_C \to \mu_R$ in the relativistic regime ($\beta_Q> 1/2$), which is important to preserve the achieved fixed-order accuracy.

\begin{figure}[!bt]
    \includegraphics[width=0.9\textwidth]{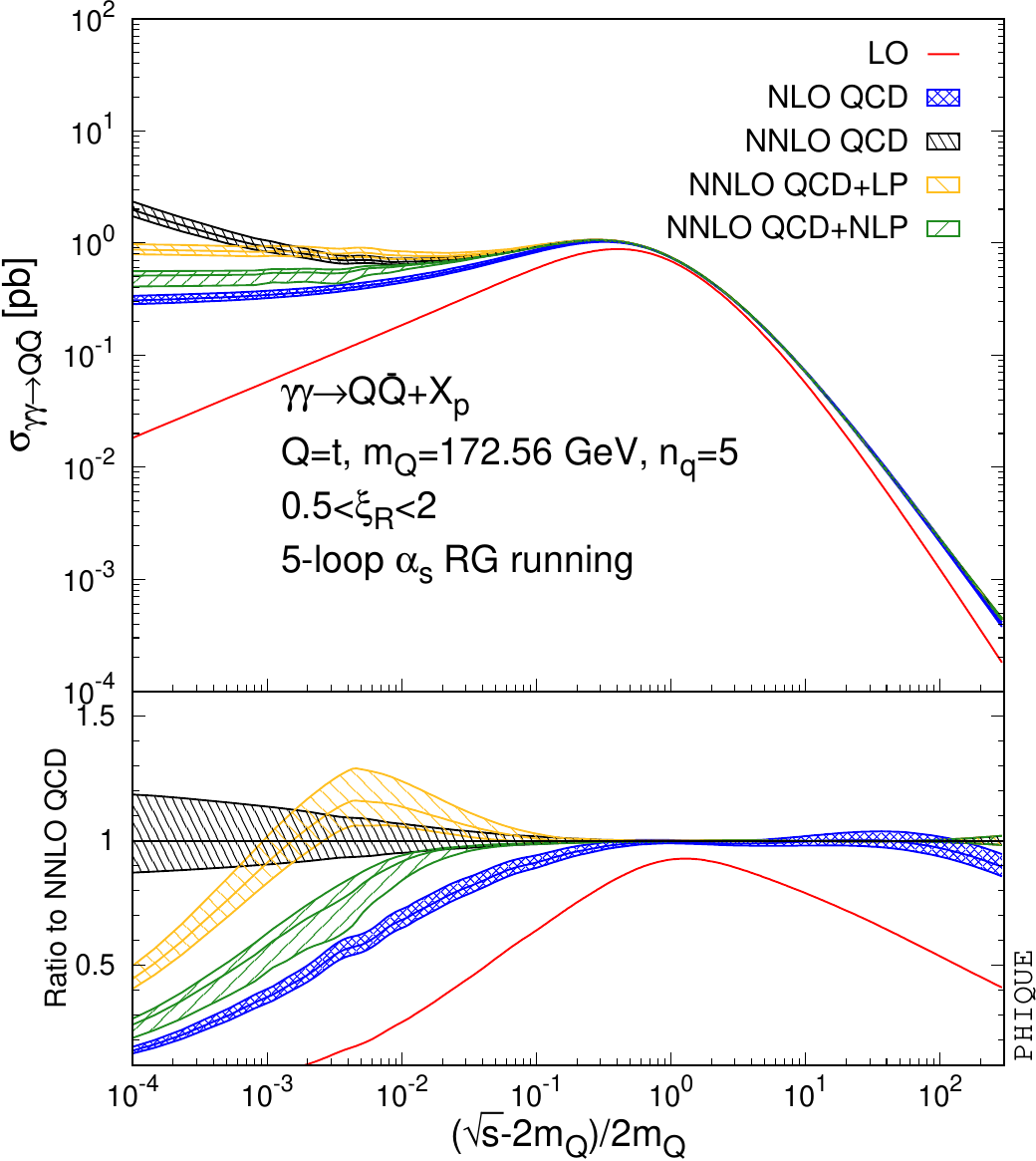}
    \caption{Comparison of the cross sections $\sigma_{\gaga\to\QQ}^{(\mathrm{\gls{lo}})}$ (red line), $\sigma_{\gaga\to\QQ}^{(\mathrm{\gls{nlo}~\gls{qcd}})}$ (blue hatched band), $\sigma_{\gaga\to\QQ}^{(\mathrm{\gls{nnlo}~\gls{qcd}})}$ (black hatched band), $\sigma_{\gaga\to\QQ}^{(\mathrm{\gls{nnlo}~\gls{qcd}+\gls{lp}})}$ (orange hatched band), and $\sigma_{\gaga\to\QQ}^{(\mathrm{\gls{nnlo}~\gls{qcd}+\gls{nlp}})}$ (green hatched band) as a function of $(\sqrt{s}-2m_Q)/(2m_Q)$ for top-quark pairs. The lower panel shows the cross section ratios relative to the central value of the \gls{nnlo} \gls{qcd} cross section.}
    \label{fig:xsCoulResum}
\end{figure}

Figure~\ref{fig:xsCoulResum} shows the comparison of cross sections $\sigma_{\gaga\to\QQ}^{(\mathrm{\gls{lo}})}$ (red line), $\sigma_{\gaga\to\QQ}^{(\mathrm{\gls{nlo}~\gls{qcd}})}$ (blue hatched band), $\sigma_{\gaga\to\QQ}^{(\mathrm{\gls{nnlo}~\gls{qcd}})}$ (black hatched band), $\sigma_{\gaga\to\QQ}^{(\mathrm{\gls{nnlo}~\gls{qcd}+\gls{lp}})}$ (orange hatched band), and $\sigma_{\gaga\to\QQ}^{(\mathrm{\gls{nnlo}~\gls{qcd}+\gls{nlp}})}$ (green hatched band) as a function of $(\sqrt{s}-2m_Q)/2m_Q$ for top-quark pairs ($Q=t$). The bands represent the scale uncertainties. The lower panel displays the cross section ratios over the central value of the \gls{nnlo} \gls{qcd} cross section. In the threshold region, the \gls{lo} cross section decreases linearly with $\beta_Q$ (cf. eq.~\eqref{eq:LOxsinbetaQexp0}), the \gls{nlo} \gls{qcd} cross section approaches a constant (cf. eq.~\eqref{eq:NLOQCDxsCoulapprox}), and the \gls{nnlo} \gls{qcd} cross section grows as $1/\beta_Q$ (cf. eq.~\eqref{eq:DNNLOCoul}). Due to Coulomb enhancements, the scale uncertainty bands do not overlap in this region. The \gls{lp} and \gls{nlp} Coulomb resummation calculations cure the $1/\beta_Q$ divergence, leading to cross sections that asymptotically approach constants. The scale uncertainty bands of the resummation-improved cross sections again do not overlap. Other effects, such as bound-state ($E<0$) and finite-width effects, are ignored here but may also be important. In the intermediate region, $\sqrt{s}-2m_Q\sim\mathcal{O}(m_Q)$, fixed-order perturbation theory works well. \gls{nlo} and \gls{nnlo} \gls{qcd} corrections only mildly enhance the cross section, and Coulomb resummation effects are small for $(\sqrt{s}-2 m_Q)/2 m_Q \gtrsim 0.1$, as seen by comparing \gls{nnlo} \gls{qcd} and \gls{nnlo} \gls{qcd}+(N)\gls{lp} results. In the high-energy region, $\sqrt{s}\gg 2m_Q$, cross sections generally decrease as $1/s$. Logarithmic enhancements in higher-order \gls{qcd} corrections (cf. eqs.~\eqref{eq:NLOHE}, \eqref{eq:NNLOnqHE}, and \eqref{eq:NNLOnonnqHE}) make the \gls{nlo} and \gls{nnlo} \gls{qcd} results significantly larger than the \gls{lo} prediction.

\clearpage
\newpage
\subsection{Top quark\label{sec:topres}}

In this section, we present our predictions for the cross sections for top-quark pair production in photon fusion in ultraperipheral hadron
collisions and at $\epem$ colliders. We use the OS top-quark mass $m_t$, the Higgs boson mass $m_h$, the $W$ boson mass $m_W$, and the $Z$ boson mass $m_Z$ with the following values~\cite{ParticleDataGroup:2024cfk}:
\begin{eqnarray}
m_t&=&172.56~\mathrm{GeV}\,,\quad m_h=125.2~\mathrm{GeV}\,,\nonumber\\
m_W&=&80.3692~\mathrm{GeV}\,,\quad m_Z=91.188~\mathrm{GeV}\,.\label{eq:mass4top}
\end{eqnarray}
All other quark masses are set to zero, and particle widths are neglected.\\
The \gls{ckm} matrix~\cite{Cabibbo:1963yz,Kobayashi:1973fv} is taken to be the identity. The \gls{rg} evolution of the strong coupling $\alpha_s(\mu_R)$ follows the five-loop \gls{qcd} beta function~\cite{Baikov:2016tgj,Herzog:2017ohr}, with the initial condition $\alpha_s(m_Z)=0.118$. The four-loop decoupling relations for the strong coupling constant~\cite{Chetyrkin:1997un,Schroder:2005hy,Chetyrkin:2005ia,Gerlach:2018hen} are employed with the OS top-quark (bottom-quark, charm-quark) mass thresholds at $173.3$ GeV ($4.7$ GeV, $1.5$ GeV)\footnote{
We note that our computation of \gls{nnlo} \gls{qcd} corrections considers $\alpha_s$ renormalisation at fixed $n_q$. This mismatch with respect to our $\alpha_s$ running can introduce small discrepancies for low values of the renormalisation scale.
}. The electromagnetic coupling constant $\alpha$ entering the external photon vertices is in the $\alpha(0)$ scheme with $\alpha^{-1}(0)=137.036$. In \gls{nlo} \gls{ew} calculations, we use the hybrid $\alpha$ renormalisation scheme described in refs.~\cite{Pagani:2021iwa,Shao:2025bma}, where the $\alpha$ value in the \gls{ew} radiative corrections is defined in the $G_\mu$ scheme with the Fermi constant $G_\mu=1.16639\times 10^{-5}$ GeV$^{-2}$, \ie\ $\alpha^{-1}_{G_\mu}\approx 132.099$. We additionally take the renormalisation and Coulomb scales as in eqs.~\eqref{eq:muRchoice} and \eqref{eq:muCchoice}, respectively. 

\begin{table}[htpb!]
\tabcolsep=1.0mm
\centering
\makebox[\textwidth][c]{%
\begin{tabular}{lr|r|r:l|r:l|r:l}
\toprule
\multicolumn{2}{l}{ $\gaga\to \ttbar$} & \multicolumn{7}{c}{\gls{nnlo} \gls{qcd} + \gls{nlp} with \phique} \\
\midrule
\multicolumn{2}{c}{} 
& \multicolumn{1}{c}{\gls{lo}} 
& \multicolumn{2}{c}{\gls{nlo} \gls{qcd}} 
& \multicolumn{2}{c}{\gls{nnlo} \gls{qcd}} 
& \multicolumn{2}{c}{\gls{nnlo} \gls{qcd}+\gls{nlp}} \\
\cmidrule{3-9}
\multicolumn{1}{c}{$\mathrm{A}_1$-$\mathrm{A}_2$}& \multicolumn{1}{c}{ $\sqrt{S}$}& \multicolumn{1}{c}{$\sigma$}  & \multicolumn{1}{c}{$\sigma$} & \multicolumn{1}{c}{$K$} & \multicolumn{1}{c}{$\sigma$} & \multicolumn{1}{c}{$K$} & \multicolumn{1}{c}{$\sigma$} & \multicolumn{1}{c}{$K$} \\
\midrule
\pp& 13~TeV   & $212.3$ ab  & $257.3^{+4.7}_{-3.5}$ ab & 1.212 & $269.9(3)^{+2.5}_{-1.9}$ ab & 1.271 & $267.2(3)^{+0.0}_{-1.1}$ ab & 1.258 \\
\pp& 13.6~TeV & $228.6$ ab  & $276.7^{+5.1}_{-3.8}$ ab & 1.211 & $290.1(3)^{+2.7}_{-2.1}$ ab & 1.269 & $287.2(3)^{+0.0}_{-1.2}$ ab & 1.257 \\
\pp& 14~TeV   & $239.5$ ab  & $289.8^{+5.3}_{-3.9}$ ab & 1.210 & $303.7(3)^{+2.8}_{-2.2}$ ab & 1.268 & $300.7(3)^{+0.0}_{-1.2}$ ab & 1.256 \\
\pp& 100~TeV  & $2.307$ fb  & $2.725^{+0.044}_{-0.033}$ fb & 1.181 & $2.827(2)^{+0.020}_{-0.016}$ fb & 1.225 & $2.805(2)^{+0.000}_{-0.008}$ fb & 1.216 \\
\pPb& 62.8~TeV& $2.822$ pb & $3.392^{+0.060}_{-0.045}$ pb & 1.202 & $3.545(3)^{+0.031}_{-0.024}$ pb & 1.256 & $3.513(3)^{+0.000}_{-0.013}$ pb & 1.244 \\
\PbPb& 39.4~TeV& $0.9017$ nb  & $1.143^{+0.025}_{-0.019}$ nb & 1.267 & $1.220(2)^{+0.016}_{-0.012}$ nb & 1.353 & $1.203(2)^{+0.001}_{-0.007}$ nb & 1.335 \\
$\epem$& 365 GeV  & $0.256$ ab  & $0.564^{+0.033}_{-0.024}$ ab & 2.208 & $0.831(2)^{+0.057}_{-0.041}$ ab & 3.250 & $0.726(2)^{+0.016}_{-0.050}$ ab & 2.840 \\
$\epem$& 500 GeV  & $42.2$ ab & $60.2^{+1.9}_{-1.4}$ ab & 1.426 & $67.6(2)^{+1.6}_{-1.2}$ ab & 1.603 & $65.9(2)^{+0.1}_{-0.8}$ ab & 1.562 \\
$\epem$& 1000 GeV & $0.739$ fb & $0.924^{+0.019}_{-0.014}$ fb & 1.250 & $0.980(1)^{+0.011}_{-0.009}$ fb & 1.326 & $0.968(1)^{+0.001}_{-0.005}$ fb & 1.309 \\
$\epem$& 3000 GeV & $4.885$ fb  & $5.825^{+0.099}_{-0.074}$ fb & 1.192 & $6.069(6)^{+0.049}_{-0.038}$ fb & 1.242 & $6.017(6)^{+0.001}_{-0.021}$ fb & 1.232 \\
\bottomrule
\end{tabular}%
}
\caption{Inclusive total cross sections and $K$ factors (as defined in eq.~\eqref{eq:xsdefs1}) for $\gaga\to \ttbar$ at \gls{lo}, \gls{nlo} \gls{qcd}, \gls{nnlo} \gls{qcd}, and \gls{nnlo} \gls{qcd}+\gls{nlp} accuracy at hadron and $\epem$ collider energies. Calculations use the ChFF $\gamma$ fluxes for hadron colliders and the iWW approximation for lepton colliders. Scale uncertainties correspond to renormalisation scale variations by a factor of $2$, and numerical errors on the last digit from Monte Carlo integration in $\hat{\sigma}_Q^{(2,2)}$ are given in parentheses.}
\label{tab:ttbarxsQCD}
\end{table}

\begin{table}[htbp]
\centering
\renewcommand{\arraystretch}{1.2}
\makebox[\textwidth][c]{%
\begin{tabular}{lr|r:l|r:l|r:l}
\toprule
\multicolumn{2}{l}{ $\gaga\to \ttbar$} & \multicolumn{6}{c}{\gls{nnlo} \gls{qcd} + \gls{nlp} with \phique \ and \gls{nlo} \gls{ew} with \mgshort} \\
\midrule
\multicolumn{2}{c}{}
& \multicolumn{2}{c}{\gls{nlo} \gls{qcd}+\gls{ew}} 
& \multicolumn{2}{c}{\gls{nnlo} \gls{qcd}+\gls{ew}} 
& \multicolumn{2}{c}{\gls{nnlo} \gls{qcd}+\gls{nlp}+\gls{ew}} \\
\cmidrule{3-8}
\multicolumn{1}{c}{$\mathrm{A}_1$-$\mathrm{A}_2$}& \multicolumn{1}{c}{$\sqrt{S}$} & \multicolumn{1}{c}{$\sigma$} & \multicolumn{1}{c}{$K$} & \multicolumn{1}{c}{$\sigma$} & \multicolumn{1}{c}{$K$} & \multicolumn{1}{c}{$\sigma$} & \multicolumn{1}{c}{$K$} \\
\midrule
\pp& 13~TeV   & $245.8^{+4.7}_{-3.5}$ ab & 1.158 & $258.4(3)^{+2.5}_{-1.9}$ ab & 1.217 & $255.7(3)^{+0.0}_{-1.1}$ ab & 1.204 \\
\pp& 13.6~TeV & $264.4^{+5.1}_{-3.8}$ ab & 1.157 & $277.7(3)^{+2.7}_{-2.1}$ ab & 1.215 & $274.9(3)^{+0.0}_{-1.2}$ ab & 1.203 \\
\pp& 14~TeV   & $276.8^{+5.3}_{-3.9}$ ab & 1.156 & $290.7(3)^{+2.8}_{-2.2}$ ab & 1.214 & $287.7(3)^{+0.0}_{-1.2}$ ab & 1.201 \\
\pp& 100~TeV  & $2.595^{+0.044}_{-0.033}$ fb & 1.125 & $2.697(2)^{+0.020}_{-0.016}$ fb & 1.169 & $2.676(2)^{+0.000}_{-0.008}$ fb & 1.160 \\
\pPb& 62.8~TeV& $3.238^{+0.060}_{-0.045}$ pb & 1.147 & $3.391(3)^{+0.031}_{-0.024}$ pb & 1.202 & $3.359(3)^{+0.000}_{-0.013}$ pb & 1.190 \\
\PbPb& 39.4~TeV& $1.094^{+0.025}_{-0.019}$ nb & 1.213 & $1.171(2)^{+0.016}_{-0.012}$ nb & 1.299 & $1.155(2)^{+0.001}_{-0.007}$ nb & 1.280 \\
$\epem$& 365~GeV & $0.554^{+0.033}_{-0.024}$ ab & 2.165 & $0.820(2)^{+0.057}_{-0.041}$ ab & 3.208 & $0.715(2)^{+0.016}_{-0.050}$ ab & 2.797 \\
$\epem$& 500~GeV & $57.8^{+1.9}_{-1.4}$ ab & 1.371 & $65.3(2)^{+1.6}_{-1.2}$ ab & 1.548 & $63.6(2)^{+0.1}_{-0.8}$ ab & 1.507 \\
$\epem$& 1000~GeV& $0.884^{+0.019}_{-0.014}$ fb & 1.196 & $0.941(1)^{+0.011}_{-0.009}$ fb & 1.272 & $0.928(1)^{+0.001}_{-0.005}$ fb & 1.255 \\
$\epem$& 3000~GeV& $5.559^{+0.099}_{-0.074}$ fb & 1.138 & $5.803(6)^{+0.049}_{-0.038}$ fb & 1.188 & $5.752(6)^{+0.001}_{-0.021}$ fb & 1.177 \\
\bottomrule
\end{tabular}}
\caption{Inclusive total cross sections and $K$ factors (as defined in eq.~\eqref{eq:xsdefs1}) for $\gaga\to \ttbar$ at \gls{nlo} \gls{qcd}+\gls{ew}, \gls{nnlo} \gls{qcd}+\gls{ew}, and \gls{nnlo} \gls{qcd}+\gls{nlp}+\gls{ew} accuracy at hadron and $\epem$ collider energies. Calculations use the ChFF $\gamma$ fluxes for hadron colliders and the iWW approximation for lepton colliders. Scale uncertainties correspond to renormalisation scale variations by a factor of $2$, and numerical errors on the last digit from Monte Carlo integration in $\hat{\sigma}_Q^{(2,2)}$ are given in parentheses.}
\label{tab:ttbarxsEW}
\end{table}

We now present our predictions for the total cross sections for $\gaga\to \ttbar$ at hadron and electron-positron colliders. Using as input the partonic cross section $\hat{\sigma}^{(2,2)}_Q$ in terms of $\sqrt{s}$ computed with \alphaLoop, we have released a new code, dubbed \phique~\cite{phique} (see appendix \ref{sec:phique}), which accompanies this paper. All physical results, except for the \gls{nlo} \gls{ew} corrections, are obtained with \phique, while the \gls{nlo} \gls{ew} corrections are computed within the \mgshort\ framework~\cite{Alwall:2014hca,Frederix:2018nkq,Shao:2025bma}. 

The total cross sections are evaluated at the \gls{lhc} (\pp\ collisions at $\sqrtsnn=13$-$14$ TeV, \pPb\ at $\sqrtsnn=8.79$ TeV, and \PbPb\ at $\sqrtsnn=5.52$ TeV), at FCC-hh~\cite{FCC:2018vvp} (\pp\ at $\sqrtsnn=100$ TeV, \pPb\ at $\sqrtsnn=62.8$ TeV, and \PbPb\ at $\sqrtsnn=39.4$ TeV), at future circular $\epem$ colliders such as FCC-ee~\cite{FCC:2018evy} and CEPC~\cite{CEPCStudyGroup:2023quu} ($\sqrt{s_{\epem}}=365$ GeV), and at future linear $\epem$ colliders like ILC~\cite{Behnke:2013xla} and CLIC~\cite{CLICdp:2018cto} ($\sqrt{s_{\epem}}=500$-$3000$ GeV). At hadron colliders, we only consider ultraperipheral collisions (\gls{upc}s), using the ChFF photon flux model implemented in the \gammaUPC\ code~\cite{Shao:2022cly}. For $\epem$ colliders, the photon flux is modelled in the iWW approximation~\cite{Frixione:1993yw} (cf. eq.~\eqref{eq:WWPDF4lepton}), with $Q_{\mathrm{max}}=1$ GeV.

We report total cross sections for $\gaga\to\ttbar$ at hadron and $\epem$ colliders in tables~\ref{tab:ttbarxsQCD} and \ref{tab:ttbarxsEW}. The \gls{lo}, \gls{nlo} \gls{qcd}, and \gls{nlo} \gls{qcd}+\gls{ew} cross sections in \gls{upc}s at the \gls{lhc} and FCC-hh have already been reported in ref.~\cite{Shao:2025bma}, albeit with a slightly different setup. The new results presented here include \gls{nnlo} \gls{qcd} corrections and \gls{nlp} Coulomb resummation, as well as predictions for $\epem$ colliders. At the \gls{lhc}, the process is hardly observable in heavy-ion collisions due to their low integrated luminosities (cf. table II in ref.~\cite{Shao:2022cly}). In contrast, its observation in \pp\ collisions is expected at the high-luminosity phase of the \gls{lhc}~\cite{CMS:2021ncv,Goncalves:2020saa,Martins:2022dfg}, provided that the effect of backgrounds, especially those arising from pile-up collisions, can be efficiently identified by intact forward proton tagging. For future colliders, the observability of this process strongly depends on yet-undecided factors, most notably the targeted integrated luminosities $\mathcal{L}_{\mathrm{int}}$. For example, the process is unlikely to be observed with only a few ab$^{-1}$ of data at the FCC-ee $\sqrt{s_{\epem}}=365$ GeV run (cf. table 1 in ref.~\cite{dEnterria:2025hbe}). The inclusion of higher-order \gls{qcd} and \gls{ew} corrections does not qualitatively change these expectations.

\begin{figure}
    \includegraphics[width=0.49\textwidth]{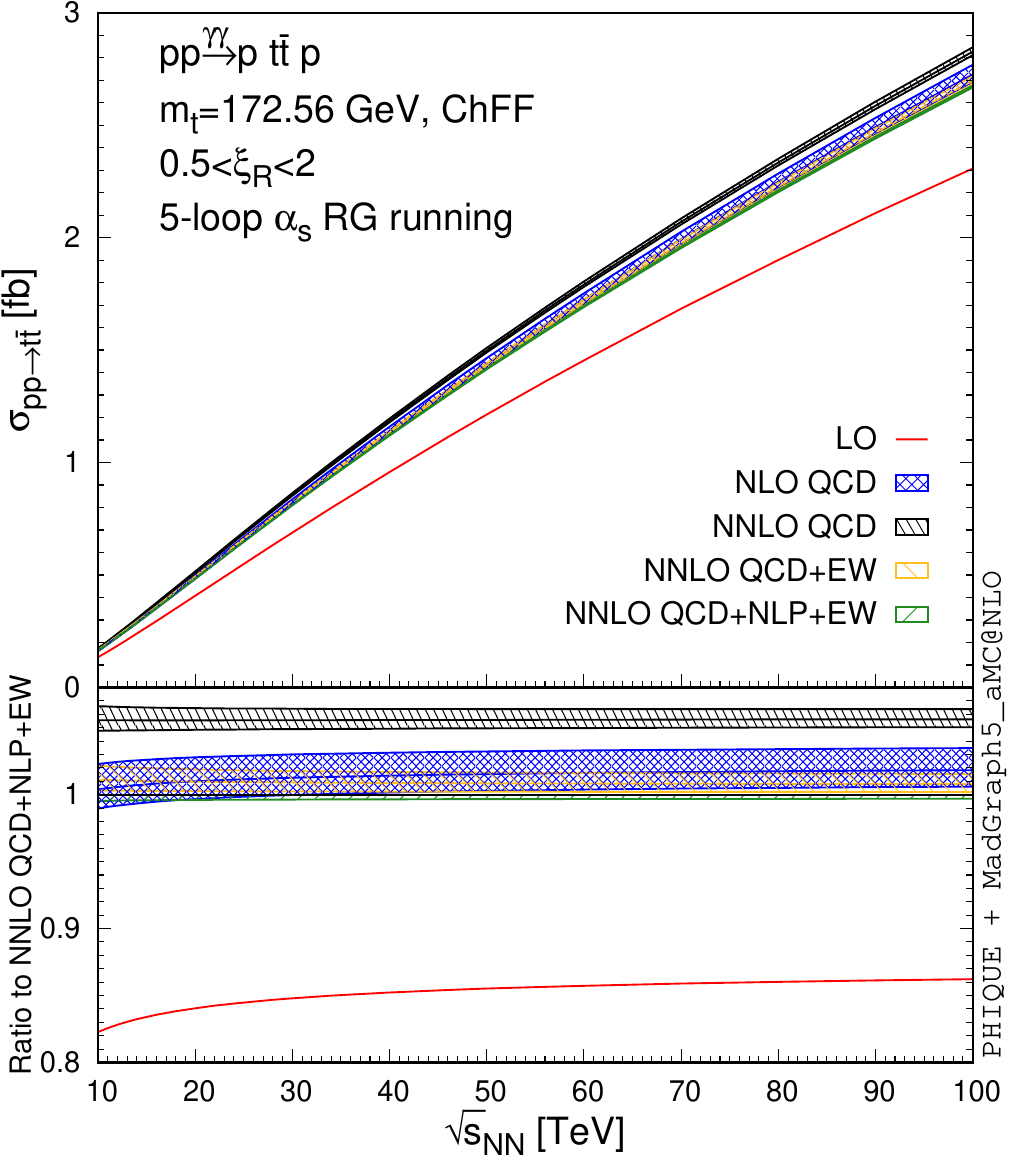}
    \includegraphics[width=0.49\textwidth]{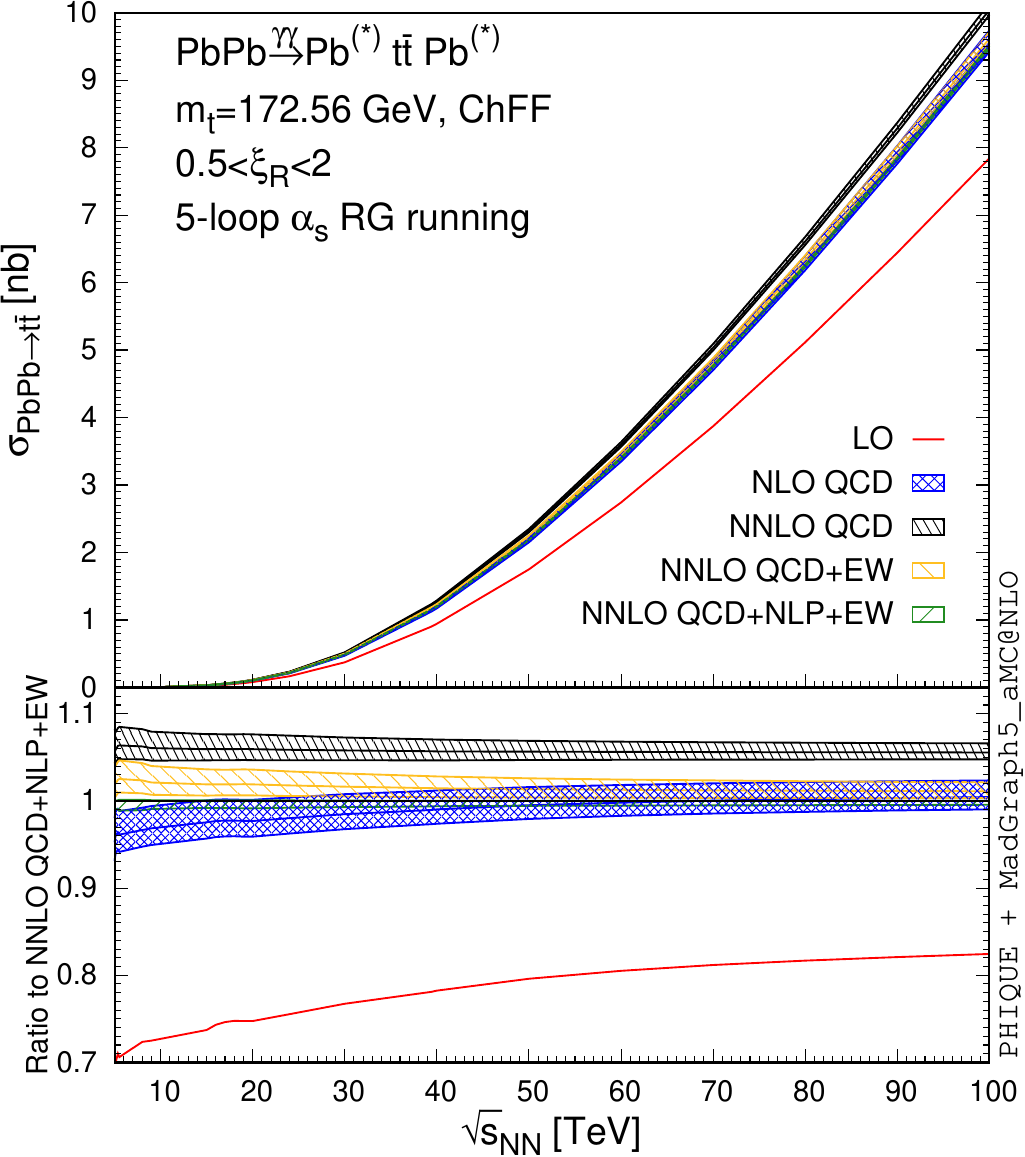}\\
    \includegraphics[width=0.49\textwidth]{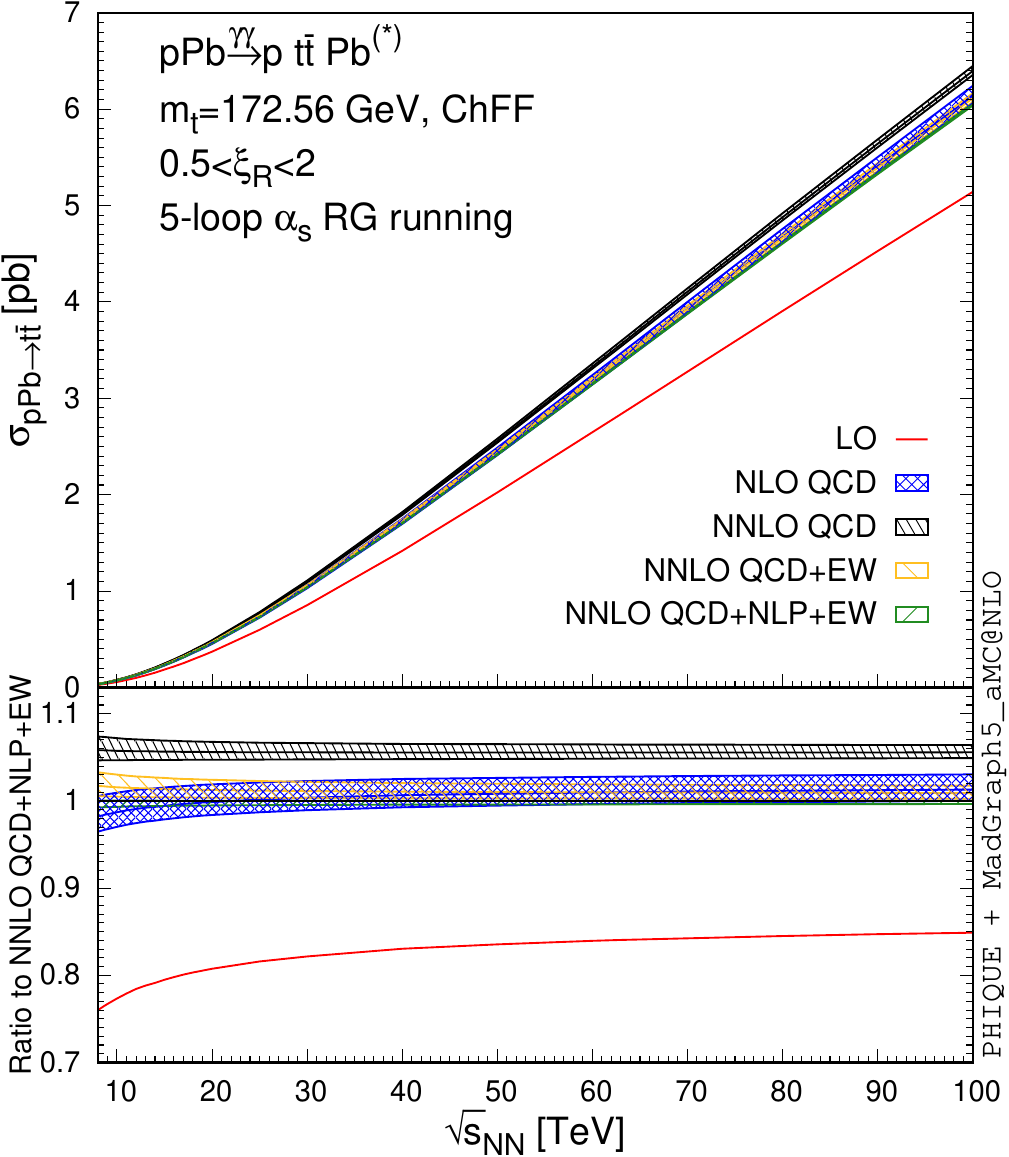}
    \includegraphics[width=0.49\textwidth]{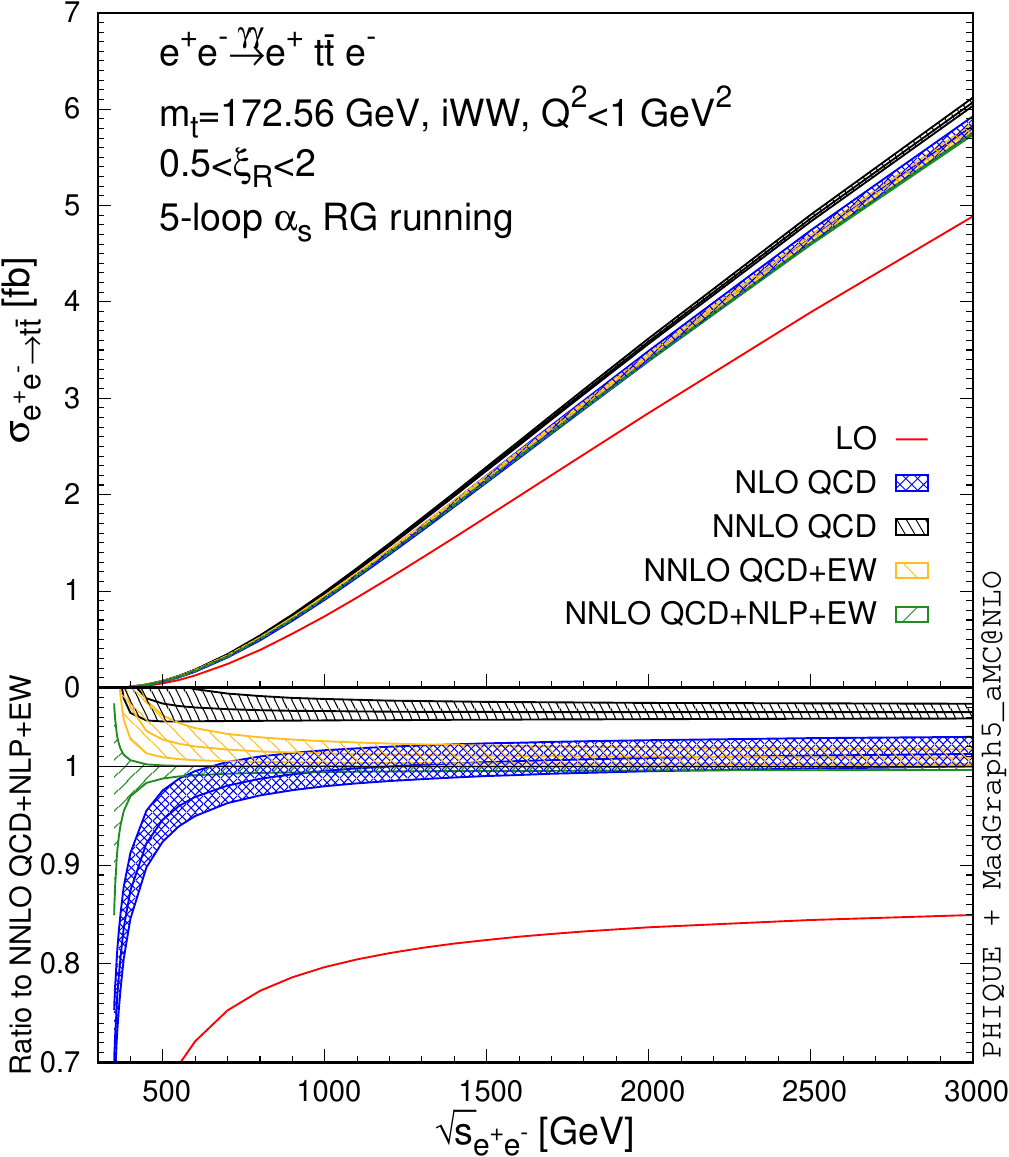}
    \caption{Total top-quark cross sections via two-photon collisions for \pp\ \gls{upc}s (upper left), \PbPb\ \gls{upc}s (upper right), \pPb\ \gls{upc}s (lower left), and $\epem$ (lower right), shown as functions of the \cm\ energy $\sqrtsnn$ or $\sqrt{s_{\epem}}$. Results are displayed at \gls{lo} (red line), \gls{nlo} \gls{qcd} (blue hatched band), \gls{nnlo} \gls{qcd} (black hatched band), \gls{nnlo} \gls{qcd}+\gls{ew} (orange hatched band), and \gls{nnlo} \gls{qcd}+\gls{nlp}+\gls{ew} (green hatched band). The lower panels show the cross section ratios relative to the central \gls{nnlo} \gls{qcd}+\gls{nlp}+\gls{ew} value. The bands indicate the scale uncertainties from varying $\xi_R$ by a factor of two.}
    \label{fig:topxsvsenergy}
\end{figure}

As in ref.~\cite{Shao:2025bma}, \gls{nlo} \gls{qcd} corrections enhance the \gls{lo} cross sections by about 21\% (18\%) in \pp\ collisions at the \gls{lhc} (FCC-hh), while \gls{nlo} \gls{ew} corrections reduce them by about $5.5\%$. \gls{nnlo} \gls{qcd} corrections further enhance the \gls{lo} cross sections by $5.8\%$ ($4.4\%$) at the \gls{lhc} (FCC-hh). These contributions are of the same order in absolute terms as the \gls{nlo} \gls{ew} corrections, consistent with the expectation that \gls{nnlo} \gls{qcd} and \gls{nlo} \gls{ew} effects are numerically comparable based on the sizes of $\alpha_s$ and $\alpha$. The \gls{nlp} Coulomb resummation lowers the \gls{lo} cross section by about $1\%$, which is subdominant in the phase-space-integrated results. This is expected since most events lie at $\beta_t>0.4$, which is much larger than $\alpha_s(m_t)$ and ensures that the fixed-order expansion in $\alpha_s$ converges sufficiently fast~\cite{Beneke:2016jpx}. As we will show later, Coulomb effects become much more pronounced when $\alpha_s$ is larger, as in the bottom- and charm-quark cases. Similarly, if $\beta_t$ is forced to be small--as in the $\epem$ 365 and 500 GeV runs listed in table \ref{tab:ttbarxsQCD}--the Coulomb resummation effects also become significant. The \cm\ energy dependence of higher-order corrections is illustrated in figure \ref{fig:topxsvsenergy}. Due to the larger photon fluxes, the cross sections increase monotonically with $\sqrtsnn$ and $\sqrt{s_{\epem}}$. The $K$ factors (shown in the lower panels) mildly depend on the collider energy, except close to the top-quark production threshold.

\begin{figure}
\centering
    \includegraphics[width=0.8\textwidth]{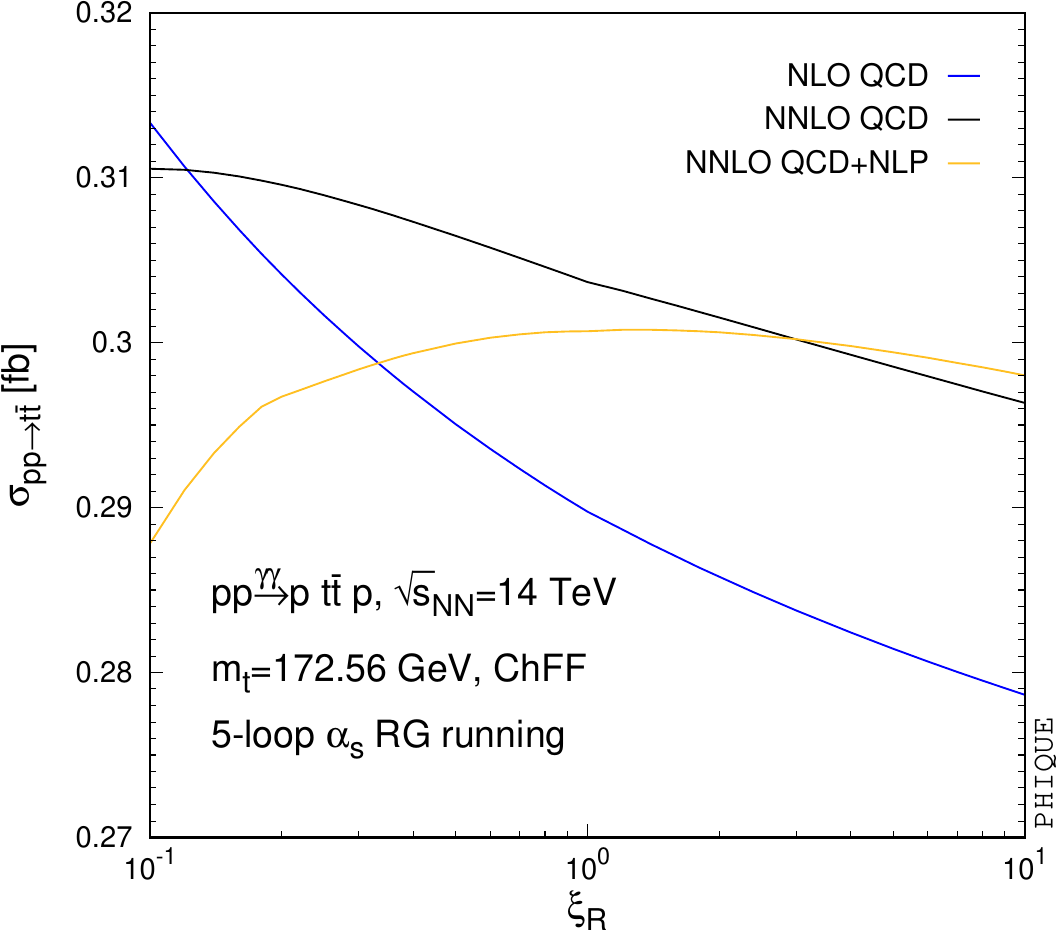}
    \caption{Scale dependence of the top-quark cross sections $\sigma_{\mathrm{pp}\to \ttbar}^{(\mathrm{\gls{nlo}~\gls{qcd}})}$ (blue line), $\sigma_{\mathrm{pp}\to \ttbar}^{(\mathrm{\gls{nnlo}~\gls{qcd}})}$ (black line), and $\sigma_{\mathrm{pp}\to \ttbar}^{(\mathrm{\gls{nnlo}~\gls{qcd}+\gls{nlp}})}$ (orange line) as a function of $\xi_R$ in \pp\ \gls{upc} at $\sqrtsnn=14$ TeV.}
    \label{fig:topscalevar4ppUPC14TeV}
\end{figure}

Concerning theoretical uncertainties on the \gls{nnlo} \gls{qcd} corrections, we consider those arising from scale variations, as well as Monte Carlo statistical errors in the computations of the \gls{nnlo} \gls{qcd} corrections with \alphaLoop. The latter are quoted as the numerical errors in parentheses in tables \ref{tab:ttbarxsQCD} and \ref{tab:ttbarxsEW}. They are typically at the permille level and can therefore be safely neglected. The main advantage of including higher-order \gls{qcd} corrections is the reduction of the residual uncertainty from missing higher orders in perturbation theory, which is customarily estimated by varying the scales by a factor of two. 

For the process $\gaga\to \QQ$, there is no scale dependence at \gls{lo}, and only a renormalisation scale $\mu_R$ dependence at fixed order. The additional Coulomb scale $\mu_C$ dependence enters when Coulomb resummation is included. It is therefore unsurprising that scale variations may underestimate the true size of missing higher-order \gls{qcd} corrections. This is clearly illustrated in figure \ref{fig:topxsvsenergy}, where for $Q=t$ the \gls{nnlo} \gls{qcd} error bands do not overlap with the \gls{nlo} \gls{qcd} ones. 

The fractional scale uncertainties from the strong interaction are reduced from about $\pm2\%$ at \gls{nlo} \gls{qcd} to $\pm1\%$ at \gls{nnlo} \gls{qcd}, and to the sub-percent level once Coulomb resummation is included. The scale uncertainties at \gls{nnlo} \gls{qcd}+\gls{nlp} and\\
\gls{nnlo} \gls{qcd}+\gls{nlp}+\gls{ew} accuracies are quite asymmetric, as seen in figure \ref{fig:topscalevar4ppUPC14TeV}. 

\noindent Figure \ref{fig:topscalevar4ppUPC14TeV} shows the scale dependence of the cross section with respect to $\xi_R$ in the range $0.1$-$10$ for \pp\ \gls{upc}s at $\sqrtsnn=14$ TeV. One sees that the scale dependence is reduced when going from \gls{nlo} \gls{qcd} to \gls{nnlo} \gls{qcd}, and further from \gls{nnlo} \gls{qcd} to \gls{nnlo} \gls{qcd}+\gls{nlp}. At \gls{nnlo} \gls{qcd}+\gls{nlp}, the central scale choice $\xi_R=1$ happens to give the maximal cross section. At very small scales ($\xi_R<0.2$), however, the scale dependence of the \gls{nnlo} \gls{qcd}+\gls{nlp} prediction grows again, because $\alpha_s$ becomes too large and resummation of $\alpha_s/\beta_t$ effects would be required, effectively reducing the perturbative accuracy. Our final best predictions for the total cross section are obtained at \gls{nnlo} \gls{qcd}+\gls{nlp}+\gls{ew} accuracy, as shown in the last column of table \ref{tab:ttbarxsEW}. Interestingly, the \gls{nnlo} \gls{qcd} and \gls{nlo} \gls{ew} corrections largely cancel each other, which underlines the importance of including both consistently in order to reach percent-level accuracy.

Before closing this subsection, we comment on the estimation of the missing \gls{nnlo} mixed \gls{qcd}$\times$\gls{ew} ($\mathcal{O}(\alpha_s\alpha^3)$) and \gls{nnlo} \gls{ew} ($\mathcal{O}(\alpha^4)$) corrections. An approximate way to estimate the size of the missing \gls{nnlo} \gls{ew} corrections in hybrid \gls{ew} renormalisation schemes is to use $\alpha(0)$ for external photons and $\alpha_{G_\mu}$ or $\alpha(m_Z)$ for internal interactions. Such a comparison indicates that the magnitude of the \gls{nnlo} \gls{ew} corrections is about $2.4\%$ of the \gls{nlo} \gls{ew} corrections. The mixed \gls{nnlo} \gls{qcd}$\times$\gls{ew} corrections, on the other hand, can be estimated by comparing additive and multiplicative schemes for combining \gls{nlo} \gls{qcd} and \gls{nlo} \gls{ew} $K$ factors.
For the example of top-quark pair production in \pp\ UPCs at the LHC, we estimate that the relative uncertainties due to the missing \gls{nnlo} mixed \gls{qcd}$\times$\gls{ew} and \gls{nnlo} \gls{ew} corrections are about $1\%$ and $0.1\%$, respectively.

\subsection{Bottom quark\label{sec:bottomres}}

For $\gaga\to \bbbar$, we adopt the same setup as for the top quark in section \ref{sec:topres}, except that the OS mass of the bottom quark is set to $m_b=4.75$ GeV, and the top-quark contribution is excluded in \gls{nnlo} \gls{qcd} calculations. Unlike the top-quark case, the bottom-quark cross sections are considerably larger. Therefore, we focus on the \gls{lhc} as a showcase of \gls{upc}s in hadron collisions, and on FCC-ee/CEPC for $\epem$ collisions, as reported in tables \ref{tab:bbbarxsQCD} and \ref{tab:bbbarxsEW}. Because of the low invariant mass in $\gaga\to\bbbar$, it would be particularly challenging to measure the process in \pp\ collisions with large pile-up. Consequently, we refrain from presenting explicit numbers for \gls{lhc} Runs 1–2 ($\sqrtsnn=7,8,13$ TeV) in the \pp\ mode. In heavy-ion runs at the \gls{lhc}, pile-up is not an issue, although the integrated luminosities $\mathcal{L}_{\mathrm{int}}$ are limited. The heavy-ion energies shown in the tables are \pPb\ at $\sqrtsnn=8.16,8.79$ TeV and \PbPb\ at $\sqrtsnn=5,5.36,5.52$ TeV. Here, the last energies for \pPb\ and \PbPb\ collisions correspond to the high-luminosity \gls{lhc}, while the others reflect existing \gls{lhc} data. 

\begin{table}[htpb!]
\tabcolsep=1.0mm

\centering
\makebox[\textwidth][c]{%
\begin{tabular}{lr|l|l:r|l:r|l:r}
\toprule

\multicolumn{2}{l}{$\gaga\to \bbbar$}&
 \multicolumn{7}{c}{\gls{nnlo} \gls{qcd} + \gls{nlp} with \phique} \\

\midrule
\multicolumn{2}{c}{}
& \multicolumn{1}{c}{\gls{lo}} 
& \multicolumn{2}{c}{\gls{nlo} \gls{qcd}} 
& \multicolumn{2}{c}{\gls{nnlo} \gls{qcd}} 
& \multicolumn{2}{c}{\gls{nnlo} \gls{qcd}+\gls{nlp}} \\
\cmidrule{3-9}
\multicolumn{1}{c}{\ $\mathrm{A}_1$-$\mathrm{A}_2$}& \multicolumn{1}{c}{ $\sqrt{S}$\ } & \multicolumn{1}{c}{$\sigma$}  & \multicolumn{1}{c}{$\sigma$} & \multicolumn{1}{c}{$K$} & \multicolumn{1}{c}{$\sigma$} & \multicolumn{1}{c}{$K$} & \multicolumn{1}{c}{$\sigma$} & \multicolumn{1}{c}{$K$}  \\
\midrule
\pp& 7~TeV   & $0.390$ pb  & $0.528^{+0.040}_{-0.023}$ pb & 1.353 & $0.597(1)^{+0.043}_{-0.026}$ pb & 1.529 & $0.558(1)^{+0.003}_{-0.066}$ pb & 1.430 \\
\pp& 8~TeV   & $0.426$ pb  & $0.576^{+0.043}_{-0.025}$ pb & 1.352 & $0.651(1)^{+0.047}_{-0.028}$ pb & 1.526 & $0.609(1)^{+0.003}_{-0.071}$ pb & 1.428 \\
\pp& 13~TeV  & $0.577$ pb  & $0.778^{+0.058}_{-0.033}$ pb & 1.348 & $0.876(2)^{+0.062}_{-0.038}$ pb & 1.518 & $0.821(2)^{+0.004}_{-0.093}$ pb & 1.423 \\
\pPb& 8.16~TeV & $1.287$ nb  & $1.756^{+0.135}_{-0.078}$ nb & 1.364 & $2.000(5)^{+0.154}_{-0.092}$ nb & 1.554 & $1.860(5)^{+0.013}_{-0.240}$ nb & 1.445 \\
\pPb& 8.79~TeV & $1.370$ nb  & $1.867^{+0.143}_{-0.083}$ nb & 1.363 & $2.125(5)^{+0.162}_{-0.098}$ nb & 1.551 & $1.978(5)^{+0.014}_{-0.253}$ nb & 1.444 \\
\PbPb& 5~TeV   & $1.683$ $\mu$b  & $2.365^{+0.197}_{-0.113}$ $\mu$b & 1.406 & $2.756(8)^{+0.247}_{-0.145}$ $\mu$b & 1.638 & $2.525(8)^{+0.029}_{-0.408}$ $\mu$b & 1.501 \\
\PbPb& 5.36~TeV& $1.853$ $\mu$b  & $2.599^{+0.215}_{-0.124}$ $\mu$b & 1.403 & $3.024(8)^{+0.268}_{-0.158}$ $\mu$b & 1.632 & $2.773(8)^{+0.031}_{-0.441}$ $\mu$b & 1.497 \\
\PbPb& 5.52~TeV& $1.929$ $\mu$b & $2.703^{+0.223}_{-0.128}$ $\mu$b & 1.401 & $3.142(9)^{+0.278}_{-0.164}$ $\mu$b & 1.629 & $2.883(9)^{+0.032}_{-0.456}$ $\mu$b & 1.495 \\
$\epem$& 90~GeV  & $0.445$ pb  & $0.615^{+0.049}_{-0.028}$ pb & 1.382 & $0.709(2)^{+0.059}_{-0.035}$ pb & 1.592 & $0.654(2)^{+0.006}_{-0.095}$ pb & 1.470 \\
$\epem$& 160~GeV & $0.787$ pb  & $1.072^{+0.082}_{-0.047}$ pb & 1.363 & $1.222(3)^{+0.094}_{-0.057}$ pb & 1.553 & $1.136(3)^{+0.008}_{-0.147}$ pb & 1.444 \\
$\epem$& 240~GeV & $1.089$ pb  & $1.475^{+0.111}_{-0.064}$ pb & 1.355 & $1.672(4)^{+0.124}_{-0.075}$ pb & 1.535 & $1.560(4)^{+0.009}_{-0.190}$ pb & 1.433 \\
$\epem$& 365~GeV & $1.454$ pb  & $1.961^{+0.146}_{-0.084}$ pb & 1.349 & $2.214(5)^{+0.159}_{-0.096}$ pb & 1.522 & $2.072(5)^{+0.011}_{-0.241}$ pb & 1.425 \\
\bottomrule
\end{tabular}%
}
\caption{Same as table~\ref{tab:ttbarxsQCD}, but for $\gaga\to \bbbar$.\label{tab:bbbarxsQCD}}
\end{table}

\begin{table}[htpb!]
\centering
\renewcommand{\arraystretch}{1.2}
\makebox[\textwidth][c]{%
\begin{tabular}{lr|r:l|r:l|r:l}
\toprule
\multicolumn{2}{l}{Process: $\gaga\to \bbbar$} & \multicolumn{6}{c}{\gls{nnlo} \gls{qcd} + \gls{nlp} with \phique \ and \gls{nlo} \gls{ew} with \mgshort} \\
\midrule
\multicolumn{2}{c}{}

& \multicolumn{2}{c}{\gls{nlo} \gls{qcd}+\gls{ew}} 
& \multicolumn{2}{c}{\gls{nnlo} \gls{qcd}+\gls{ew}} 
& \multicolumn{2}{c}{\gls{nnlo} \gls{qcd}+\gls{nlp}+\gls{ew}} \\
\cmidrule{3-8}
\multicolumn{1}{c}{$\mathrm{A}_1$-$\mathrm{A}_2$} & \multicolumn{1}{c}{$\sqrt{S}$}  & \multicolumn{1}{c}{$\sigma$} & \multicolumn{1}{c}{$K$} & \multicolumn{1}{c}{$\sigma$} & \multicolumn{1}{c}{$K$} & \multicolumn{1}{c}{$\sigma$} & \multicolumn{1}{c}{$K$} \\
\midrule
\pp & 7~TeV   & $0.528^{+0.040}_{-0.023}$ pb & 1.354 & $0.597(1)^{+0.043}_{-0.026}$ pb & 1.530 & $0.558(1)^{+0.003}_{-0.066}$ pb & 1.430 \\
\pp & 8~TeV   & $0.577^{+0.043}_{-0.025}$ pb & 1.353 & $0.651(1)^{+0.047}_{-0.028}$ pb & 1.528 & $0.609(1)^{+0.003}_{-0.071}$ pb & 1.429 \\
\pp & 13~TeV  & $0.778^{+0.058}_{-0.033}$ pb & 1.349 & $0.876(2)^{+0.062}_{-0.038}$ pb & 1.520 & $0.821(2)^{+0.004}_{-0.093}$ pb & 1.424 \\
\pPb & 8.16~TeV & $1.757^{+0.135}_{-0.078}$ nb & 1.365 & $2.001(5)^{+0.154}_{-0.092}$ nb & 1.555 & $1.862(5)^{+0.013}_{-0.240}$ nb & 1.447 \\
\pPb & 8.79~TeV & $1.869^{+0.143}_{-0.083}$ nb & 1.364 & $2.127(5)^{+0.162}_{-0.098}$ nb & 1.553 & $1.979(5)^{+0.014}_{-0.253}$ nb & 1.445 \\
\PbPb & 5~TeV   & $2.367^{+0.197}_{-0.113}$ $\mu$b & 1.407 & $2.758(8)^{+0.247}_{-0.145}$ $\mu$b & 1.639 & $2.527(8)^{+0.029}_{-0.408}$ $\mu$b & 1.502 \\
\PbPb & 5.36~TeV& $2.601^{+0.215}_{-0.124}$ $\mu$b & 1.404 & $3.026(8)^{+0.268}_{-0.158}$ $\mu$b & 1.633 & $2.776(8)^{+0.031}_{-0.441}$ $\mu$b & 1.498 \\
\PbPb & 5.52~TeV& $2.705^{+0.223}_{-0.128}$ $\mu$b & 1.403 & $3.145(9)^{+0.278}_{-0.164}$ $\mu$b & 1.630 & $2.886(9)^{+0.032}_{-0.456}$ $\mu$b & 1.496 \\
$\epem$ & 90~GeV   & $0.616^{+0.049}_{-0.028}$ pb & 1.384 & $0.709(2)^{+0.059}_{-0.035}$ pb & 1.593 & $0.655(2)^{+0.006}_{-0.095}$ pb & 1.471 \\
$\epem$ & 160~GeV  & $1.073^{+0.082}_{-0.047}$ pb & 1.364 & $1.223(3)^{+0.094}_{-0.057}$ pb & 1.554 & $1.137(3)^{+0.008}_{-0.147}$ pb & 1.445 \\
$\epem$ & 240~GeV  & $1.476^{+0.111}_{-0.064}$ pb & 1.356 & $1.673(4)^{+0.124}_{-0.075}$ pb & 1.537 & $1.561(4)^{+0.009}_{-0.190}$ pb & 1.434 \\
$\epem$ & 365~GeV  & $1.963^{+0.146}_{-0.084}$ pb & 1.350 & $2.215(5)^{+0.159}_{-0.096}$ pb & 1.524 & $2.073(5)^{+0.011}_{-0.241}$ pb & 1.426 \\
\bottomrule
\end{tabular}}
 \caption{Same as table \ref{tab:ttbarxsEW}, but for the bottom quark.}
\label{tab:bbbarxsEW}
\end{table}

\begin{figure}
    \includegraphics[width=0.49\textwidth]{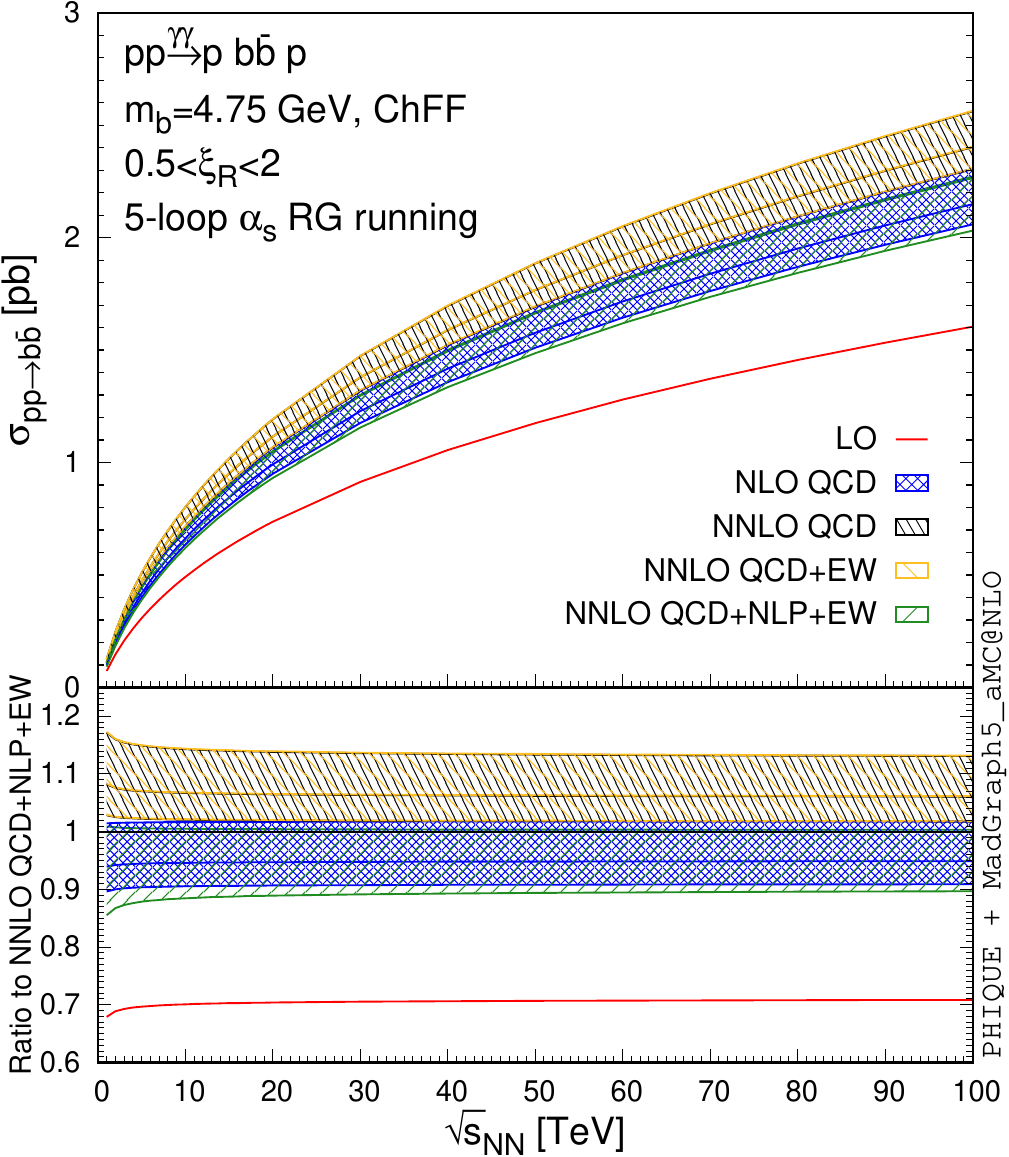}
    \includegraphics[width=0.49\textwidth]{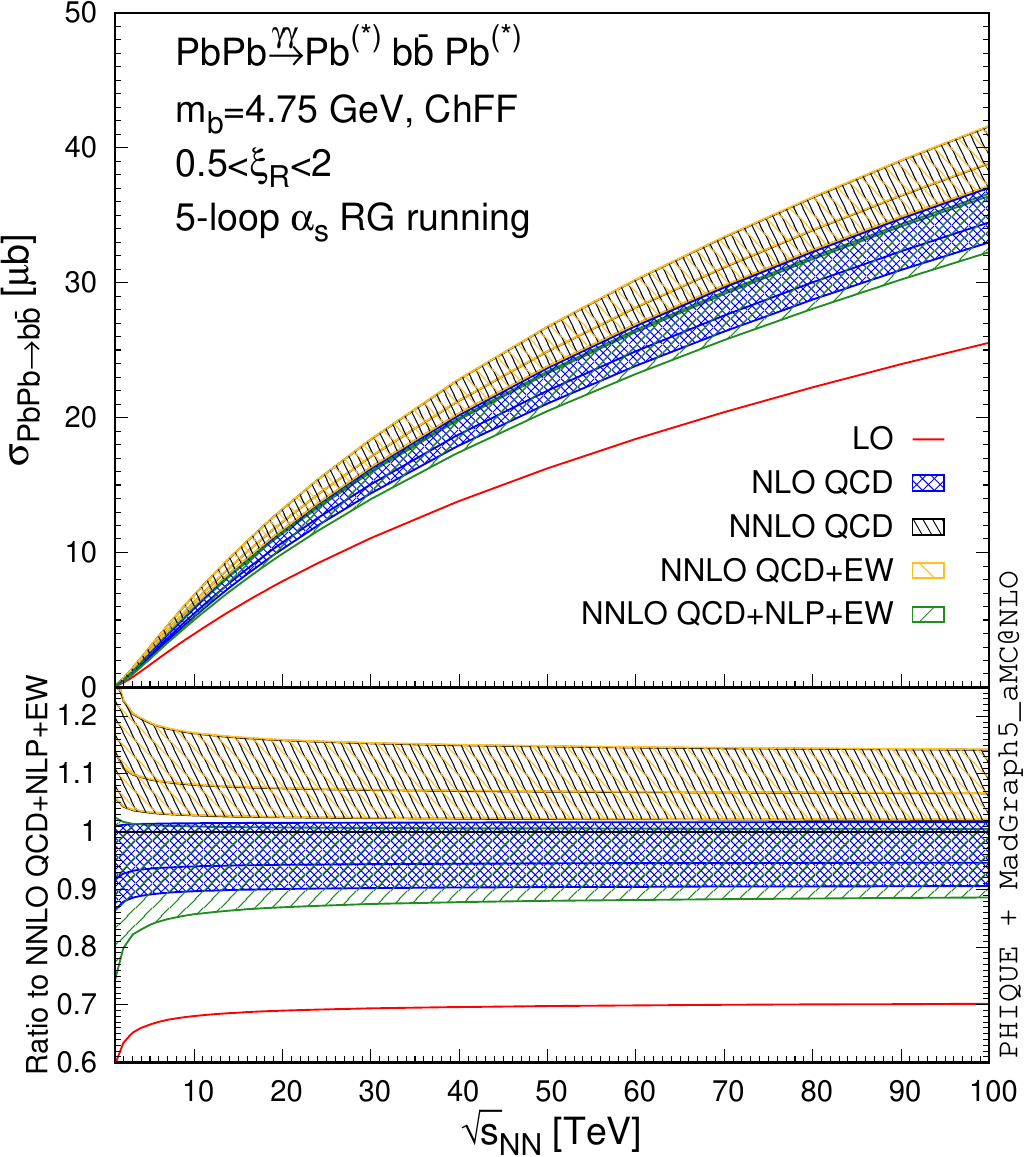}\\
    \includegraphics[width=0.49\textwidth]{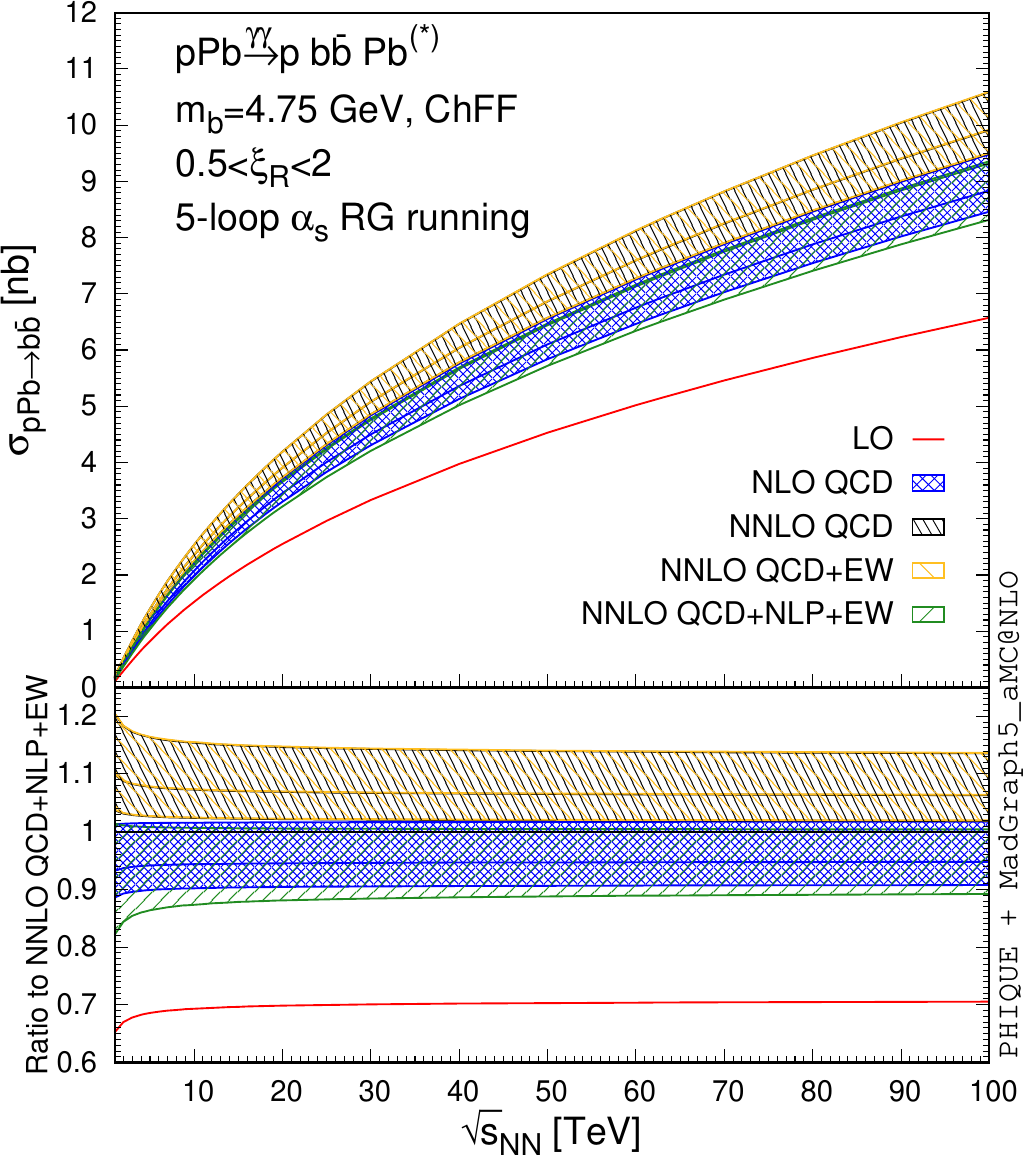}
    \includegraphics[width=0.49\textwidth]{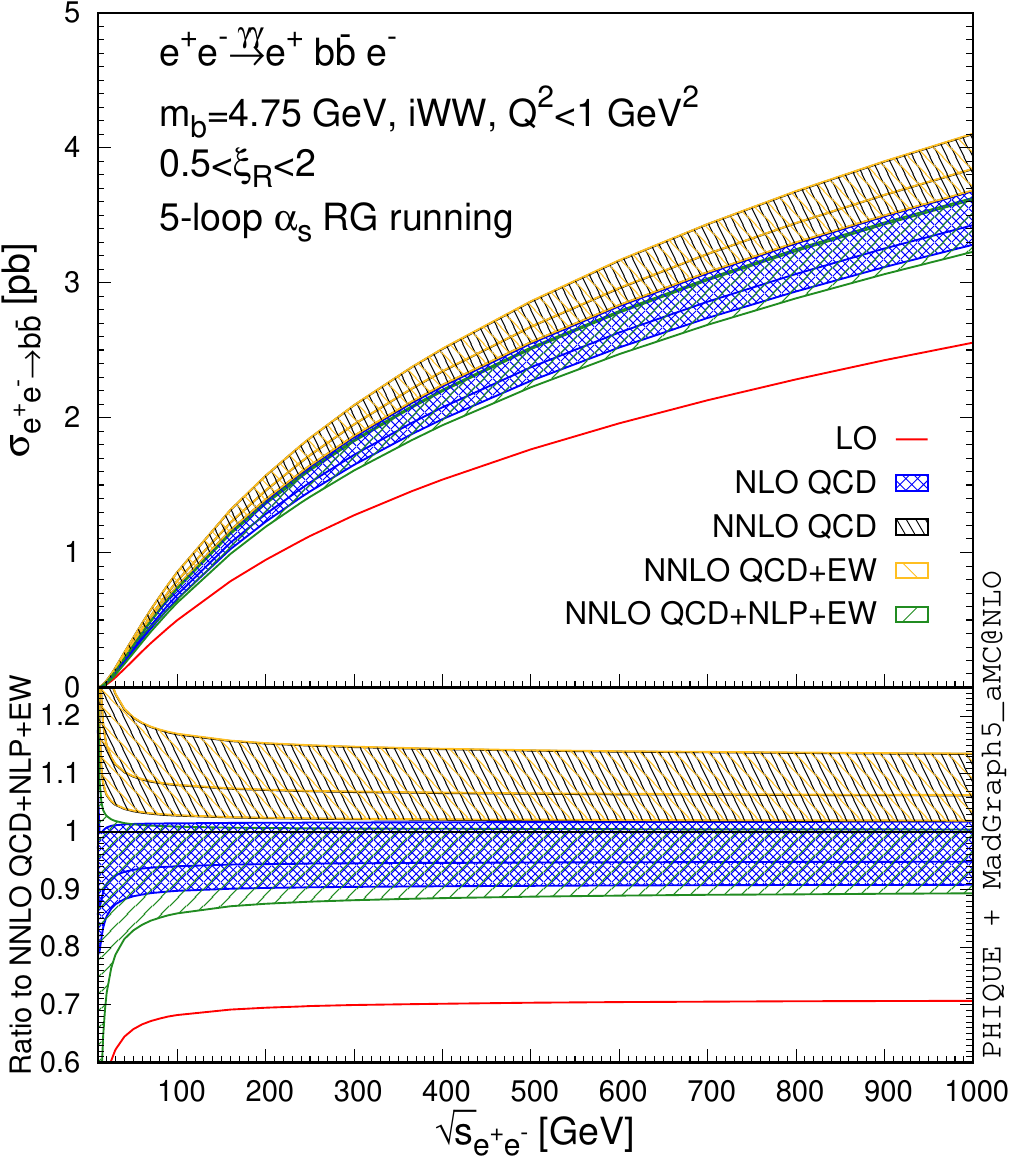}
    \caption{Same as figure~\ref{fig:topxsvsenergy}, but for the bottom quark.}
    \label{fig:bottomxsvsenergy}
\end{figure}

As mentioned in the introduction, it would be particularly interesting to measure $\gaga\to \bbbar$ at future $\epem$ colliders, given the discrepancy between \gls{nlo} \gls{qcd} predictions and the L3 measurements. We therefore also present our results for $\epem$ collisions at four different nominal energies at future circular $\epem$ colliders. The \cm\ energy dependence of bottom-quark cross sections at both hadron and $\epem$ colliders is shown in figure \ref{fig:bottomxsvsenergy}.

Due to the smaller probed scale, \gls{qcd} corrections in $\gaga\to\bbbar$ are more significant than in the top-quark case. \gls{nlo} \gls{qcd} corrections enhance the \gls{lo} cross sections by $35$-$40\%$, while \gls{nnlo} \gls{qcd} corrections further increase them by $17$-$23\%$. In contrast, the Coulomb resummation reduces the \gls{lo} cross sections by $10$-$14\%$. The larger value of $\alpha_s$ in the bottom-quark case, which is closer to the typical velocity $\beta_b\sim 0.4$, makes the Coulomb resummation effects more pronounced. Moreover, the Coulomb resummation corrections above the $2m_b$ threshold, as considered here, are of the same order of magnitude but opposite in sign compared to the cross sections of bottomonia (cf. Table V in ref.~\cite{Shao:2022cly}). \gls{ew} corrections, in contrast, are quite small, amounting to only $+0.1\%$ of the \gls{lo} cross sections. The \gls{nlo} \gls{ew} corrections for the bottom quark differ significantly from those for the top quark, even in sign. This is because the weak corrections are suppressed by $m_b^2/m_W^2$ relative to the QED corrections in $\gaga\to\bbbar$. The latter are further suppressed by the factor $e_b^2=1/9$ relative to the $+1\%$ QED corrections reported in $\gaga\to\tau^+\tau^-$~\cite{Shao:2024dmk,Jiang:2024dhf,Shao:2025bma,Dittmaier:2025ikh}, yielding an overall effect of approximately $+0.1\%$. Unlike the top-quark case, there is a partial cancellation between the \gls{nnlo} \gls{qcd} and Coulomb corrections in $\gaga\to\bbbar$.

\begin{figure}
    \includegraphics[width=0.9\textwidth]{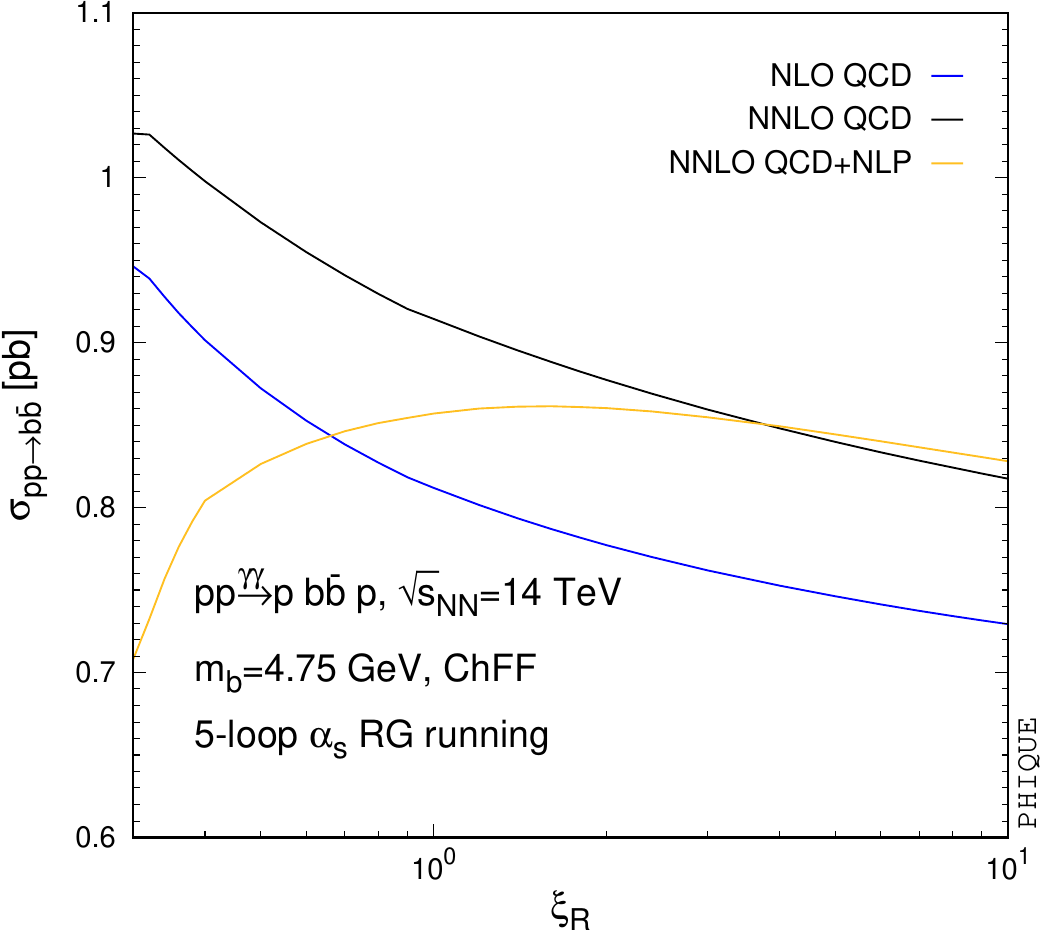}
    \caption{Same as figure~\ref{fig:topscalevar4ppUPC14TeV}, but for the bottom quark.}
    \label{fig:bottomscalevar4ppUPC14TeV}
\end{figure}

In contrast to the usual expectation in perturbation theory, the scale uncertainties at \gls{nnlo} \gls{qcd} and \gls{nnlo} \gls{qcd}+\gls{nlp} accuracies are of similar size to those at \gls{nlo} \gls{qcd}. As shown in figure \ref{fig:bottomscalevar4ppUPC14TeV}, the estimated scale uncertainties of the \gls{nnlo} \gls{qcd}+\gls{nlp} results are quite sensitive to the choice of the central scale. Choosing $\xi_R\approx 2$ as the central scale for $\gaga\to\bbbar$ significantly reduces these uncertainties, a behaviour similar to that observed for the total cross sections of the inclusive reaction $pp\to \bbbar+X$ at \gls{nnlo} \gls{qcd}~\cite{Catani:2020kkl}. On the other hand, the Monte Carlo integration uncertainties in the \gls{nnlo} \gls{qcd} corrections, quoted as errors in parentheses in tables \ref{tab:bbbarxsQCD} and \ref{tab:bbbarxsEW}, are negligible.

\clearpage
\newpage
\subsection{Charm quark\label{sec:charmres}}

Predictions for charm-quark production are challenging because the charm-quark mass, $m_c \sim 1.5$ GeV, is close to the intrinsic nonperturbative \gls{qcd} scale, $\Lambda_{\mathrm{\gls{qcd}}} \sim 0.3$ GeV. As a result, the perturbative series converges slowly, leading to typically large theoretical uncertainties. Increasing the perturbative accuracy of charm-quark cross sections does not always improve the reliability of the prediction~\cite{Bonino:2023icn}. Therefore, it is interesting to examine the perturbative series pattern in the total cross section of $\gaga\to\ccbar$.

\begin{table}[htpb!]
\tabcolsep=1.0mm
\centering
\renewcommand{\arraystretch}{1.2}
\makebox[\textwidth][c]{%
\begin{tabular}{lr|r|r:l|r:l|r:l}
\toprule
\multicolumn{2}{l}{$\gaga\to \ccbar$} & \multicolumn{7}{c}{\gls{nnlo} \gls{qcd} + \gls{nlp} with \phique} \\
\midrule
\multicolumn{2}{c}{}
& \multicolumn{1}{c}{\gls{lo}}
& \multicolumn{2}{c}{\gls{nlo} \gls{qcd}}
& \multicolumn{2}{c}{\gls{nnlo} \gls{qcd}}
& \multicolumn{2}{c}{\gls{nnlo} \gls{qcd}+\gls{nlp}} \\
\cmidrule(lr){3-9}
\multicolumn{1}{c}{$\mathrm{A}_1$-$\mathrm{A}_2$} & \multicolumn{1}{c}{$\sqrt{S}$} & \multicolumn{1}{c}{$\sigma$} & \multicolumn{1}{c}{$\sigma$} & \multicolumn{1}{c}{$K$} & \multicolumn{1}{c}{$\sigma$} & \multicolumn{1}{c}{$K$} & \multicolumn{1}{c}{$\sigma$} & \multicolumn{1}{c}{$K$} \\
\midrule
\pp   & 7~TeV     & $0.126$ nb  & $0.194^{+0.026}_{-0.018}$ nb & 1.548 & $0.250(1)^{+0.053}_{-0.033}$ nb & 1.992 & $0.195(1)^{+0.013}_{-0.013}$ nb & 1.555 \\
\pp   & 8~TeV     & $0.135$ nb  & $0.208^{+0.027}_{-0.019}$ nb & 1.547 & $0.268(1)^{+0.056}_{-0.035}$ nb & 1.988 & $0.209(1)^{+0.013}_{-0.014}$ nb & 1.555 \\
\pp   & 13~TeV    & $0.172$ nb  & $0.266^{+0.035}_{-0.025}$ nb & 1.543 & $0.341(1)^{+0.071}_{-0.044}$ nb & 1.977 & $0.267(1)^{+0.016}_{-0.018}$ nb & 1.553 \\
\pPb  & 8.16~TeV  & $0.488$ $\mu$b & $0.760^{+0.101}_{-0.071}$ $\mu$b & 1.557 & $0.987(4)^{+0.214}_{-0.132}$ $\mu$b & 2.023 & $0.762(4)^{+0.054}_{-0.055}$ $\mu$b & 1.561 \\
\pPb  & 8.79~TeV  & $0.512$ $\mu$b & $0.796^{+0.106}_{-0.075}$ $\mu$b & 1.556 & $1.034(4)^{+0.224}_{-0.138}$ $\mu$b & 2.020 & $0.799(4)^{+0.056}_{-0.058}$ $\mu$b & 1.561 \\
\PbPb & 5~TeV     & $1.02$ mb   & $1.62^{+0.22}_{-0.16}$ mb & 1.588 & $2.15(1)^{+0.50}_{-0.31}$ mb & 2.113 & $1.61(1)^{+0.14}_{-0.13}$ mb & 1.581 \\
\PbPb & 5.36~TeV  & $1.09$ mb   & $1.73^{+0.24}_{-0.17}$ mb & 1.586 & $2.29(1)^{+0.53}_{-0.33}$ mb & 2.107 & $1.72(1)^{+0.15}_{-0.14}$ mb & 1.579 \\
\PbPb & 5.52~TeV  & $1.12$ mb   & $1.77^{+0.24}_{-0.17}$ mb & 1.585 & $2.35(1)^{+0.55}_{-0.33}$ mb & 2.104 & $1.77(1)^{+0.15}_{-0.14}$ mb & 1.579 \\
$\epem$ & 10.58~GeV & $15.5$ pb  & $27.0^{+4.3}_{-3.0}$ pb & 1.741 & $38.9(2)^{+11.1}_{-6.6}$ pb & 2.511 & $26.1(2)^{+3.9}_{-2.8}$ pb & 1.686 \\
$\epem$ & 90~GeV    & $0.198$ nb & $0.308^{+0.041}_{-0.029}$ nb & 1.559 & $0.402(2)^{+0.088}_{-0.054}$ nb & 2.033 & $0.309(2)^{+0.023}_{-0.023}$ nb & 1.561 \\
$\epem$ & 160~GeV   & $0.285$ nb & $0.442^{+0.058}_{-0.041}$ nb & 1.548 & $0.570(2)^{+0.122}_{-0.075}$ nb & 2.000 & $0.443(2)^{+0.030}_{-0.031}$ nb & 1.555 \\
$\epem$ & 240~GeV   & $0.357$ nb & $0.551^{+0.072}_{-0.051}$ nb & 1.544 & $0.708(3)^{+0.148}_{-0.092}$ nb & 1.984 & $0.554(3)^{+0.035}_{-0.038}$ nb & 1.552 \\
$\epem$ & 365~GeV   & $0.439$ nb & $0.677^{+0.088}_{-0.062}$ nb & 1.540 & $0.866(3)^{+0.179}_{-0.111}$ nb & 1.971 & $0.681(3)^{+0.041}_{-0.045}$ nb & 1.550 \\
\bottomrule
\end{tabular}%
}
\caption{Same as table~\ref{tab:ttbarxsQCD}, but for $\gaga\to \ccbar$.\label{tab:ccbarxsQCD}}
\end{table}

\begin{table}[htpb!]
\centering
\renewcommand{\arraystretch}{1.2}
\makebox[\textwidth][c]{%
\begin{tabular}{lr|r:l|r:l|r:l}
\toprule
\multicolumn{2}{l}{$\gaga\to \ccbar$} & \multicolumn{6}{c}{\gls{nnlo} \gls{qcd} + \gls{nlp} with \phique \ and \gls{nlo} \gls{ew} with \mgshort} \\
\midrule
\multicolumn{2}{c}{} 
& \multicolumn{2}{c}{\gls{nlo} \gls{qcd}+\gls{ew}} 
& \multicolumn{2}{c}{\gls{nnlo} \gls{qcd}+\gls{ew}} 
& \multicolumn{2}{c}{\gls{nnlo} \gls{qcd}+\gls{nlp}+\gls{ew}} \\
\cmidrule{3-8}
\multicolumn{1}{c}{$\mathrm{A}_1$-$\mathrm{A}_2$} & \multicolumn{1}{c}{$\sqrt{S}$} & \multicolumn{1}{c}{$\sigma$} & \multicolumn{1}{c}{$K$} & \multicolumn{1}{c}{$\sigma$} & \multicolumn{1}{c}{$K$} & \multicolumn{1}{c}{$\sigma$} & \multicolumn{1}{c}{$K$} \\
\midrule
\pp   & 7~TeV   & $0.195^{+0.026}_{-0.018}$ nb & 1.552 & $0.250(1)^{+0.053}_{-0.033}$ nb & 1.996 & $0.196(1)^{+0.013}_{-0.013}$ nb & 1.559 \\
\pp   & 8~TeV   & $0.209^{+0.027}_{-0.019}$ nb & 1.551 & $0.268(1)^{+0.056}_{-0.035}$ nb & 1.992 & $0.210(1)^{+0.013}_{-0.014}$ nb & 1.559 \\
\pp   & 13~TeV  & $0.267^{+0.035}_{-0.025}$ nb & 1.547 & $0.341(1)^{+0.071}_{-0.044}$ nb & 1.981 & $0.268(1)^{+0.016}_{-0.018}$ nb & 1.557 \\
\pPb  & 8.16~TeV& $0.762^{+0.101}_{-0.071}$ $\mu$b & 1.561 & $0.989(4)^{+0.214}_{-0.132}$ $\mu$b & 2.027 & $0.764(4)^{+0.054}_{-0.055}$ $\mu$b & 1.565 \\
\pPb  & 8.79~TeV& $0.799^{+0.106}_{-0.075}$ $\mu$b & 1.561 & $1.036(4)^{+0.224}_{-0.138}$ $\mu$b & 2.024 & $0.801(4)^{+0.056}_{-0.058}$ $\mu$b & 1.565 \\
\PbPb & 5~TeV   & $1.62^{+0.22}_{-0.16}$ mb & 1.592 & $2.16(1)^{+0.50}_{-0.31}$ mb & 2.117 & $1.62(1)^{+0.14}_{-0.13}$ mb & 1.585 \\
\PbPb & 5.36~TeV& $1.73^{+0.24}_{-0.17}$ mb & 1.590 & $2.30(1)^{+0.53}_{-0.33}$ mb & 2.111 & $1.72(1)^{+0.15}_{-0.14}$ mb & 1.583 \\
\PbPb & 5.52~TeV& $1.78^{+0.24}_{-0.17}$ mb & 1.589 & $2.36(1)^{+0.55}_{-0.33}$ mb & 2.109 & $1.77(1)^{+0.15}_{-0.14}$ mb & 1.583 \\
$\epem$ & 10.58~GeV & $27.1^{+4.3}_{-3.0}$ pb & 1.746 & $39.0(2)^{+11.1}_{-6.6}$ pb & 2.516 & $26.2(2)^{+3.9}_{-2.8}$ pb & 1.691 \\
$\epem$ & 90~GeV    & $0.309^{+0.041}_{-0.029}$ nb & 1.563 & $0.403(2)^{+0.088}_{-0.054}$ nb & 2.037 & $0.309(2)^{+0.023}_{-0.023}$ nb & 1.565 \\
$\epem$ & 160~GeV   & $0.443^{+0.058}_{-0.041}$ nb & 1.552 & $0.572(2)^{+0.122}_{-0.075}$ nb & 2.004 & $0.445(2)^{+0.030}_{-0.031}$ nb & 1.559 \\
$\epem$ & 240~GeV   & $0.552^{+0.072}_{-0.051}$ nb & 1.548 & $0.709(3)^{+0.148}_{-0.092}$ nb & 1.988 & $0.555(3)^{+0.035}_{-0.038}$ nb & 1.556 \\
$\epem$ & 365~GeV   & $0.679^{+0.088}_{-0.062}$ nb & 1.544 & $0.868(3)^{+0.179}_{-0.111}$ nb & 1.975 & $0.683(3)^{+0.041}_{-0.045}$ nb & 1.554 \\
\bottomrule
\end{tabular}}
\caption{Same as table \ref{tab:ttbarxsEW}, but for the charm quark.\label{tab:ccbarxsEW}}
\end{table}

\begin{figure}
    \includegraphics[width=0.49\textwidth]{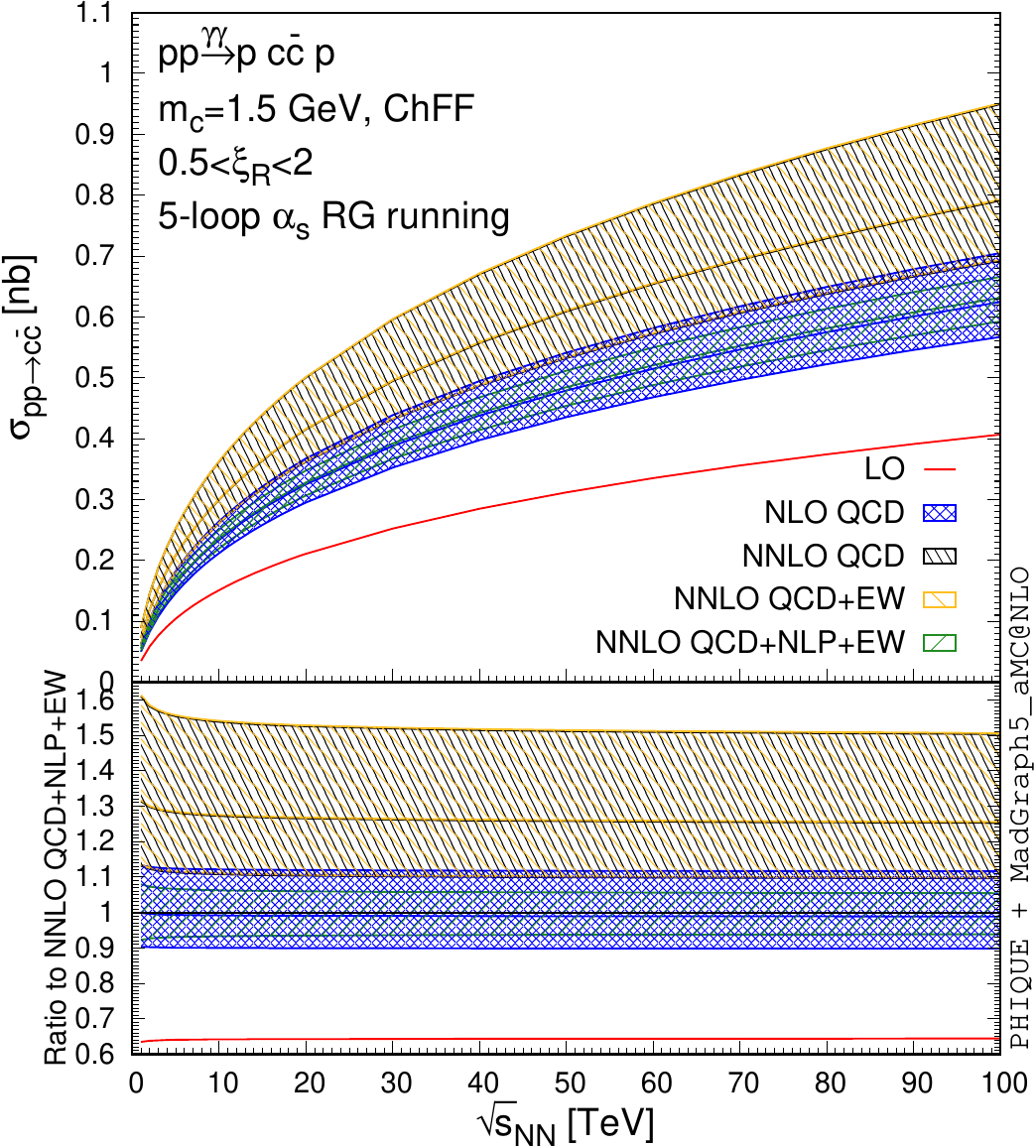}
    \includegraphics[width=0.49\textwidth]{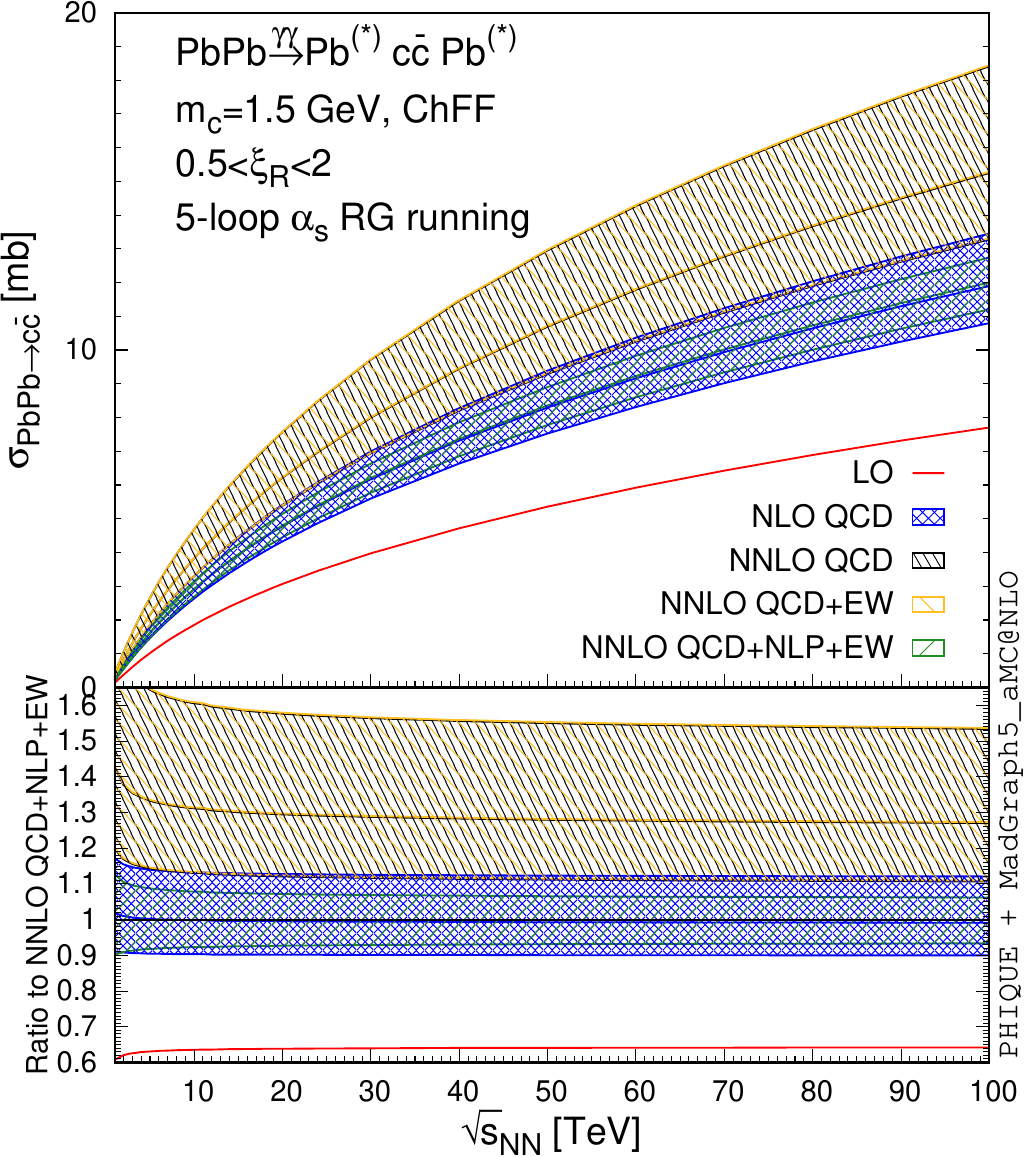}\\
    \includegraphics[width=0.49\textwidth]{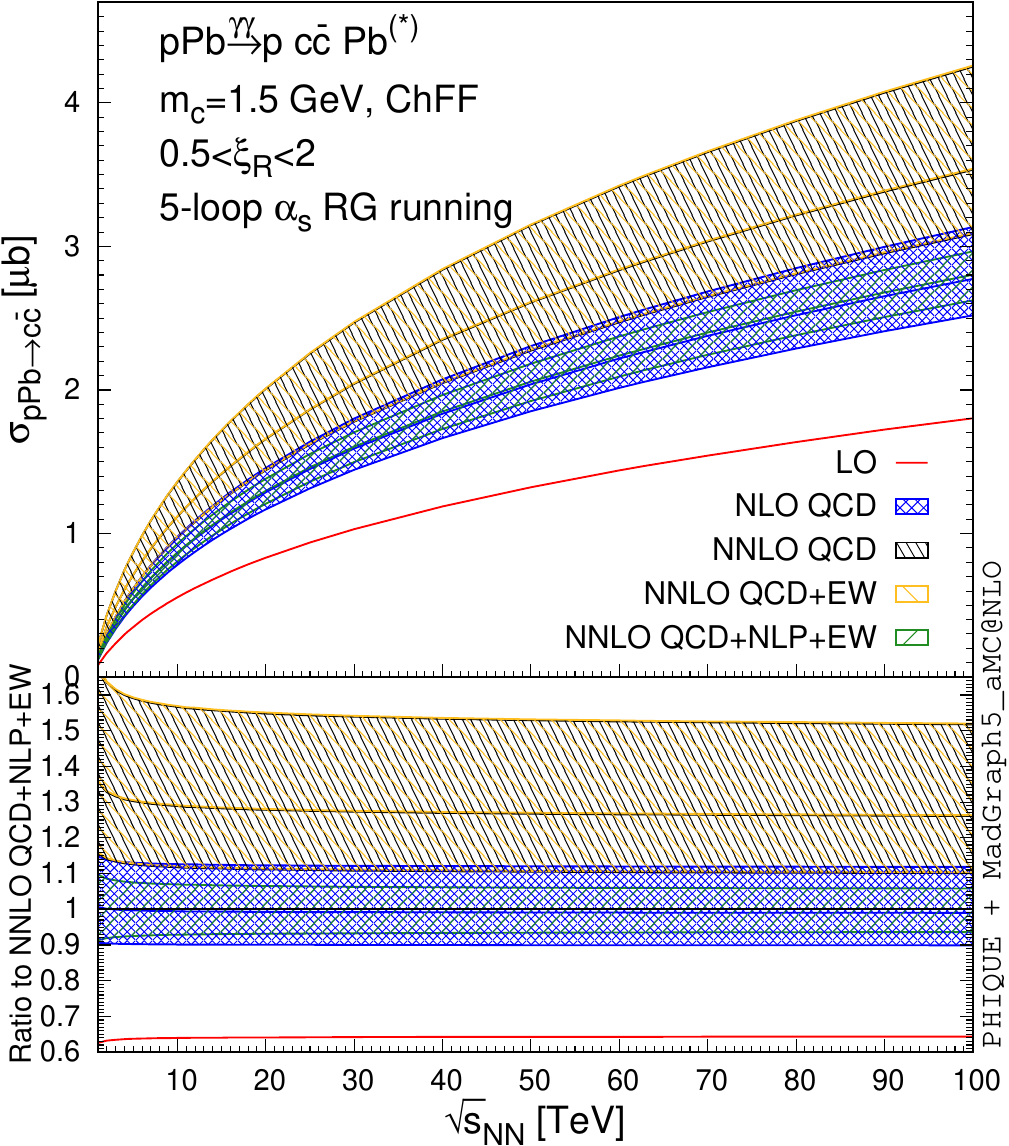}
    \includegraphics[width=0.49\textwidth]{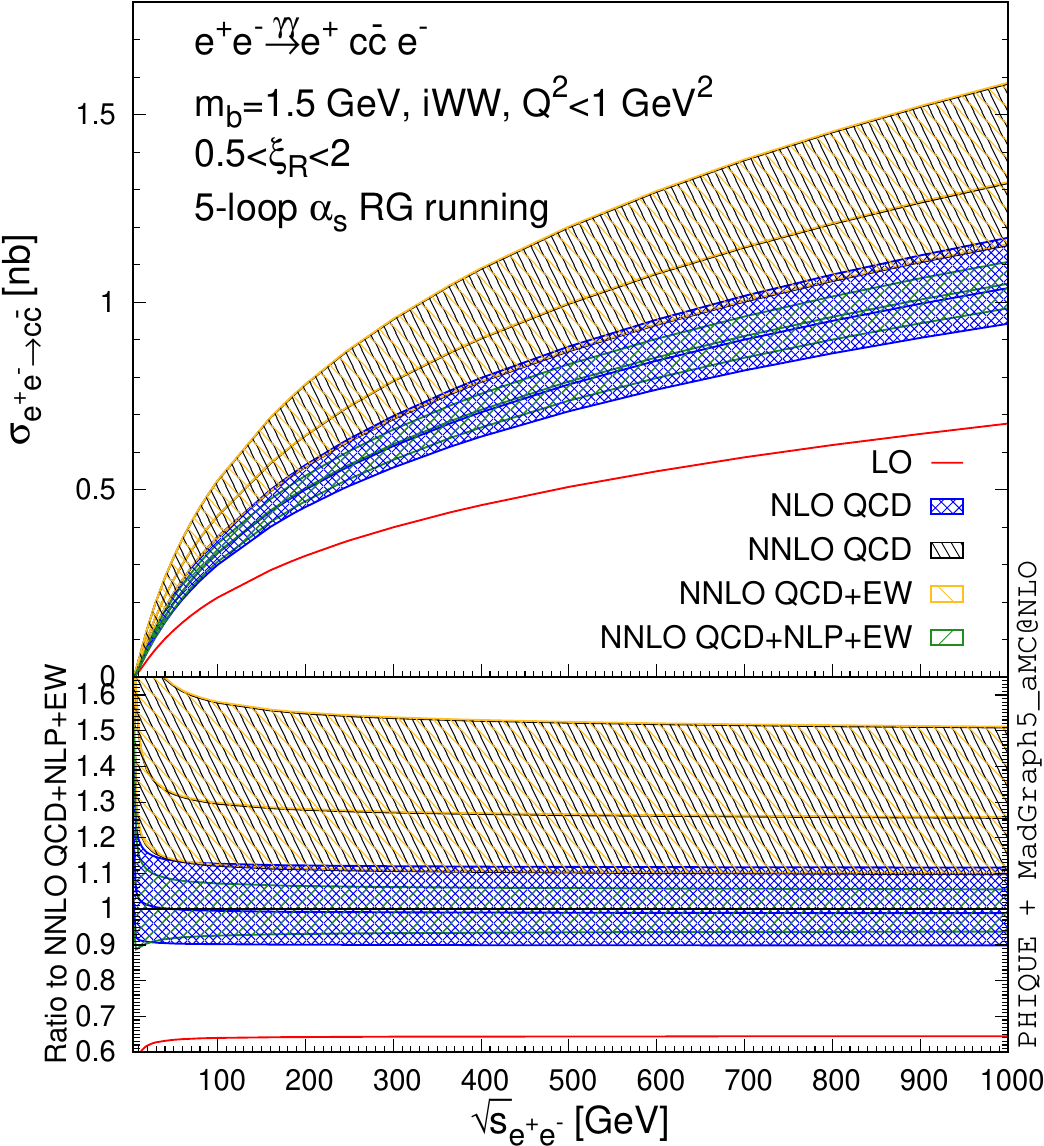}
    \caption{Same as figure~\ref{fig:topxsvsenergy}, but for the charm quark.}
    \label{fig:charmxsvsenergy}
\end{figure}

In our numerical calculations, we adopt the same setup as for the top quark in section \ref{sec:topres}, with the exceptions that
the OS mass of the charm quark is set to $m_c=1.5$ GeV, and the bottom- and top-quark contributions are ignored in the \gls{nnlo} \gls{qcd} corrections. The total cross sections for charm-quark production through photon fusion in ultraperipheral \pp, \pPb, and \PbPb\ collisions, as well as at $\epem$ colliders, are presented in tables~\ref{tab:ccbarxsQCD}, \ref{tab:ccbarxsEW}, and figure \ref{fig:charmxsvsenergy}. In addition, we show results at the Belle II energy $\sqrt{s_{\epem}}=10.58$ GeV. 

We find that the scale uncertainties at \gls{nnlo} \gls{qcd} are almost twice as large as those at \gls{nlo} \gls{qcd} and are only reduced after including \gls{nlp} Coulomb resummation. The fractional scale uncertainties at \gls{nlo} \gls{qcd}, \gls{nnlo} \gls{qcd}, and \gls{nnlo} \gls{qcd}+\gls{nlp} are approximately $\pm11\%$, $\pm17\%$, and $\pm7\%$, respectively, indicating that scale variation clearly underestimates the true size of higher-order effects. The $K$ factors in table \ref{tab:ccbarxsQCD} show that \gls{nlo} and \gls{nnlo} \gls{qcd} corrections increase the \gls{lo} cross sections by roughly $55\%$ and more than $40\%$, respectively. In contrast, Coulomb corrections reduce the cross sections by around $40\%$, nearly compensating the increase from \gls{nnlo} \gls{qcd} corrections. Analogous to the bottom-quark case discussed in section \ref{sec:bottomres}, the \gls{ew} corrections to the inclusive total cross sections of charm-quark production (table \ref{tab:ccbarxsEW}) are dominated by QED contributions, resulting in a $+0.4\%$ enhancement relative to \gls{lo}. This value is a factor of four larger than the QED corrections in $\gaga\to\bbbar$, consistent with the quark-charge ratio $e_c^2/e_b^2=4$.

\begin{figure}
    \centering
    \includegraphics[width=0.8\textwidth]{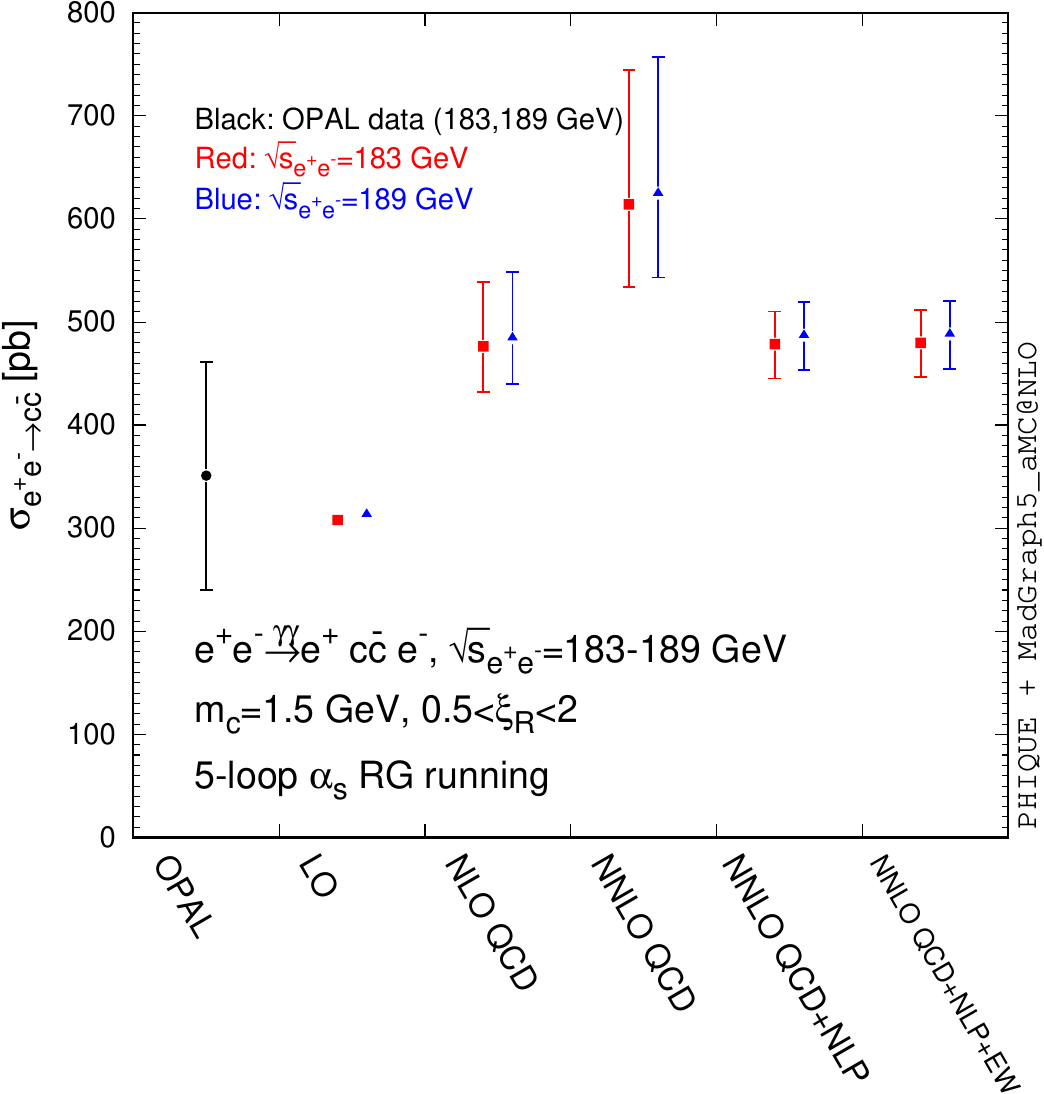}
    \caption{Comparison of the open-charm cross section between the OPAL measurement~\cite{OPAL:1999peo} and perturbative calculations at LEP.}
    \label{fig:charmvsOPALdata}
\end{figure}

Figure \ref{fig:charmvsOPALdata} compares the direct two-photon production cross section of $\gaga\to\ccbar$ measured by the OPAL collaboration at LEP~\cite{OPAL:1999peo} with our theoretical predictions. Using data collected at $\sqrt{s_{\epem}}=183$ and $189$ GeV, OPAL reports
\begin{equation}
\sigma_{\mathrm{OPAL}}(\gaga\to \ccbar)=351 \pm 40~(\mathrm{stat}) \pm 79~(\mathrm{sys}) \pm 66~(\mathrm{extr})~\mathrm{pb} \,,\label{eq:OPALccbarxs}
\end{equation}
where the first, second, and third uncertainties are statistical, systematic, and due to the phase-space extrapolation, respectively. Since the measured cross section combines data at two different \cm\ energies, we show our results for the two energies separately in figure \ref{fig:charmvsOPALdata}. All theoretical predictions, except those at \gls{nnlo} \gls{qcd}, agree with the OPAL data within one standard deviation. 
Our study highlights the importance of including \gls{nnlo} \gls{qcd} and Coulomb corrections for precise predictions of bottom- and charm-quark pair production in two-photon collisions. The effects of soft-gluon threshold resummation in these processes are also of particular interest and could be investigated in a future work.

%\subsection{Physical cross-sections at the \gls{lhc} and FCC-hh \label{sec:physical_cross_sections_UPC}}
%{\color{red} TODO}

%\subsection{Physical cross-sections at $e^+e^-$ colliders\label{sec:physical_cross_sections_ee}}

%{\color{red} TODO}

\clearpage
\newpage
\section{Conclusion\label{sec:conclusion}}

In this paper, we have presented the first \gls{nnlo} \gls{qcd} accurate predictions for the total cross sections proportional to $e_Q^4$ of a heavy-quark pair direct production in photon-photon collisions (\ie\ $\gaga \to \QQ$) at hadron and lepton colliders. This work constitutes the first application of the Local Unitarity approach to this previously unknown \gls{nnlo} \gls{qcd} correction, highlighting its potential as a powerful tool for tackling cutting-edge problems in high-energy physics. In particular, unlike standard approaches, \gls{lu} treats real and virtual contributions on equal footing, circumventing the need for intricate IR-subtraction algorithms and restricting the use of dimensional regularisation to UV renormalisation only. As a result, we could avoid the analytic calculation of the two-loop amplitude for $\gaga \to \QQ$, which involves complicated special functions including those related to elliptic curves, as well as the construction of intricate real-emission phase-space counter-terms typically required for \gls{nnlo} cross-section computations with massive coloured particles. Although \gls{lu} is devised for fully differential distributions, its implementation in $\alphaLoop$ only has limited support for threshold regularisations. For this reason, in this work we present only total cross sections computed with $\alphaLoop$, since threshold singularities are easier to handle in this case (see section \ref{sec:thresinginLU}). Our ongoing efforts for a new implementation of \gls{lu} in $\gammaLoop$~\cite{gammaloop} will lift this limitation.

We have also released a public standalone code, \phique\ (see appendix \ref{sec:phique}), for calculating the total cross sections of $\gaga \to \QQ$ including both \gls{nnlo} \gls{qcd} and Coulomb-resummation effects. Combining it with the \gls{nlo} \gls{ew} corrections from \mgshort\ and integrating over incoming photon fluxes, we obtain the total cross-section for the production of a pair of top, bottom, and charm quarks in ultraperipheral hadron collisions and at $\epem$ colliders. We find strikingly different perturbative behaviours for these three heavy quarks. For top-quark production, perturbative convergence from \gls{nlo} \gls{qcd} to \gls{nnlo} \gls{qcd} is clearly observed, driven by the large mass scale. The \gls{nlo} \gls{ew} corrections are of similar magnitude but opposite in sign to the \gls{nnlo} \gls{qcd} corrections, consistent with the na\"ive expectation from $\alpha \sim \alpha_s^2$, while the Coulomb-resummation effects are marginal. In contrast, for bottom- and charm-quark production, the \gls{nlo} \gls{ew} corrections are negligible, and the positive \gls{nnlo} \gls{qcd} corrections are largely canceled by the negative Coulomb contributions. In the charm-quark case, because of the very low scale and the large value of $\alpha_s$, including the \gls{nnlo} \gls{qcd} corrections even increases the scale uncertainties, which are significantly reduced only after combining with the Coulomb corrections. Our analysis further indicates that choosing a central scale larger than $m_Q$, such as $2m_Q$, improves the perturbative convergence of total cross sections. Cross-section computations at alternative scale choices and different collider setups can be readily performed by the reader using our public code \phique.

\newpage
\begin{acknowledgments}
HSS is grateful to Ekta Chaubey, David d'Enterria, Michael Fucilla, and Guoxing Wang for useful discussions. This work is supported by the European Research Council (grant 101041109, ``BOSON"), the French National Research Agency (grant
ANR-20-CE31-0015, ``PrecisOnium"), and the Swiss National Science Foundation (grant PCEFP2\_203335). Views and opinions expressed are however those of the authors only and do not necessarily reflect those of the European Union or the European Research Council Executive Agency. Neither the European Union nor the granting authority can be held responsible for them.
\end{acknowledgments}

\appendix

\begin{figure}[!bt]
\centering
    \includegraphics[width=0.89\textwidth]{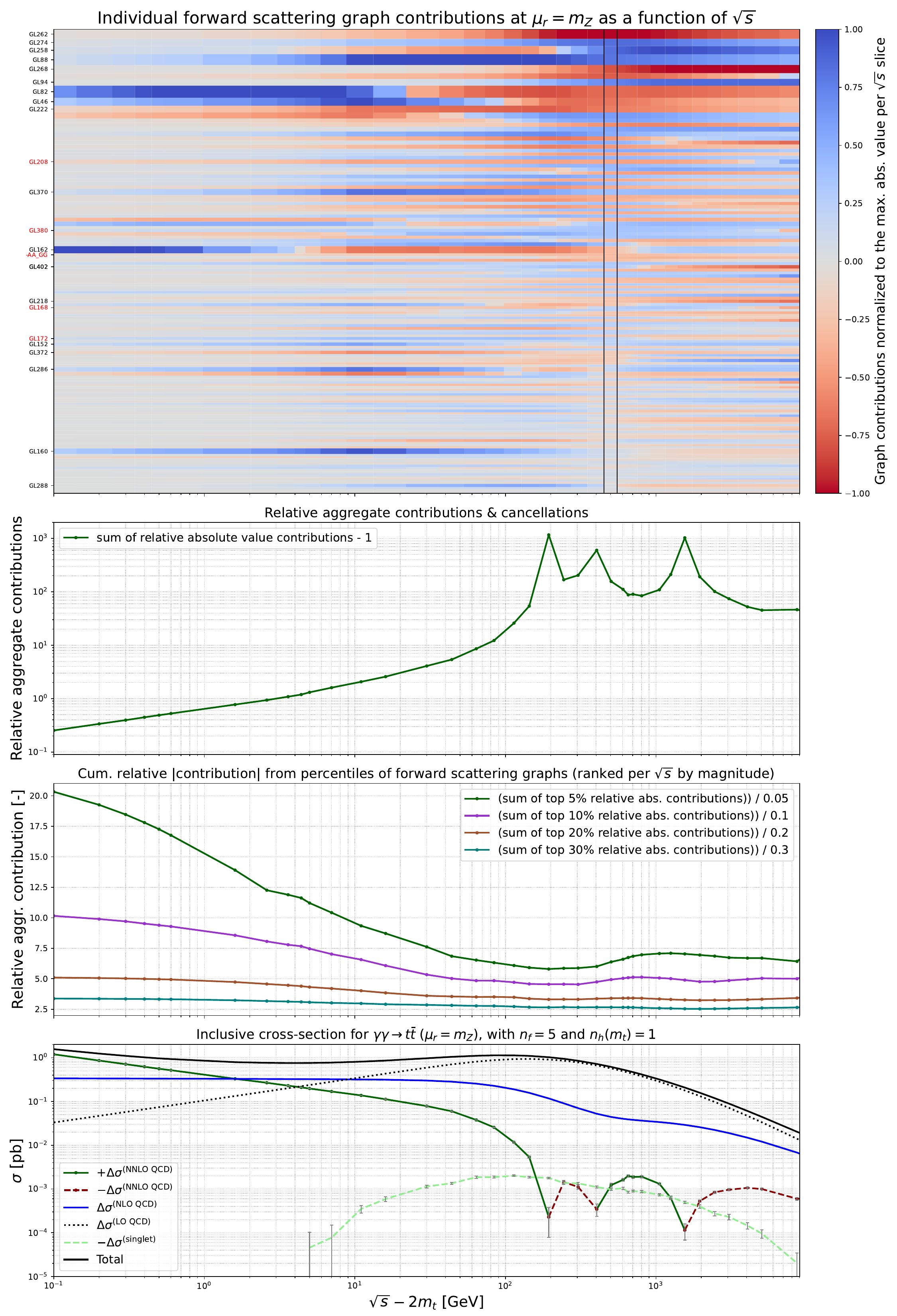}
    \caption{Forward-scattering graph contributions hierarchy. See text in appendix~\ref{sec:graph_distribution} for details.}
    \label{fig:SupergraphHierarchy}
\end{figure}

\section{Details of the Monte Carlo integration\label{sec:graph_distribution}}

\subsection{Relative contributions from individual forward-scattering diagrams}

In figure~\ref{fig:SupergraphHierarchy}, we visualise the breakdown of the total cross section across its contributing graphs, $\hat{\sigma}_{t}^{(2,2)} = \sum_{G\in\Gamma_{\text{proc}}^{\mathrm{\gls{fs}}}} \hat{\sigma}_{t,G}^{(2,2)}$,  as a function of the photon scattering energy in excess of the production threshold: $\sqrt{s} - 2 m_t$ with $m_t=173$ GeV.
We stress that many aspects presented in figure~\ref{fig:SupergraphHierarchy} significantly depend on our choice of the renormalisation scale (set to $\mu_R = m_Z$ for this visualisation) since individual \gls{fs} graphs can receive contributions proportional to $\log^2{(\mu_R^2)}$, even though we have eq.~\eqref{eq:RGeq4sigma22}.
We now first detail and then discuss each inset shown in figure~\ref{fig:SupergraphHierarchy}:
\begin{itemize}
\item The upper panel shows a heatmap with each of the 138 \gls{fs} graphs on the $y$-axis, with a colour scheme representing
\begin{equation}
\mathcal{F}_{t,G}\equiv\frac{\hat{\sigma}^{(2,2)}_{t,G}}{\max_{G\in \Gamma^{\mathrm{\gls{fs}}}_{\text{proc}}}{\left\{\left|\hat{\sigma}^{(2,2)}_{t,G}\right|\right\}}}\,,
\end{equation}
computed individually for each chosen value of the scattering energy so that each column necessarily contains at least one entry maximally blue or red. The graphs are sorted according to $\left|\hat{\sigma}^{(2,2)}_{t,G}\right|$ evaluated at $\sqrt{s} - 2 m_t = 500\;\text{GeV}$, with the corresponding column marked with black sides on the heatmap. The height of each row is also adjusted to enhance the readability of the more relevant graph contributions over the range of energies considered, of which a selection of 18 non-singlet graphs are marked in black on the $y$-axis and shown in figure~\ref{fig:NNLO_other_graphs_selection}. Additionally, the four singlet graphs are marked in red and shown in figure~\ref{fig:NNLO_singlet_graphs}. Note that the row labelled `\texttt{\red{-AA\_GG}}' does not refer to any specific \gls{fs} graph, but instead it captures the contribution from the loop-induced process $\gaga \to g g$, computed using \mgshort~\cite{Alwall:2014hca,Hirschi:2015iia}, which needs to be subtracted from the overall singlet contribution as per the procedure discussed in eq.~\eqref{eq:singlet_unitarity_relation}.

Individual \gls{fs} graph topologies are not gauge-invariant, but they probe different specific kinematic configurations, which makes the study of the hierarchy of their contributions interesting. The ability to compute the finite cross-section contribution stemming from each graph is particular to \gls{lu}, and such a breakdown of the cross section is typically not available.
The first observation is that graphs with the {\it{direct graphs}} topology tend to dominate, especially at low (\eg\ for \texttt{GL82}, \texttt{GL162} and \texttt{GL46}), and high (\eg\ for \texttt{GL268} and \texttt{GL94}) energies. However, the ones with the {\it{crossed graphs}} topology, such as the triple box \texttt{GL160}, can also be relevant but typically only at intermediate scattering energies. In general, this visualisation reveals an intricate structure with non-trivial dependencies on the scattering energy, which is reflected in the complicated shape of $\hat{\sigma}_t^{(2,2)}$ shown in the bottom panel.

\item The first inset highlights the severity of cancellations among \gls{fs} graphs by plotting the quantity 
\begin{equation}
\left|\frac{\sum_{G\in\Gamma^{\mathrm{\gls{fs}}}_{\text{proc}}} \left|\hat{\sigma}^{(2,2)}_{t,G}\right|}{\hat{\sigma}^{(2,2)}_{t}}\right| - 1\,,
\end{equation}
which would evaluate to zero for a sum of contributions with definite sign.

One potential challenge in computing physical cross section through the IR-finite contributions of gauge-dependent individual \gls{fs} graphs is the possibility of large cancellations among graphs. Such an issue occurred in the \gls{lu} computation of the light-by-light scattering~\cite{AH:2023kor} $\gaga \to \gaga$, and could be successfully addressed by projecting out the unphysical degrees of freedom of external photons.
The problem appears to be much milder in the context of $\gaga \to \ttbar$. Indeed, the first inset reveals that cancellation happens at the level of less than two digits, even for $\sqrt{s}-2 m_t \gtrsim 100\;\text{GeV}$, except for when $\hat{\sigma}_t^{(2,2)}$ crosses zero where the absolute value of the error remains small anyway.
At lower values of the scattering energy, the situation is even better since the cross section is dominated by a couple of graphs only, with their sum being mostly coherent.

\item The second inset gives a sense of the distribution of the relative importance in absolute value of \gls{fs} graphs by showing
\begin{equation}
\frac{\sum_{G \in S_p} \left| \hat{\sigma}^{(2,2)}_{t,G} \right |}{p \sum_{G\in \Gamma^{\mathrm{\gls{fs}}}_{\text{proc}}} \left|\hat{\sigma}^{(2,2)}_{t,G}\right|}\,,
\end{equation}
where $S_p$ is the subset of the top $p\%$ of graphs, sorted according to their contributions in absolute value at that scattering energy. It can be interpreted as the multiplier of the absolute contribution of the first $p\%$ graphs compared to what a uniform distribution would yield.

The severity of large cancellations does not inform about the distribution of the magnitude of individual graph contributions. This is however relevant for assessing the benefits of discrete importance sampling over \gls{fs} graph contributions~\footnote{Note that discrete sampling densities are typically adjusted according to the importance of contributions with respect to the Monte-Carlo error, and not the integral directly. However, the variance of graphs scales like the magnitude of its contribution to the cross section, so our comments in this appendix regarding discrete importance sampling remain qualitatively valid.}.
At intermediate scattering energies, we observe that graph contributions are fairly uniformly distributed, with the top $10\%$ of them contributing to only about five times more than what they would in the case of an exactly uniform distribution.
At lower energies, the situation is different, and the cross section is dominated by the two Coulomb-enhanced graphs, \texttt{GL82} and \texttt{GL162}, so that discrete importance sampling greatly speeds up integration in that regime, which happens to also be the region most relevant to the inclusive cross section.

\item Finally, the bottom panel shows the pure \gls{lo}, \gls{nlo} \gls{qcd}, \gls{nnlo} \gls{qcd}, and singlet cross sections for $\mu_R = m_Z$, $n_q = 5$ and $n_h=1$, with $\sqrt{s}-2 m_t$ on the aligned $x$-axis to provide context to the insets above. Negative cross-section contributions are indicated with dashed lines.

At low energies, the pure \gls{nnlo} \gls{qcd} cross section $\hat{\sigma}_t^{(2,2)}$ exhibits the expected Coulomb enhancement discussed in section~\ref{sec:threshold_region}, which makes it the dominant contribution to the fixed-order total cross section.
At scattering energies $\sqrt{s}-2 m_t \;>\; \sim 100\;\text{GeV}$, the pure \gls{nnlo} \gls{qcd} cross section undergoes a sharp transition and becomes significantly smaller than its pure \gls{lo} and \gls{nlo} counterparts, even crossing zero at values of $\sqrt{s}-2 m_t$ around $250\;\text{GeV}$ and $2000\;\text{GeV}$.
For that reason, \gls{nnlo} \gls{qcd} corrections to the physical inclusive cross section are mostly insensitive to this energy regime. We however stress that this pattern is particularly manifest to our choice of $\mu_R = m_Z$, and it may also not hold true when studying differential corners of the top-quark pair production phase-space.
The singlet contribution is at a couple permille level for  $\sqrt{s}-2 m_t \;>\; \sim 200\;\text{GeV}$. It however seems to dominate the pure \gls{nnlo} \gls{qcd} corrections for the choice of $\mu_R = m_Z$ shown, but it is only about 10\% of the latter for the choice of $\mu_R = m_t$.

\end{itemize}

\newpage
\subsection{Timings and usage of computational resources}

Given the novelty of the Local Unitarity computational approach considered for our computation of the \gls{nnlo} \gls{qcd} corrections to $\gaga \rightarrow \QQ$, it is important to detail runtime performance in order to assess its viability as an alternative approach to the traditional (semi-)analytical techniques. It is also useful as a benchmark point for future refined implementation of Local Unitarity in our upcoming code \gammaLoop~\cite{gammaloop} or by other groups.

\begin{figure}[!h]
    \centering
    \includegraphics[trim=0cm 0.8cm 0cm 0.6cm,clip,width=\textwidth]{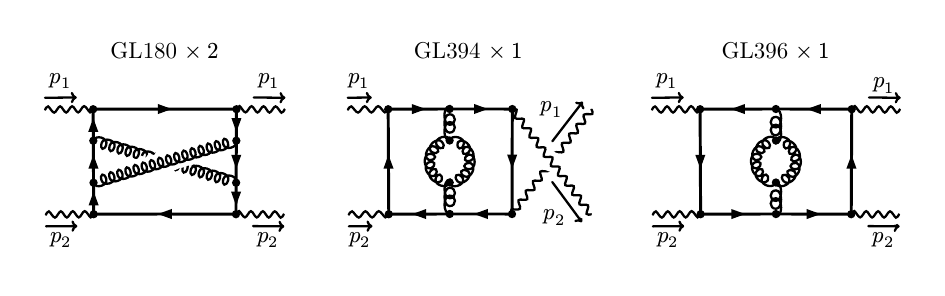}
    \caption{Fastest \gls{nnlo} \gls{qcd} \gls{fs} graph \texttt{GL180} (6.9 $\mu$s), containing a single double real-emission cut, and slowest (715.4 $\mu$s) \gls{fs} graph \texttt{GL396} involving complicated cuts with two-loop amplitudes requiring derivatives in their UV subtraction counter-terms. The \gls{fs} graph \texttt{GL394} is the crossed version, but evaluates faster (430.3 $\mu$s). Multiplicity factors arise from the combination of isomorphic graphs when accounting for the symmetry stemming from swapping initial states as well as complex conjugation symmetry.}
    \label{fig:nnlo_timing_diags}
\end{figure}

We start by discussing the simplest performance figure of merit, which is also the easiest to compare across implementations, namely the evaluation times of individual \gls{fs} graphs presented in table~\ref{tab:LU_timings}. For each perturbative order, we report these timings for the complete list of \gls{fs} graphs sorted by evaluation speed, also indicating their number of contributing Cutkosky cuts.
The timing presented corresponds to the evaluation of the Local Unitarity representation of each \gls{fs} graph, as given in eq.~\eqref{eq:LU_representation}, for a single sample point $\vec{\mathbf{K}}$ specifying both loop and phase-space momenta depending on the Cutkosky cut.
We find a similar average evaluation time of about 5 $\mu$s at both \gls{lo} and \gls{nlo}.
At \gls{nnlo}, results greatly vary across \gls{fs} graphs. 
The fastest \gls{fs} graph \texttt{GL180} evaluates in only 7 $\mu$s because it contains a single Cutkosky cut of double real-emission type (with final-state $\QQ gg$, see figure~\ref{fig:nnlo_timing_diags}), so that it is effectively a simple \gls{lo} type of evaluation. Conversely, the slowest \gls{fs} graph \texttt{GL396} contains multiple Cutkosky cuts involving two-loop amplitudes, which are moreover slow to evaluate because the nested gluonic self-energy it contains involves the complicated UV subtraction counter-terms described in section 4.4 of ref.~\cite{Capatti:2022tit} and the computation of derivative terms (obtained exactly and numerically using dual numbers) as described in sections 2.3, 2.4, and 2.5 of ref.~\cite{Capatti:2022tit}.
We note that the \gls{fs} graphs \texttt{GL394} and \texttt{GL396} are identical, as shown in figure~\ref{fig:nnlo_timing_diags}, except that the former exhibits the \textit{crossed} external ordering discussed in section~\ref{sec:FSGenumerate} and the latter corresponds to the \textit{direct} ordering. There is therefore no good reason that their respective evaluation timings should differ by almost 40\% like table~\ref{tab:LU_timings} reveals. This is due to the fact our generated code (which uses the manifestly causal \gls{ltd} representation of ref.~\cite{Capatti:2020ytd}) and its resulting efficiency can be significantly affected by minor changes in the loop momentum basis selected for generation as well as the resulting linear combination of momenta forming the momentum carried by each edge.
This will no longer be the case in our \gammaLoop~\cite{gammaloop} implementation of Local Unitarity which is based on the Cross-Free Family \gls{ltd} representation of ref.~\cite{Capatti:2022mly}.
In general, the \gls{fs} evaluation time mostly depends on the number of contributing Cutkosky cuts, the number of loops in the amplitudes on each side, the complexity of their local UV subtraction and, when required by the observable considered, the structure of the non-pinched thresholds necessitating regularisation.

When actually performing the numerical integration of the \gls{fs} graphs listed in table~\ref{tab:LU_timings}, many factors affect their translation into the final quantity of interest: the total number of CPU-hours to complete the Monte-Carlo integration of the complete \gls{nnlo} \gls{qcd} corrections:
\newpage
\begin{itemize}

\item First, numerical stability must be assessed and potentially rescued with higher-accuracy arithmetics, following what we describe in section 5.3.3 of ref.~\cite{Capatti:2020xjc}. For this work, we considered a stability test involving two additional rotated samples (resulting in an evaluation three times slower overall), although a single one would have been sufficient. We found that a negligible number of points (far less than one permille in general) were numerical unstable (less that 5 digits accurate). Additional integration overhead beyond the \gls{fs} graph evaluation time is insignificant in our calculation, but it may be relevant in other scenarios, such as when complicated threshold regularisation procedures are necessary or when complicated observables are considered. 

\item Second, we remove integrable singularities arising from prefactors involving on-shell energies of gluons using the multi-channelling procedure over loop momentum basis discussed in section 5.4.1 of ref.~\cite{Capatti:2020xjc}. A curated selection of bases was automatically built for each \gls{fs} graph. The resulting slowdown as part of this procedure stemming from evaluating correlated samples in different bases is only apparent given that these additional correlated samples do retain statistical relevance. Future implementation of \gammaLoop~\cite{gammaloop} will instead consider tropical sampling in momentum space as discussed in ref.~\cite{Borinsky:2025asc} to remove the need for correlated samples and further reduce the variance of the \gls{lu} integrand.
In general, the choice of parameterisation significantly impacts the variance of the integrand. For this work, we limited ourselves to spherical coordinates, with $r = \sqrt{s}\;x_r / (1-x_r)$ and $x_r \in (0,1)$, for each loop momentum of a given basis, and the \texttt{Havana} integrator of \symbolica~\cite{Symbolica}, which implements the original \texttt{VEGAS}~\cite{Lepage:1980dq} algorithm\footnote{We note that we observe almost no benefit from adaptive importance sampling over the input parameter of our spherical parameterisations. This is expected in light of the fact that the features of the \gls{lu} integrands are not expected to be aligned with these integration variables.}, without stratified sampling~\cite{Lepage:2020tgj}.
In future work, we plan to consider machine-learning based integration approaches, such as normalising flows as implemented in, \eg\ \texttt{MadNIS}~\cite{Heimel:2022wyj,Heimel:2023ngj,Heimel:2024wph}.

\item Third, the total integration time typically depends quadratically on the accuracy target. We considered a target of 1\% on the total \gls{nnlo} \gls{qcd} correction. Because this quantity can become zero for our parameter choices and specific values of $\sqrt{s}$ (see figure~\ref{fig:SupergraphHierarchy}) we also enforced a termination condition when the Monte Carlo error falls below 0.005\% of the total cross-section.

\item Finally, a key feature of \gls{lu} is that it can yield \textit{finite} cross-section results for individual \gls{fs} graph (albeit gauge-dependent, and thus unphysical). For this reason, we can use \texttt{Havana} to implement discrete importance sampling over their sum, adjusted so that each \gls{fs} graph ends up contributing equally to the total \textit{error}\footnote{In practice, we consider a safety parameter enforcing a minimum fraction of points to be assigned to each \gls{fs} graph so as to make sure to not undersample any. As a result of this, some very stable \gls{fs} graphs contribute less to the Monte Carlo integration error than others.}. Depending on the hierarchy of \gls{fs} graph contributions (see figure~\ref{fig:SupergraphHierarchy}), this can yield significant speedups. For instance, within less than 1 GeV of the top quark production threshold of $2 m_t$ (the dominant region for the inclusive cross-section computation), more than 60\% of the points are assigned to only two \gls{fs} graphs: \texttt{GL82} and \texttt{GL162}. Their combined evaluation time of 548 $\mu$s is thirty times less than the total evaluation time of 16.2 ms of the complete set of 138 \gls{nnlo} graphs.
This underscores the potential of future work aiming at minimising the number of dominant \gls{fs} graphs by, for example, optimising the gauge choice~\cite{Chen:2022xlg} or sampling efficiently over them~\cite{Borinsky:2025stz}.
\end{itemize}
For all the reasons listed above, it is not particularly relevant to quote exact total integration times and we prefer to only precisely report on the individual \gls{fs} graph evaluation performance in table~\ref{tab:LU_timings}.

Our implementation computes the \gls{nlo} \gls{qcd} corrections to $\gaga \to t\bar{t}$ at $\sqrt{s}=500$ GeV at a relative accuracy of 1\% about 30 seconds and $\sim$1M sample points on one CPU core, whereas \mgshort\ takes about 1 second. It is however clear that we did no optimise \alphaLoop\ for fast \gls{nlo} calculations of simple processes.

Reaching our target error for the \gls{nnlo} \gls{qcd} correction required a similar number of sample points (each applying to a single \gls{fs} graph) across the energy range considered: between 200M and 2B samples (see details in the folder \texttt{local\_unitarity\_raw\_data} of the supplementary materials discussed in appendix~\ref{sec:supplementary_material}). However, the resulting integration time varied significantly between $\sim$ 1000 CPU-hours for individual values of the scattering energy $\sqrt{s}$ close to the production threshold of $2 m_t$, up to 50,000 CPU-hours in the higher energy region of $\sqrt{s} \geq 500~\text{GeV}$.
This implies that if one is only interested in the total inclusive cross-section for $\mathrm{A}_1 \mathrm{A}_2 \overset{\gaga}{\to} \QQ+X$, which is dominated by the threshold region, then the resulting integration times from our Local Unitarity method are particularly competitive.

\begin{table}[htpb!]
\tabcolsep=1.0mm
\centering
\renewcommand{\arraystretch}{0.95}
\makebox[\textwidth][c]{%
\begin{tabular}{lcrlcrlcrlcrlcrlcr}
\toprule

\multicolumn{12}{l}{Timing of the \textbf{2 \gls{lo}} forward-scattering graphs in [ $\mathbf{\mu}$s ]} & 
\multicolumn{6}{r}{Total time: $\mathbf{9.6}$} \\
\midrule
\multicolumn{18}{c}{
Avg. t: $\mathbf{4.8}$\ \ \ 
Min. t: $\mathbf{4.8}$\ \ \ 
Max. t: $\mathbf{4.8}$\ \ \ 
Avg number of cuts: $\mathbf{1.0}$\ \ \ 
Avg t per cut: $\mathbf{4.8}$} \\
%\multicolumn{2}{c}{Avg. t: } & $124.4$ &
%\multicolumn{2}{c}{Min. t: } & $124.4$ &
%\multicolumn{2}{c}{Max. t: } & $124.4$ &
%\multicolumn{3}{c}{Avg |cuts|: } & 3 &
%\multicolumn{3}{c}{Avg t per cut: } & $124.4$ \\
\midrule
\texttt{GL0} & $1$ & $4.8$ | &
\texttt{GL2} & $1$ & $4.8$ |&
& & &
& & & 
& & & 
& & \\
\midrule

\multicolumn{12}{l}{Timing of the \textbf{10 \gls{nlo}} forward-scattering graphs in [ $\mathbf{\mu}$s ]} &
\multicolumn{6}{r}{Total time: $\mathbf{105.5}$} \\
\midrule
\multicolumn{18}{c}{
Avg. t: $\mathbf{10.6}$\ \ \ 
Min. t: $\mathbf{5.7}$\ \ \ 
Max. t: $\mathbf{17.7}$\ \ \ 
Avg number of cuts: $\mathbf{2.0}$\ \ \ 
Avg t per cut: $\mathbf{5.3}$} \\
%\multicolumn{2}{c}{Avg. t: } & $124.4$ &
%\multicolumn{2}{c}{Min. t: } & $124.4$ &
%\multicolumn{2}{c}{Max. t: } & $124.4$ &
%\multicolumn{3}{c}{Avg |cuts|: } & 3 &
%\multicolumn{3}{c}{Avg t per cut: } & $124.4$ \\
\midrule
\texttt{GL16 } & $1 $ & $5.7  $      |&
\texttt{GL18 } & $1 $ & $5.7  $      |&
\texttt{GL4  } & $1 $ & $6.0  $      |&
\texttt{GL6  } & $1 $ & $6.2  $      |&
\texttt{GL20 } & $2 $ & $9.4  $      |&
\texttt{GL22 } & $2 $ & $9.7  $     \\
\texttt{GL0  } & $2 $ & $13.4 $     | &
\texttt{GL2  } & $2 $ & $14.4 $      |&
\texttt{GL8  } & $4 $ & $17.3 $      |&
\texttt{GL10 } & $4 $ & $17.7 $      |&
& & & 
& & \\
\midrule

\multicolumn{12}{l}{Timing of the \textbf{138 \gls{nnlo}} forward-scattering graphs in [ $\mathbf{\mu}$s ]} &
\multicolumn{6}{r}{Total time: $\mathbf{16198.1}$} \\
\midrule
\multicolumn{18}{c}{
Avg. t: $\mathbf{117.4}$\ \ \
Min. t: $\mathbf{7.0}$\ \ \
Max. t: $\mathbf{717.9}$\ \ \
Avg number of cuts: $\mathbf{3.7}$\ \ \
Avg t per cut: $\mathbf{32.0}$} \\
%\multicolumn{2}{c}{Avg. t: } & $124.4$ &
%\multicolumn{2}{c}{Min. t: } & $124.4$ &
%\multicolumn{2}{c}{Max. t: } & $124.4$ &
%\multicolumn{3}{c}{Avg |cuts|: } & 3 &
%\multicolumn{3}{c}{Avg t per cut: } & $124.4$ \\
\midrule
\texttt{GL180} & $1 $ & $6.9  $     | &
\texttt{GL176} & $1 $ & $7.0  $     | &
\texttt{GL178} & $1 $ & $7.1  $     | &
\texttt{GL182} & $1 $ & $7.7  $     | &
\texttt{GL130} & $2 $ & $8.6  $     | &
\texttt{GL94 } & $1 $ & $8.9  $    \\
\texttt{GL132} & $2 $ & $9.8  $     | &
\texttt{GL96 } & $1 $ & $9.9  $     | &
\texttt{GL218} & $2 $ & $11.8 $     | &
\texttt{GL268} & $2 $ & $11.8 $     | &
\texttt{GL148} & $1 $ & $12.0 $     | &
\texttt{GL8  } & $1 $ & $12.9 $    \\
\texttt{GL150} & $1 $ & $12.9 $     | &
\texttt{GL238} & $2 $ & $13.4 $     | &
\texttt{GL270} & $2 $ & $13.7 $     | &
\texttt{GL12 } & $1 $ & $13.8 $     | &
\texttt{GL116} & $1 $ & $14.1 $     | &
\texttt{GL362} & $1 $ & $14.1 $    \\
\texttt{GL282} & $1 $ & $14.8 $     | &
\texttt{GL212} & $3 $ & $14.9 $     | &
\texttt{GL298} & $1 $ & $14.9 $     | &
\texttt{GL118} & $1 $ & $15.0 $     | &
\texttt{GL224} & $3 $ & $15.2 $     | &
\texttt{GL9  } & $1 $ & $15.3 $    \\
\texttt{GL300} & $1 $ & $15.5 $     | &
\texttt{GL13 } & $1 $ & $16.5 $     | &
\texttt{GL284} & $1 $ & $16.7 $     | &
\texttt{GL202} & $4 $ & $17.8 $     | &
\texttt{GL290} & $1 $ & $17.8 $     | &
\texttt{GL156} & $1 $ & $18.3 $    \\
\texttt{GL36 } & $2 $ & $19.3 $     | &
\texttt{GL292} & $1 $ & $20.4 $     | &
\texttt{GL200} & $3 $ & $20.7 $     | &
\texttt{GL230} & $4 $ & $23.1 $     | &
\texttt{GL98 } & $2 $ & $23.4 $     | &
\texttt{GL134} & $3 $ & $24.5 $    \\
\texttt{GL382} & $4 $ & $24.5 $     | &
\texttt{GL374} & $4 $ & $24.8 $     | &
\texttt{GL364} & $1 $ & $26.2 $     | &
\texttt{GL106} & $2 $ & $27.2 $     | &
\texttt{GL220} & $3 $ & $27.5 $     | &
\texttt{GL226} & $4 $ & $27.6 $    \\
\texttt{GL272} & $3 $ & $27.7 $     | &
\texttt{GL188} & $3 $ & $28.2 $     | &
\texttt{GL138} & $3 $ & $28.8 $     | &
\texttt{GL158} & $1 $ & $29.5 $     | &
\texttt{GL380} & $3 $ & $29.7 $     | &
\texttt{GL32 } & $2 $ & $30.0 $    \\
\texttt{GL198} & $4 $ & $30.3 $     | &
\texttt{GL384} & $4 $ & $30.3 $     | &
\texttt{GL376} & $6 $ & $31.1 $     | &
\texttt{GL104} & $2 $ & $31.4 $     | &
\texttt{GL42 } & $2 $ & $31.8 $     | &
\texttt{GL196} & $3 $ & $32.0 $    \\
\texttt{GL256} & $4 $ & $32.7 $     | &
\texttt{GL122} & $5 $ & $34.2 $     | &
\texttt{GL276} & $3 $ & $34.6 $     | &
\texttt{GL244} & $4 $ & $35.1 $     | &
\texttt{GL378} & $6 $ & $35.1 $     | &
\texttt{GL190} & $4 $ & $35.6 $    \\
\texttt{GL136} & $5 $ & $40.1 $     | &
\texttt{GL402} & $2 $ & $41.5 $     | &
\texttt{GL260} & $5 $ & $42.4 $     | &
\texttt{GL80 } & $3 $ & $44.0 $     | &
\texttt{GL100} & $2 $ & $44.2 $     | &
\texttt{GL274} & $5 $ & $55.8 $    \\
\texttt{GL24 } & $2 $ & $64.3 $     | &
\texttt{GL448} & $2 $ & $65.4 $     | &
\texttt{GL38 } & $2 $ & $65.8 $     | &
\texttt{GL424} & $2 $ & $74.4 $     | &
\texttt{GL44 } & $4 $ & $80.5 $     | &
\texttt{GL144} & $5 $ & $85.7 $    \\
\texttt{GL26 } & $2 $ & $88.1 $     | &
\texttt{GL246} & $6 $ & $89.0 $     | &
\texttt{GL426} & $2 $ & $90.8 $     | &
\texttt{GL228} & $6 $ & $92.1 $     | &
\texttt{GL146} & $5 $ & $92.8 $     | &
\texttt{GL450} & $2 $ & $94.5 $    \\
\texttt{GL46 } & $4 $ & $95.9 $     | &
\texttt{GL222} & $6 $ & $98.1 $     | &
\texttt{GL86 } & $3 $ & $107.5$     | &
\texttt{GL204} & $6 $ & $107.5$     | &
\texttt{GL404} & $2 $ & $109.3$     | &
\texttt{GL84 } & $3 $ & $110.2$    \\
\texttt{GL112} & $5 $ & $111.3$     | &
\texttt{GL28 } & $3 $ & $112.7$     | &
\texttt{GL452} & $3 $ & $114.9$     | &
\texttt{GL428} & $3 $ & $115.1$     | &
\texttt{GL4  } & $3 $ & $117.1$     | &
\texttt{GL114} & $5 $ & $119.0$    \\
\texttt{GL0  } & $3 $ & $119.1$     | &
\texttt{GL34 } & $4 $ & $124.5$     | &
\texttt{GL232} & $8 $ & $129.5$     | &
\texttt{GL208} & $5 $ & $132.2$     | &
\texttt{GL206} & $8 $ & $142.2$     | &
\texttt{GL296} & $3 $ & $143.0$    \\
\texttt{GL388} & $7 $ & $143.4$     | &
\texttt{GL370} & $2 $ & $146.3$     | &
\texttt{GL386} & $7 $ & $146.5$     | &
\texttt{GL5  } & $3 $ & $146.9$     | &
\texttt{GL406} & $3 $ & $147.9$     | &
\texttt{GL278} & $3 $ & $149.0$    \\
\texttt{GL1  } & $3 $ & $151.3$     | &
\texttt{GL294} & $3 $ & $157.0$     | &
\texttt{GL280} & $3 $ & $161.2$     | &
\texttt{GL152} & $3 $ & $164.2$     | &
\texttt{GL172} & $7 $ & $169.5$     | &
\texttt{GL168} & $7 $ & $176.9$    \\
\texttt{GL286} & $3 $ & $185.1$     | &
\texttt{GL288} & $3 $ & $186.2$     | &
\texttt{GL124} & $10$ & $196.1$     | &
\texttt{GL160} & $9 $ & $197.1$     | &
\texttt{GL120} & $10$ & $204.1$     | &
\texttt{GL262} & $10$ & $206.2$    \\
\texttt{GL162} & $9 $ & $213.2$     | &
\texttt{GL258} & $10$ & $213.2$     | &
\texttt{GL30 } & $3 $ & $225.5$     | &
\texttt{GL454} & $3 $ & $229.6$     | &
\texttt{GL430} & $3 $ & $230.7$     | &
\texttt{GL102} & $3 $ & $241.5$    \\
\texttt{GL408} & $3 $ & $256.6$     | &
\texttt{GL366} & $4 $ & $257.9$     | &
\texttt{GL88 } & $6 $ & $262.1$     | &
\texttt{GL164} & $11$ & $262.4$     | &
\texttt{GL40 } & $4 $ & $285.6$     | &
\texttt{GL166} & $11$ & $288.4$    \\
\texttt{GL368} & $4 $ & $321.3$     | &
\texttt{GL82 } & $6 $ & $329.9$     | &
\texttt{GL372} & $2 $ & $330.7$     | &
\texttt{GL154} & $3 $ & $346.5$     | &
\texttt{GL440} & $6 $ & $390.1$     | &
\texttt{GL16 } & $6 $ & $393.9$    \\
\texttt{GL416} & $6 $ & $406.7$     | &
\texttt{GL394} & $6 $ & $430.3$     | &
\texttt{GL418} & $6 $ & $585.9$     | &
\texttt{GL442} & $6 $ & $636.3$     | &
\texttt{GL18 } & $6 $ & $648.1$     | &
\texttt{GL396} & $6 $ & $715.4$    \\
\bottomrule
\end{tabular}%
}
\caption{\label{tab:LU_timings}Runtime performance for the evaluation of all \gls{lo}, \gls{nlo} and \gls{nnlo} forward-scattering within Local Unitarity for a single sample point (specifying both external and loop spatial momenta) without stability test and on a single core of a \texttt{AMD EPYC 9754} CPU, at an approximate clock frequency of 3.1 GHz. The three elements in each column are, in order, the graph identifier, the number of contributing Cutkosky cuts it contains and its total evaluation time in $\mu$s.}
\end{table}

\newpage
\section{Supplementary material \label{sec:supplementary_material}}

The Local Unitarity implementation of our computation in $\alpha\text{Loop}$ is not mature enough to be easily reproducible directly from our path-finder public code \alphaLoop~\cite{alphaloopGitHub}. This is the objective of ongoing work on its successor: $\gamma\text{Loop}$~\cite{gammaloop}.
We nonetheless provide the raw data from this study as ancillary material, along with \phique~\cite{phique} (see appendix~\ref{sec:phique}), allowing the reader to independently reproduce the cross-sections presented in this work, as well as additional ones for other collider settings of interest. Our supplementary material consists of the following five top-level folders:
\begin{itemize}
\item "\texttt{figs\_and\_tables\_raw\_data}" contains the raw data from all our tables and figures, in an easily parseable format.

\item "\texttt{MG5aMC\_EW\_generation}" contains the \ttt{bash} scripts, input cards, and other resources to reproduce our computation of the \gls{nlo} \gls{ew} contributions to $\gaga \to \QQ$. 

\item "\texttt{forward\_scattering\_graphs\_contributions\_visualisation}" contains the \\
\texttt{matplotlib} script and raw data for reproducing the visualisation of figure~\ref{fig:SupergraphHierarchy}. Run with "\texttt{python3 plotter.py raw\_data\_with\_singlet.py}", which generates the file \texttt{supergraph\_hierarchy.pdf}.

\item "\texttt{drawings}" contains the drawings of all \gls{lo}, \gls{nlo} and \gls{nnlo} \gls{fs} graphs considered in this work. Moreover, the corresponding source graph files in the "\texttt{dot}" format provide all the necessary information to fully characterise the graphs, including the loop momentum basis selected for the generation of the code and solving of the Cutkosky cut energy Dirac delta function with the causal flow discussed in section~\ref{sec:LU_FS}.

\item "\texttt{local\_unitarity\_raw\_data/}" contains files detailing the raw integration results obtained from our Local Unitarity method for $\gaga \to t\bar{t}$ selected interpolating values of $\sqrt{s}$ in the range $[ 346.002, 9000.0 ]$ GeV.
In particular the file\\
\texttt{forward\_scattering\_graphs\_raw\_results.json} details our integration metrics for each \gls{fs} graph, and the file \texttt{integration\_results\_for\_PHIQUE.txt} provides the combined results that served as input for the interpolating grids built by \phique\ for reproducing the \gls{nnlo} \gls{qcd} cross section for $\gaga \to \QQ$ for any final-state flavour.
Finally, the subfolder \texttt{runtime\_performance} contains data and additional details about the runtime performance presented in table~\ref{tab:LU_timings}.

\end{itemize}

\newpage
\section{PHIQUE: PHoton-Induced heavy QUark pair procEss\label{sec:phique}}

All inclusive total cross sections of heavy-quark pair production in photon fusion reported in this paper are computed with a newly built tool dubbed \phique~\cite{phique}, which takes care of computing and convolving two-photon fluxes from various collider setups, of building and interpolating \gls{nnlo} \gls{qcd} $K$-factor grids from Local Unitarity results at selected values of $\sqrt{s}/m_Q$, and of performing Coulomb resummation. However, \gls{nlo} \gls{ew} corrections are obtained from \mgshort~\cite{Alwall:2014hca,Frederix:2018nkq,Shao:2025bma}, using a separate public branch of the code not yet released (see appendix~\ref{sec:supplementary_material}). 

As a side remark on the \gls{nnlo} \gls{qcd} grids from \gls{lu}, we note that the total cross section $\hat{\sigma}_{Q}^{(2,2)}$ for $\gaga \to \QQ$ at \gls{nnlo} \gls{qcd} accuracy depends only on three scales: $m_Q$, $\sqrt{s}$, and $\mu_R$. Whereas the $\mu_R$ dependence is fully determined by the \gls{rg} equation (cf. eq.~\eqref{eq:RGeq4sigma22}), we performed one-dimensional scan of the integrated cross-section, computing with Local Unitarity in \alphaLoop\ the quantities $\hat{\sigma}_{Q,\mathrm{non}n_q}^{(2,2)}$ and $\hat{\sigma}_{Q,n_q}^{(2,2)}$ (as defined in eq.~\eqref{eq:sigmaQ22intonq}) for 41 different values of $\sqrt{s}/m_Q$ (see appendix~\ref{sec:supplementary_material} for detailed results of this scan). 
These results are given as input to \phique\ which builds interpolating functions to be used for integrating any \gls{nnlo} \gls{qcd} correction to $\gaga \to \QQ$. We validated this approach at \gls{nlo} \gls{qcd} by comparing results from two different interpolating functions built by \phique\ from the two independent scans to verify that no significant interpolation error is introduced.

We now present the main features and usage of \phique, written in {\tt Fortran90}, which we made publicly available through the following git repository:

~~~~~~~~~~~~~~~~~~~~~~\url{https://github.com/alphal00p/phique}\\
which can be cloned using:
\vskip 0.25truecm
\noindent
~~~{\tt git clone https://github.com/alphal00p/phique.git}

\vskip 0.25truecm
\noindent
The code can then be compiled with:
\vskip 0.25truecm
\noindent
~~~{\tt cd phique; ./config}

\vskip 0.25truecm
\noindent
The file {\tt config} is a {\tt bash} script that generates a {\tt makefile}, creates a few new folders, and runs {\tt make} to compile all {\tt Fortran} and {\tt C++} files with the {\tt gfortran} and {\tt gcc} compilers. The third-party \gammaUPC\ code~\cite{Shao:2022cly} has been included in \phique\ in order to enable the calculations in \gls{upc}s. The main code is in the {\tt src/} subdirectory, where the $\alpha_s$ \gls{rg} evolution related files are stripped from the {\tt HOPPET} program~\cite{Salam:2008qg} and are extended to include five-loop \gls{qcd} running. After the compilation, an executable {\tt RunPHIQUE} is produced in the {\tt bin/} subdirectory. A symbolic link to {\tt bin/RunPHIQUE} is also created in main directory.

The use of the program is straightforward as long as the run setup has been implemented in the {\tt input/run.inp} file. It can be simply run via executing
\vskip 0.25truecm
\noindent
~~~{\tt ./RunPHIQUE}

\vskip 0.25truecm

\newpage
\noindent
The integrated cross section will be displayed on the screen and some output files will be collected in the {\tt output/} subdirectory. As a validation of the installation, let us consider the $\gaga\to \ttbar$ process in \pp\ \gls{upc}s at $\sqrtsnn=14$ TeV with the default setup in {\tt input/run.inp}. The total cross section at \gls{nnlo} \gls{qcd}+\gls{nlp} accuracy that \phique\ should report the following upon invoking {\tt ./RunPHIQUE}:
\begin{equation}
\sigma^{(\mathrm{\gls{nnlo}~\gls{qcd}}+\mathrm{\gls{nlp}})}_{\mathrm{pp}\to \ttbar}=3.00694(1)\times 10^{-4}~\mathrm{pb}\,.
\end{equation}

We now discuss the parameters in the {\tt input/run.inp} file. The format of the file is analogous to the {\tt user.inp} file in \helaconia~\cite{Shao:2012iz,Shao:2015vga}. The beam configuration contains the following parameters:
\begin{itemize}
\item {\tt colpar} is an integer that specifies the colliding type, which can be $1$ (\gls{upc} at hadron colliders), $2$ ($e^\pm$-proton two-photon collisions), $3$ ($\epem$ colliders), and $4$ ($\gaga$ colliders).
\item In the case of hadron colliders ({\tt colpar=1}), one needs to specify the atomic mass and atomic numbers of the initial hadrons for the values of {\tt nuclearA\_beam1}, {\tt nuclearA\_beam2}, {\tt nuclearZ\_beam1}, and {\tt nuclearZ\_beam2}.
\item The energies, in units of GeV, of the two beam particles should be decided with the parameters {\tt energy\_beam1} and {\tt energy\_beam2}. If the beam particle is a nuclear, the energy parameter corresponds to the energy per nucleon inside the nuclear.
\end{itemize}
In ultraperipheral hadron collisions ({\tt colpar=1}), one can change the photon flux type through the parameter {\tt \gls{upc}\_photon\_flux}, whose value can be $1$ (ChFF), $2$ (EDFF), and $3$ (iWW only for protons). Furthermore, the bool parameter {\tt use\_MC\_Gluaber} decides whether use Monte Carlo ({\tt T}) or optical ({\tt F}) Glauber modelling for the nuclear thickness $T_{\mathrm{A}}(b)$ and overlap $T_{\mathrm{AB}}(b)$ functions in calculating the soft survival probability. When using the iWW approximation for photon fluxes of $e^\pm$ or protons, the user needs to tell the code the value of $Q_{\mathrm{max}}^2$ through the parameter {\tt q2max}. The perturbative order can be specified through the integer parameters {\tt order} and {\tt coulomb}. The former one can be $0$ (\gls{lo}), $1$ (\gls{nlo} \gls{qcd}), and $2$ (\gls{nnlo} \gls{qcd}), and the latter one can be $0$ (no Coulomb resummation), $1$ (\gls{lp} Coulomb resummation), and $2$ (\gls{nlp} Coulomb resummation). The heavy-quark type and the quark mass $m_Q$ can be determined through parameters {\tt quark} and {\tt qmass}. Additionally, there are the following parameters for the coupling constants:
\begin{itemize}
\item {\tt alphaemm1} is the value of $\alpha(0)$, and {\tt alphasMZ} is the value of $\alpha_s(m_Z)$.
\item {\tt alphas\_nloop} is the parameter for the perturbative order of \gls{qcd} beta function.
\end{itemize}
Other parameters are less important and can be easily understood from the comments in the {\tt input/run.inp} file, which reads:

\begin{minipage}{\textwidth}
\fontsize{10pt}{10pt}\selectfont\vspace{-0.4cm}\hspace{-0.4cm}
\begin{verbatim}
# beam configuration
colpar 1              # colliding particles: 1=hadron-hadron UPC,
                      #  2=electron/positron-proton, 3=e-e+, 
                      # 4=gamma-gamma
energy_beam1 7000d0   # beam 1 energy per nucleon/electron/positron/gamma (GeV)
energy_beam2 7000d0   # beam 2 energy per nucleon/electron/positron/gamma (GeV)
nuclearA_beam1  0     # A of nuclear of beam 1, 0: it is a proton
nuclearA_beam2  0     # A of nuclear of beam 2, 0: it is a proton
nuclearZ_beam1  0     # Z of nuclear of beam 1, 0: it is a proton
nuclearZ_beam2  0     # Z of nuclear of beam 2, 0: it is a proton

# gamma-UPC (2207.03012) options
UPC_photon_flux 1     # only colpar=1. 1: ChFF (2207.03012); 2: EDFF (2207.03012); 
                      # 3: improved Weizsacker-Williams (only for proton beams)
use_MC_Glauber F      # whether or not to use MC-Glauber TAA for the survival probability
                      # when colpar=1 and UPC_photon_flux=1 or 2.
# q2max in electron or proton iWW photon fluxes
q2max 1d0   # the maximal q2 cut for iWW photon flux [in unit of GeV**2]

# order
# 0: LO
# 1: NLO QCD
# 2: NNLO QCD
order 2

# Coulomb resummation
# 0: no Coulomb resummation
# 1: Leading-power Coulomb resummation
# 2: Next-to-leading power Coulomb resummation
coulomb 2

# quark type (4: c; 5: b; 6: t)
quark 6
# quark mass (in unit of GeV)
qmass 172.56d0   # quark mass (pole mass)

# couplings
alphaemm1     137.036d0    # alpha(0)**(-1)
alphasMZ      0.118d0      # alpha_s(MZ) for the QCD corrections
alphas_nloop  5            # how many loops run of alpha_s
                           # in the MSbar scheme (maximal 5)
Scale         0            # renormalisation scale.
                           #  0: fixed scale;
                           #  1: invariant mass of initial-state photons;
FScaleValue   172.56d0     # the scale value in the fixed scale (Scale = 0)
muR_over_ref  1            # the true renorm central scale is muR_over_ref*scale 
                           # or muR_over_ref*FScaleValue (when Scale = 0)
reweight_Scale T           # reweight to get (renorm) scale dependence
rw_RScale_down 0.5d0       # lower bound for renormalisation scale variations
rw_RScale_up   2.0d0       # upper bound for renormalisation scale variations

# event generation setup
nmc      1000000           # number of Monte Carlo points or integrating points

# histogram output
topdrawer_output F         # topdrawer output file (T) or not (F)
gnuplot_output   F         # gnuplot output file (T) or not (F)
root_output      F         # root output file (T) or not (F)
hwu_output       T         # hwu output file (T) or not (F)
\end{verbatim}  
\end{minipage}

\vskip 0.5truecm
\noindent 

\newpage

\bibliographystyle{myutphys}
\bibliography{reference}

\end{document}